\documentclass[twoside,11pt]{article}

\def\BibTeX{{\rm B\kern-.05em{\sc i\kern-.025em b}\kern-.08em
    T\kern-.1667em\lower.7ex\hbox{E}\kern-.125emX}}

\newtheorem{mydef}{Def.}

\DeclareMathAlphabet\mathbfcal{OMS}{cmsy}{b}{n}
\newcommand{\Mod}[1]{\ \mathrm{mod}\ #1}

\usepackage{jmlr2e}
\usepackage{amsmath}

\usepackage{booktabs} 
\usepackage{multirow}
\usepackage{balance} 

\firstpageno{1}

\begin{document}

\title{Accurate and Efficient Suffix Tree Based Privacy-Preserving
       String Matching}

\author{\name Sirintra Vaiwsri \email sirintra.vaiwsri@anu.edu.au \\
  \name Thilina Ranbaduge \email thilina.ranbaduge@anu.edu.au \\
  \name Peter Christen \email peter.christen@anu.edu.au \\
  \name Kee Siong Ng \email keesiong.ng@anu.edu.au \\
  \addr School of Computing, The Australian National University,
  Canberra, Australia}

\editor{}


\maketitle

\begin{abstract}
The task of calculating similarities between strings held by
different organizations without revealing these strings is an
increasingly important problem in areas such as health informatics,
national censuses, genomics, and fraud detection. Most existing
privacy-preserving string comparison functions are either based on
comparing sets of encoded character q-grams, allow only exact
matching of encrypted strings, or they are aimed at long genomic
sequences that have a small alphabet. The set-based
privacy-preserving similarity functions commonly used to compare
name and address strings in the context of privacy-preserving record
linkage do not take the positions of sub-strings into account. As a
result, two very different strings can potentially be considered as
an exact match leading to wrongly linked records. Existing set-based
techniques also cannot identify the length of the longest common
sub-string across two strings.
In this paper we propose a novel approach for accurate and efficient
privacy-preserving string matching based on suffix trees that are
encoded using chained hashing. We incorporate a hashing based
encoding technique upon the encoded suffixes to improve privacy 
against frequency attacks such as those exploiting Benford's law.
Our approach allows various operations to be performed without the
strings to be compared being revealed: the length of the longest
common sub-string, do two strings have the same beginning, middle
or end, and the longest common sub-string similarity between two
strings. These functions allow a more accurate comparison of, for
example, bank account, credit card, or telephone numbers, which
cannot be compared appropriately with existing privacy-preserving
string matching techniques. Our evaluation on several data sets
with different types of strings validates the privacy and accuracy
of our proposed approach.
\end{abstract}



\section{Introduction}

In application domains such as banking, health, bioinformatics, and
national security, it has become an increasingly important aspect in
decision making activities to integrate information from multiple
data sources. Integrating databases can help to identify and link
similar records that correspond to the same entity across different
databases, a task known as \emph{record linkage}~\citep{Chr12}. This
in turn can facilitate efficient and effective data analysis not
possible on an individual database.
%

Increasingly, record linkage needs to be conducted across databases 
held by different organizations~\citep{Vat17}, where the complementary
information held by these organizations can for example help identify
patient groups that are susceptible to certain adverse drug reactions
(linking doctors, hospital, and pharmacy databases), or detect
welfare cheats (linking taxation with employment and social security
databases). However, in many of these applications the databases to
be linked contain private or confidential information which cannot
be shared between the organizations involved in a
linkage~\citep{Vat17}. Similarly, the comparison of genomic data
often raises privacy concern as genome sequences might contain
proprietary information and because such data are highly
confidential in nature~\citep{Shi16}.

\emph{Privacy-preserving record linkage} (PPRL)~\citep{Vat13}
research aims to develop techniques that can link databases that
contain sensitive information without the need of any private or
confidential information to be shared between the organizations
involved in the linkage process. In PPRL, the attribute values of
records are usually encoded in some form before they are being
compared. Any encoding used must ensure that similarities can
still be calculated between encoded values without the need of
sharing the corresponding plain-text attribute values.
PPRL is conducted in such a way that only limited information
about the record pairs classified as matches is revealed to the
participating organizations. The techniques used in PPRL must
guarantee no participating party, nor any external party, can
compromise the privacy of the entities in the databases that are
linked.

Popular techniques to allow privacy-preserving string comparison
are based on converting strings into sets of q-grams (sub-strings
of length $q$ characters) and encoding these sets for example into
Bloom filters~\citep{Sch09}.
Bloom filters are bit arrays where multiple independent hash
functions are used to encode the elements of a set by setting those
bit positions to $1$ that are hit by a hash function. Bloom filters
can be compared using set-based similarity functions such as the
Dice coefficient~\citep{Chr12}. It has been shown that Bloom filter
based PPRL is both efficient and it can achieve accurate linkage
results comparable to non privacy-preserving record
linkage~\citep{Sch09}.

\begin{table}[t!]
  \centering
  \caption{Example string pairs from a real US voter
    database~\citep{Chr14c} that have the same set of bigrams
    (\emph{q\,=\,}2) and therefore Jaccard or Dice similarities of
    1.0 (same strings), but low edit distance
    similarities~\citep{Chr12}.}
  \label{tab:q-gram=example}
  \begin{scriptsize}
  \begin{tabular}{ccccc} \hline\noalign{\smallskip}
    Attribute~& First string & Second string & Bigram set &
      \begin{tabular}[c]{@{}l@{}}Edit dist. \\ similarity
      \end{tabular} \\
      \noalign{\smallskip} \hline \noalign{\smallskip}
    Zipcode & 27828 & 28278 & (27, 28, 78, 82) & 0.600 \\
    First name & amira & ramir & (am, ir, mi, ra) & 0.600 \\
    First name   & geroge & roger & (er, ge, og, ro) & 0.500 \\
    First name   & jeane & jeaneane & (an, ea, je, ne) & 0.625 \\  
    Last name & avera & raver & (av, er, ra, ve) & 0.600  \\
    Last name & einstein & steins & (ei, in, ns, st, te) & 0.500 \\
    Last name & gering & ringer & (er, ge, in, ng, ri) & 0.333 \\ 
      \noalign{\smallskip}\hline
  \end{tabular}
  \end{scriptsize}
\end{table}

\begin{table*}[th!]
  \centering
  \caption{Overview of related privacy-preserving string matching
    techniques, where we show the complexity for encoding and
    matching one string. \emph{l} is the string length,
    $|\Sigma|$ the size of the alphabet, \emph{h} the
    number of hash functions used, \emph{b} the length of a Bloom
    filter or bit array, and \emph{t} the number of hash tables.}
    \label{tab:comparison}
  \begin{scriptsize}
  \begin{tabular}{lccccc} \hline\noalign{\smallskip}
    Methods / Authors & Data type & Match type & Encoding
       & Matching & Application \\
      \noalign{\smallskip} \hline \noalign{\smallskip}
    \textbf{Chained hash encoded} & String &
    Exact & $O(l^2)$ & $O(l \times \log~l)$ & PPRL \\
    ~~~ \textbf{suffix tree} (our work) & \\
    Bloom filter & String & Approx & $O(l \times h) $ & $O(b)$ & PPRL \\
    ~~~ \citep{Sch09} & \\
    Tabulation hashing & String & Approx &
    $O(l \times t \times h)$ & $O(b)$ & PPRL \\
    ~~~ \citep{Smi17} \\
    Bloom filter tree & String &
      Exact & $O(l^2 \times h)$ & $O(l \times \log~l)$ & Cloud
      comp. \\
    ~~~ \citep{Bez15} \\
    Symmetric encrypted suffix
      & String & Exact & $O(l \times b)$ & $O(l \times b)$ & Cloud
    comp. \\
    ~~~ tree \citep{Cha14} \\
    Oblivious RAM suffix array &
      String & Exact & $O(l \times \log~l)$ & $O(l + \log~l)$ &
      Cloud comp. \\
    ~~~ \citep{Moa15} \\
    Burrows-Wheeler transformation &
      Genomes & Exact & $O(l \times \sqrt{l \times |\Sigma|})$ &
      $O(l^2 \times |\Sigma|)$ & Genomics \\
    ~~~ \citep{Shi16} \\
    \noalign{\smallskip}\hline
  \end{tabular}
  \end{scriptsize}
\end{table*}

One drawback of set-based comparisons is however that the sequence
of characters of a string is lost when it is converted into a
q-gram set. As shown in Table~\ref{tab:q-gram=example},
two different strings can result in the same q-gram set and thus
the same encoded Bloom filter, and therefore 
can potentially identify the strings to be the same. This can lead
to falsely matched record pairs because of too high similarities
between rather different string values~\citep{Chr12}.

A second drawback of set-based comparison functions is that
they only allow the calculation of an overall similarity between
two strings. However, identifying the longest common sub-string
between two strings can be crucial in certain applications.
For example, Financial Intelligence Units around the world,
including FinCEN (US), the National Crime Agency (UK), and AUSTRAC
(Australia), 
collect financial information to help identify tax evasion, money
laundering, and terrorism financing. This involves linking records
from different reporting entities such as banks, casinos, and money
remitters such as Western Union, and requires finding matches in a
privacy-preserving way where bank identifiers such as SWIFT/BIC
codes need to be paired with bank account numbers. 
Sub-string matching is crucial because leading zeros are often
omitted, such that `DK54000074491162' would be the same account as
`DK5474491162'.

The likelihood of two different strings sharing the same or a highly
similar q-gram set increases if the size of the alphabet (the number
of unique characters) used to generate the strings becomes smaller,
because less unique q-grams can be generated. Therefore, strings
made from digits only (alphabet of size 10) will more likely result
in increased q-gram set similarities compared to strings that
contain letters (alphabet of size 26).


\smallskip

\textbf{Contributions}:
In this paper we propose a novel approach to privacy-preserving
string matching that is based on secure chained hash encoded suffix
trees. In our approach each input string in a database is first 
converted into a suffix tree and then encoded by the database owner
(DO). These encoded suffix trees are then sent to a linkage unit
(LU)~\citep{Vat17}. The LU compares the encoded suffix trees it
receives from two or more DOs to identify those pairs of trees that
correspond to two strings that have (1) a sub-string of a certain
minimum length in common, (2) a certain minimum similarity, (3)
the same beginning, (4) same middle, or (5) same ending. The LU
however cannot learn the actual input strings. To improve the
privacy against frequency attacks, such as exploiting Benford's
law~\citep{Ben38}, we propose a hash based encoding for each suffix
which does not allow the LU to learn the actual input strings. We
analyze the complexity, accuracy, as well as privacy characteristics
of our approach, and we experimentally evaluate the approach using
several data sets with different string types (only letters,
only digits, and mixed) and compare the approach to Bloom
filter encoding~\citep{Sch09} and tabulation hashing~\citep{Smi17} based
privacy-preserving string matching.


\section{Related Work}
\label{sec:rel_works}


The privacy-preserving comparison of values (such as strings or
numbers) across databases is a common problem for many application
domains, and therefore a variety of techniques and algorithms have
been proposed, as illustrated in Table~\ref{tab:comparison}.

String matching is often used in a PPRL context where encoded values
of quasi-identifying attributes of individuals (such as their names
and addresses) need to be compared across two or more databases to
link records~\citep{Vat17}. Bloom filter (BF) encoding is widely used
in PPRL because it is efficient and supports approximate matching of
both strings~\citep{Sch09,Vat17} and numerical
values~\citep{Kar17,Vat16b}.
However, BFs cannot be used to identify longest common sub-strings,
because they require values to be converted into q-gram sets whereby
positional information is lost. Furthermore, the hashing functions
used in BF encoding likely lead to collisions (several q-grams hashed
to the same bit position) and therefore the similarities between
BFs are approximations and can be higher than the actual similarity
between their corresponding q-gram sets, as we experimentally show
in Sect.~\ref{sec:experiments}.

 
Privacy-preserving matching of sequences is increasingly required
in bioinformatics applications where the aim is to find the longest
matching sub-sequences for a query sequence in large genome
databases~\citep{Shi16,Wan14}. The algorithms used in such
applications often have high computational complexities. \citet{Shi16}
recently proposed an approach for searching
similar string patterns in a genome database. The approach uses a
recursive oblivious transfer protocol based on additive homomorphic
encryption to query sequences in the genome database while ensuring
each query does not lead to the identification of other similar
strings in the database. However, this approach does not scale to
queries of longer sequences because they incur high computational
and communication costs due to the complex cryptographic functions
used.

Suffix trees~\citep{Mcc76} are often used in bioinformatics
to search for patterns in genome or protein
sequences~\citep{Wan16}. A suffix tree allows searching for a given
pattern with a linear complexity in terms of the length of the
query string being searched.
\citet{Ukk93} showed how suffix trees can be used for string
matching, however his approach requires more space to hold a suffix
tree than the original string collection.
\citet{Cha03b} proposed pruning techniques to reduce the
size of suffix trees generated from large string databases. Their
approach aims to improve the querying of strings by pruning
infrequent sub-string patterns and duplicate paths in a tree.
However, pruning shorter sub-strings results in some string patterns
not being matched. Similarly, \citet{Pat13} proposed a
method that combines length and position filtering techniques for
pruning suffix trees and inverted lists of q-grams which results
in a reduction of the query time. 

\citet{Kim13} proposed a string matching approach
based on suffix and longest common prefix arrays of q-grams. In
their approach, sub-strings in the database are extracted, where
sub-strings with frequencies higher than a given threshold and of
a minimum length are used as indexes for sub-strings matching. The
processing time of this approach crucially depends upon the
frequency and length threshold parameters used, where longer
minimum string length will reduce the success of sub-string matching.

\citet{Bab08} proposed two algorithms that
use suffix arrays combined with longest common prefix arrays to
facilitate longest common sub-string searching in suffix trees.
These algorithms merge the two strings to be compared using a
special character (\$) and employ either a sliding window or tree
based approach over the sorted arrays, achieving a linear time
complexity in the lengths of the two strings being compared.

\citet{Wan16} recently proposed a string matching
protocol based on suffix trees and edit distance constraints. This
approach finds all similar sub-strings for a given query in a 
collection of strings, such that their edit distance with the query
is within a given threshold. To improve the efficiency of suffix
tree generation the approach employs the Burrows-Wheeler
Transformation (BWT) to index the string collection. Query strings are
first partitioned into segments where each segment is queried to
find exactly matching sub-strings to generate a group of candidate
strings. Due to the partitioning of query strings some segments
can however result in higher edit distances which potentially can
lead to missed matching strings. 


A suffix-tree based method to find the shortest unique sub-string
query for constant time online applications was proposed by
\citet{Pei13}. They employed suffix trees as they can be
used to get left-bound shortest unique sub-strings in constant time
which helps to improve the efficiency of online query application.


The use of suffix trees in privacy-preserving sub-string matching
has been investigated by \citet{Cha14}. Their
proposed approach constructs a queryable encryption scheme for
finding all occurrences of a query string in a long encrypted
string stored on a server. The approach uses symmetric encryption
over a generated suffix tree to identify all matching sub-string
patterns.
However, this approach reveals information about user queries to
the server which compromises the privacy of a user's data.
\citet{Moa15} investigated the applicability of oblivious
suffix tree search over encrypted string data. Their approach
provides privacy on the user search patterns from the server but it
incurs large communication overhead for each query. 


\citet{Bez15} proposed a protocol based on a
pattern aware secure search tree where each tree node contains a
Bloom filter that encodes a set of the encrypted strings. The
approach is aimed at cloud environments for two parties to compare
strings securely, where the parties only learn if their strings are
matched but not the actual matching sub-strings. This approach
therefore does not allow the privacy-preserving identification of
longest common sub-strings.

The approaches discussed above mostly allow a user to query a
database of strings or sequences for similar patterns, while the
problem we aim to address involves the identification of similar
sub-strings in two databases owned by different parties without
each party having to reveal their input strings.
In contrast to most existing techniques, our approach allows the
efficient and accurate privacy-preserving comparison of strings
from two databases to identify those string pairs that share a
sub-string with a certain minimum length.


\section{Privacy-Preserving Suffix Tree Matching}
\label{sec:overview}

\begin{figure}[!t]
  \centering
  \includegraphics[width=0.69\textwidth]{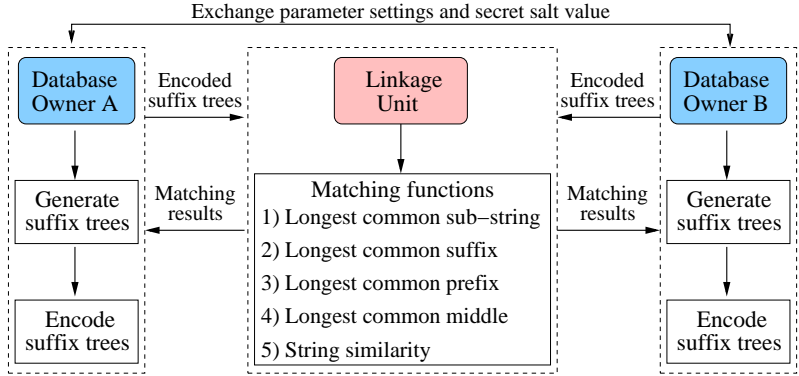}
  \caption{Overview of our proposed secure chained hash encoded
    suffix tree based privacy-preserving string matching protocol.}
    \label{fig:pp-sub-string-protocol}
\end{figure}

As outlined in Fig.~\ref{fig:pp-sub-string-protocol}, we now
describe our protocol to match strings across two databases in a
privacy-preserving way using encoded suffix trees. We assume two
database owners (DOs), each having a database of sensitive private
string values they want to compare with each other without revealing
their actual strings. As in common with many other PPRL
approaches~\citep{Vat17}, our protocol makes use of a linkage unit
(LU), a third party that will conduct the comparison of strings as
converted into encoded suffix trees by the two DOs. As we discuss
in more detail in the privacy analysis in Sect.~\ref{sec:privacy},
we assume the DOs and the LU are semi-honest and follow the
honest-but-curious (HBC) adversary model without any
collusion~\citep{Lin09}. We now define the problem we aim to solve
more formally: 

\begin{mydef}{\textbf{Privacy-preserving string matching:}}
Without loss of generality, we assume two DOs with their respective
databases, $\mathbf{D}_A$ and $\mathbf{D}_B$, that wish to identify,
through the use of a LU, all pairs of matching strings $(s_1,s_2)$,
with $s_1 \in \mathbf{D}_A$ and $s_2 \in \mathbf{D}_B$, such that
$lcs(s_1,s_2) \ge m$, where $lcs()$ is a function that returns the
longest common sub-string, and $m \ge 1$ is the minimum length
required of a matching sub-string for $s_1$ and $s_2$ to be included
in the set of matching string pairs. The two DOs do not wish to
reveal their actual strings with each other nor with any other
party, and the only information the LU can learn are the lengths
and positions of the matching sub-strings but not their actual
characters.\label{def:ppsm}
\end{mydef}

As we describe in Sect.~\ref{sec:matching}, our encoding approach
can also identify if two strings have the same beginning, middle,
or end.

For the remainder of this paper we use the following notation. We
assume all strings $s$ are sequences of characters from a given
alphabet $\Sigma$, such as digits, letters, special characters, or a
mix of them, where $s = \Sigma^*$ is a string of arbitrary length
and $l = |s|$ is the length of a string. We use $\$$ to denote the
special terminal character that indicates the end of a string, where
$\$ \notin \Sigma$ and \$ is not included in the length of a
string (for example, $|123\$| = 3$). Each string $s_1 \in
\mathbf{D}_A$ and $s_2 \in \mathbf{D}_B$ is then converted into one
suffix tree, $\mathcal{T}_{s_1}$ and $\mathcal{T}_{s_2}$,
respectively, as we describe below.

To encode the sub-strings in all edges of a suffix tree
$\mathcal{T}_{s_1}$, we use a secure hash function, denoted by $h()$,
such as SHA256~\citep{Sch96}, resulting in a corresponding encoded
tree $\mathcal{T}_{s_1}^e$. We use a secret salt value, $r$, that is
only known to the DOs, for all encodings to prevent dictionary
attacks by the LU.
We next describe how we generate and encode suffix trees, in
Sect.~\ref{sec:first_encoding} propose a method to overcome
frequency attacks by special encoding of the first characters in
suffixes, and in Sect.~\ref{sec:matching} discuss how we calculate
the longest common sub-string, as well as other matching functions,
between encoded suffix trees in a privacy-preserving way.


\subsection{Suffix Tree Construction and Encoding}
\label{sec:encrypted suffix tree}

\begin{figure}[!t]
  \centering
  \includegraphics[width=0.25\textwidth]{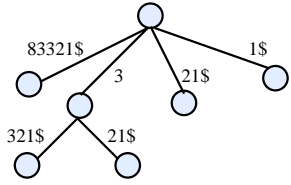}
  \caption{Example suffix tree generated from string `83321',
    where $\$$ is used to indicate the end of each suffix.}
    \label{fig:suffix-tree}
\end{figure}

We follow Ukkonen's algorithm~\citep{Ukk93} to construct one suffix
tree for each string $s_1 \in \mathbf{D}_A$ and $s_2 \in
\mathbf{D}_B$. As an example, Fig.~\ref{fig:suffix-tree} shows the
suffix tree generated from string `83321'. Note that we do not
store the terminal character $\$$ in any edges of a suffix tree.

Following Def.~\ref{def:ppsm}, if one is only interested in matching
sub-strings of minimum length $m > 1$, then only suffixes of length
$m$ and longer need to be included in a suffix tree because suffixes
shorter than $m$ can never be part of a longest common suffix with a
minimum length of $m$. For example, if $m=3$, then the suffixes
`1\$' and `21\$' in Fig.~\ref{fig:suffix-tree} will not be included.

Encoding a suffix tree to allow the calculation of longest common
sub-strings with other suffix trees requires an encoding that allows
privacy-preserving matching of individual characters in a sub-string
without revealing these characters. However, the LU needs to know
which encoded characters are matching at what positions (i.e.\
correspond to the same unencoded character) in order to be able to
identify the longest common sub-string.

Since we assume the LU is semi-honest~\citep{Lin09}, it can attempt
to re-identify the original values that were encoded into the
encoded suffix trees it receives from the DOs. One common approach
to attack such encodings are frequency
attacks~\citep{Chr18a,Kuz11,Nie14}, where frequent encodings are
mapped to frequent plain-text values or frequent q-grams. A
character based encoding, as we require in our approach, will
potentially allow a frequency analysis of hash codes and thus
likely lead to information leakage. 

To overcome such attacks, we propose a chained hash encoding
approach inspired by Blockchain~\citep{merkle80} combined with
salting~\citep{Nie14}. The \emph{salt}, $r$, is a secret string
value agreed by the DOs that they do not share with the LU or any
other party.

Algorithm~1 outlines the steps we use to encode each string in a
database. In line 1, we first initialize two lists, $\mathbf{T}$
and $\mathbf{T}^e$, to store unencoded and encoded suffix trees,
respectively. Next we iterate over each string value $s$ in
database $\mathbf{D}$ in line 2 and use function $\mathit{genSuffixTree}()$
to generate a suffix tree $\mathcal{T}$ for $s$ (line 3). The
function $\mathit{getSuffixes}()$ in line 4 generates a list of
suffixes, $L$, of the suffix tree $\mathcal{T}$. In lines 6 to 14
we encode each character in each suffix $x$ in the list $L$ using
a chained hash encoding method as described next.

We denote the character at position $p$ in a suffix $x$ as $c_p$,
with $1 \le p \le |x|$. Note that these positions are counted within
a suffix (a tree edge) but not within the full string. To encode the
suffix $x= c_1c_2 \ldots c_l$, with $l=|x|$, assigned to an edge in
a suffix tree, we propose the following chained encoding scheme to
generate the encoded suffix $E=[e_1, e_2, \ldots, e_l]$:
\begin{align}
&e_1 = encode(c_1, r) = h(c_1 + r), \\ 
&e_p = encode(c_p, e_{p-1}, r) 
 = h(c_p + e_{p-1} + r), ~~ p > 1,
\end{align}

\begin{figure}[!t]
  \centering
  \includegraphics[width=0.55\textwidth]{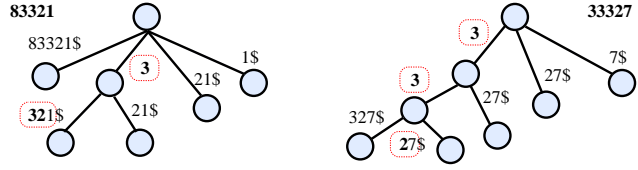}
  \caption{The suffix trees for two strings where their longest
    common sub-string `332' is highlighted in bold and red circles.
    The corresponding chained encodings of these two suffixes
    (paths) are described in
    Sect.~\ref{sec:encrypted suffix tree}.}
  \label{fig:encoded-suffix}
\end{figure}

\begin{figure}[ht]
\begin{center}
  \label{algo:basic-encoding}
  \begin{footnotesize}
  \begin{tabular}{ll} \hline \noalign{\smallskip}
  \multicolumn{2}{l}{\textbf{Algorithm~1: \emph{Basic Encoding of
    Suffix Trees}}} \\ 
  \noalign{\smallskip} \hline \noalign{\smallskip}
  \multicolumn{2}{l}{Input:} \\
  \multicolumn{2}{l}{- $\mathbf{D}$: \hspace{1.1mm}A database of strings} \\
  \multicolumn{2}{l}{- $m$: \hspace{1mm}Minimum suffix length} \\
  \multicolumn{2}{l}{- $r$: \hspace{2mm}Secret salt value} \\
  \multicolumn{2}{l}{- $h()$: Hash function} \\
  \multicolumn{2}{l}{Output:} \\
  \multicolumn{2}{l}{- $\mathbf{T}$:
    \hspace{2mm}List of suffix trees} \\
  \multicolumn{2}{l}{- $\mathbf{T}^e$: \hspace{0.1mm}
    List of encoded suffix trees} \\ \noalign{\smallskip}
  1:  & $\mathbf{\mathbf{T}} = []$, $\mathbf{T}^e  = []$
        \hspace*{\fill} // Initialize the lists of suffix and encoded
        suffix trees \\
  2:  & \textbf{for} $s \in \mathbf{D}$ \textbf{do:}
        \hspace*{\fill} // Loop over all strings in the database \\
  3:  & \hspace{2mm} $\mathcal{T} = \mathit{genSuffixTree}(s)$
        \hspace*{\fill} // Generate the suffix tree for the string \\
  4:  & \hspace{2mm} $L = \mathit{getSuffixes}(\mathcal{T})$
        \hspace*{\fill} // Get the list of suffix values \\
  5:  & \hspace{2mm} $L^e = []$ 
		\hspace*{\fill} // Initialize a list to keep encoded
		suffixes \\
  6:  & \hspace{2mm} \textbf{for} $x \in L$ \textbf{do:} 
        \hspace*{\fill} // Loop over all suffixes \\
  7:  & \hspace{4mm} \textbf{if} $|x| \ge m$ \textbf{do}:
        \hspace*{\fill} // Check if suffix is long enough \\
  8:  & \hspace{6mm} E = [] \hspace*{\fill} // Initialize the list
        of encodings for this suffix \\
  9:  & \hspace{6mm} \textbf{for} $p \in [1, |x|]$ \textbf{do:}
        \hspace*{\fill} // Loop over all characters in the suffix \\
  10: & \hspace{8mm} \textbf{if} $p = 1$ \textbf{do}: \\
  11: & \hspace{10mm} $e_p = h(c_1 + r)$ 
		\hspace*{\fill} // Encode the first character with salt \\
  12: & \hspace{8mm} \textbf{else}: \\
  13: & \hspace{10mm} $e_p = h(c_p + e_{p - 1} + r)$
        \hspace*{\fill} // Chained hash encoding with salt \\
  14: & \hspace{8mm} $E.append(e_p)$ 
        \hspace*{\fill} // Append encoding to encoded suffix \\
  15: & \hspace{6mm} $L^e.add(E)$ 
		\hspace*{\fill} // Add encoded
        suffix to the list of encoded suffixes \\
  16: & \hspace{2mm} $\mathcal{T}^e = \mathit{genEncSuffixTree}(L^e,
        \mathcal{T})$ 
        \hspace*{\fill} // Generate an encoded suffix
        tree \\
  17: & \hspace{2mm} $\mathbf{T}^e.add(\mathcal{T}^e)$
        \hspace*{\fill} // Add encoded tree to list of encoded
        suffix trees \\
  18: & \hspace{2mm} $\mathbf{T}.add(\mathcal{T})$
        \hspace*{\fill} // Add unencoded tree to list of suffix
        trees \\
  19: & return $\mathbf{T}$, $\mathbf{T}^e$ \\
      \hline
  \end{tabular}
  \end{footnotesize}
\end{center}
\end{figure}

\begin{figure*}[!ht]
  \centering
  \includegraphics[width=0.99\textwidth]{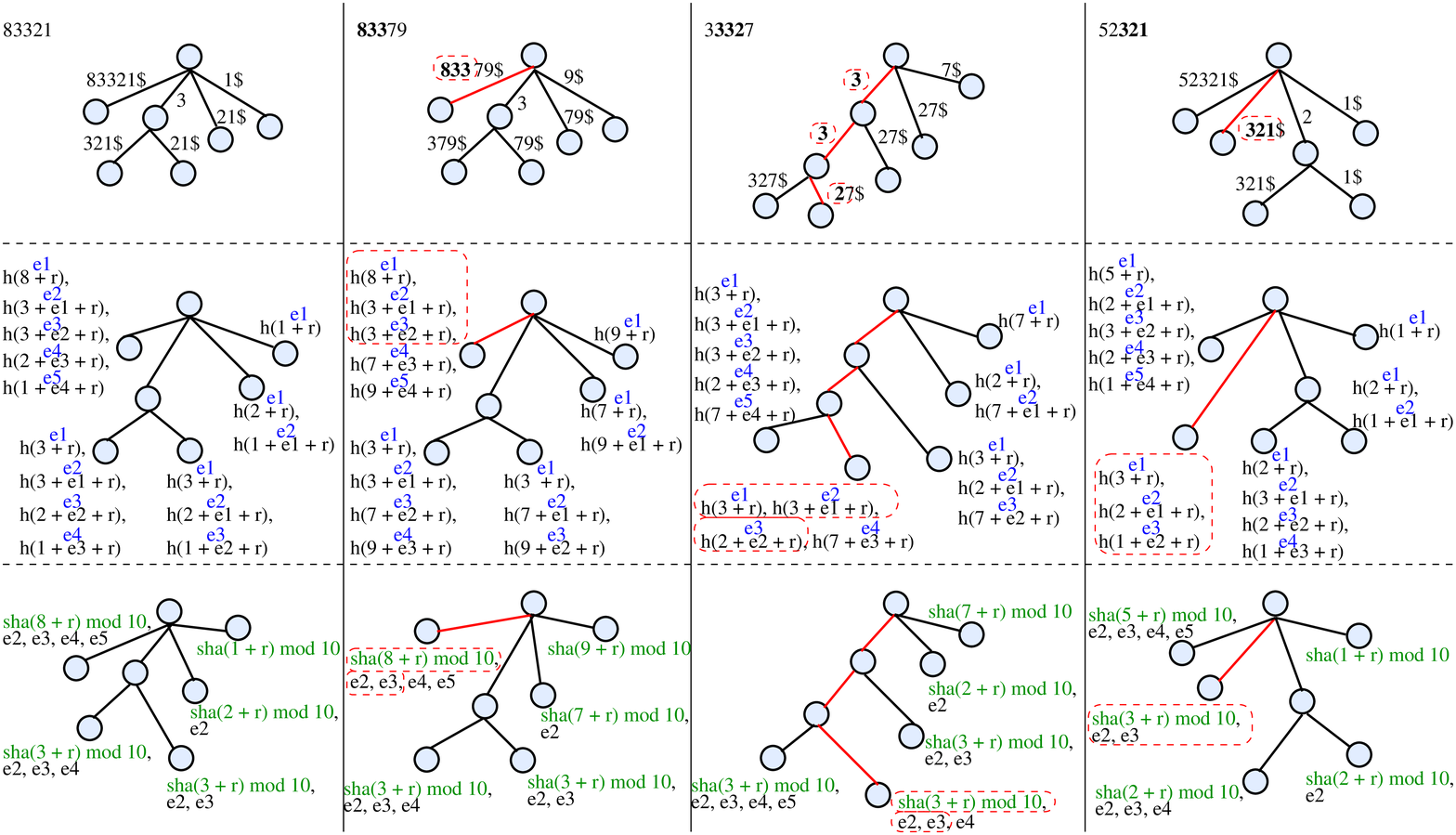}
  \caption{Examples of three matching string pairs where the top row
  shows the original strings and their suffix trees, the middle row
  shows the basic encodings from
  Sect.~\ref{sec:encrypted suffix tree}, and the bottom row shows the
  first character encoding described in Sect.~\ref{sec:first_encoding}.
  The red circles and paths show the matching sub-strings, where
  the second column shows matching beginnings, the third column
  matching middles, and the last column matching ends. In the third
  row the first character encodings (which replace the basic
  encodings for the first characters in all suffixes) are shown in
  green, and \emph{r} denotes the salt value.}
  \label{fig:encoded-matches}
\end{figure*}

\noindent
where $+$ indicates the string concatenation operation, $r$ is the
secret salt value (known only to the DOs but not the LU), and
$h()$ is a hash function from the SHA family~\citep{Sch96}. To
generalize the encoding function for a suffix tree, we encode the
sub-string on each edge as above, but using the last encoded
character in its parent edge (if one exists) as the salt for the
first character, unless the edge has no parent, in which case we use
the original salt. Each edge in $\mathcal{T}$ therefore leads to one
or more hash encodings which are added in a list $L^e$ of encoded
suffixes. 
%
%

For example, for the two strings illustrated in
Fig.~\ref{fig:encoded-suffix}, their highlighted longest common
sub-string `332' when using $r=$~`z' as the secret salt value, will
be encoded as:
\smallskip

\noindent(1) For string `8\textbf{332}1':

~{\bf [}$h$(`3z\text{'}){\bf ]},
~{\bf [}$h$(`3\text{'}+$h$(`3z\text{'})+`z'),
 $h$(`2\text{'}+$h$(`3\text{'}+$h$(`3z\text{'})+`z')+`z'), \\
~\indent{}\hspace*{8mm}~$h$(`1\text{'}+$h$(`2\text{'}+$h$(`3\text{'}+$h$(`3z\text{'})+`z')+`z')+`z'){\bf ]} = 

~{\bf [$h$(`3z\text{'})]}, {\bf [$h$(`3\text{'}+e$_1$+`z')},
{\bf $h$(`2\text{'}+e$_2$+`z')}, $h$(`1\text{'}+e$_3$+`z'){\bf ]}
\smallskip

\noindent(2) For string `3\textbf{332}7':

~{\bf [}$h$(`3z\text{'}){\bf ]},
~{\bf [}$h$(`3\text{'}+$h$(`3z\text{'})+`z'){\bf ]},
~{\bf [}$h$(`2\text{'}+$h$(`3\text{'}+$h$(`3z\text{'})+`z')+`z'),\\
~\indent{}\hspace*{8mm}~h(`7\text{'}+$h$(`2\text{'}+$h$(`3\text{'}+$h$(`3z\text{'})+`z')+`z')+`z'){\bf ]} = 

~{\bf [$h$(`3z\text{'})]}, {\bf [$h$(`3\text{'}+e$_1$+`z')]},
{\bf [$h$(`2\text{'}+e$_2$+`z')}, $h$(`7\text{'}+e$_3$+`z'){\bf ]}
\smallskip


As can be seen from the highlighted bold encodings, these chained
hash encodings allow the privacy-preserving identification of the
longest common sub-string by the LU without it learning what the
characters in the two input strings are.

Back to Algo.~1, in line 16, using $\mathcal{T}$ and the encoded
suffixes in $L^e$, the function $\mathit{genEncSuffixTree}()$
generates an encoded suffix tree, $\mathcal{T}^e$, from
$\mathcal{T}$. Finally, in lines 17 and 18, the generated encoded
and unencoded suffix trees are added to the lists
$\mathbf{T}^e$ and $\mathbf{T}$, respectively. 

%
%

A result of our encoding is that different occurrences of the same
character in a suffix, in fact, across a database, will be assigned
different hash codes depending upon what comes before the character,
thereby making a frequency attack more challenging. In
Fig.~\ref{fig:encoded-suffix}, the same digit in different tree
edges will be encoded differently, such that every encoding in a
tree is unique. This is discussed in detail in the privacy analysis
in Sect.~\ref{sec:privacy}.

%


\subsection{Secure First Character Encoding}
\label{sec:first_encoding}

As we discuss in more detail in Sect.~\ref{sec:privacy}, the
distribution of the first character in values can follow a specific
distribution law, such as Benford's law~\citep{Ben38} for telephone
numbers and Zipf's law~\citep{Zip49} for surnames. This potentially
allows the LU to analyze if the first character encodings of strings
follow a specific distribution law which would allow the
identification of corresponding plain-text characters.
To prevent such frequency-based attacks, we apply an extra encoding
to the first characters of every suffix (path) in a suffix tree. Our
first character encoding aims to make the frequency distribution of
the encodings of the first characters close to a uniform
distribution.

\begin{figure}[t]
\begin{center}
  \label{algo:first-char-encoding}
  \begin{footnotesize}
  \begin{tabular}{ll} \hline \noalign{\smallskip}
  \multicolumn{2}{l}{\textbf{Algorithm~2: \emph{Secure First
    Character Encoding}}} \\ 
  \noalign{\smallskip} \hline \noalign{\smallskip}
  \multicolumn{2}{l}{Input:} \\
  \multicolumn{2}{l}{- $\mathbf{T}$: \hspace{2.1mm}List of
    suffix trees} \\
  \multicolumn{2}{l}{- $\mathbf{T}^e$:
    \hspace{0.7mm}List of encoded suffix trees} \\
  \multicolumn{2}{l}{- $k$: \hspace{2.2mm}Number of characters to
    use to re-encode first character} \\
  \multicolumn{2}{l}{- $r$: \hspace{2.4mm}Secret salt value} \\
  \multicolumn{2}{l}{- $n$: \hspace{2.3mm}Modulo value for
    encoding} \\
  \multicolumn{2}{l}{- $h()$: \hspace{0.09mm} Hash function} \\
  \multicolumn{2}{l}{Output:} \\
  \multicolumn{2}{l}{- $\mathbf{T}^f$: \hspace{0.7mm} List of first
    character encoded suffix trees} \\
    \noalign{\smallskip}
  1:  & $\mathbf{T}^f = []$ \hspace*{\fill} // Initialize
        the list of first character encoded suffix trees \\
  2:  & \textbf{for} $\mathcal{T}^e \in \mathbf{T}^e$
        \textbf{do:} \hspace*{\fill} // Loop over all encoded suffix
        trees \\
  3:  & \hspace{2mm} \textbf{for} $E \in \mathcal{T}^e$ \textbf{do:}
        \hspace*{\fill} // Loop over each encoded suffix \\        
  4: & \hspace{4mm} $x = \mathit{getSuffix}(E, \mathcal{T}^e,
       \mathbfcal{T})$ \hspace*{\fill} // Get the corresponding
       original suffix \\ 
  5: & \hspace{4mm} $e'_1 = \mathit{genEncFirstChar}(x,k,r,n,h)$
       \hspace*{\fill} // Get first character encoding \\ 
  6: & \hspace{4mm} $\mathcal{T}^e.\mathit{replace}(e'_1, E)$
       \hspace*{\fill} // Replace the original first character
       encoding \\
  7: & \hspace{2mm} $\mathbf{T}^f.add(\mathcal{T}^e)$ \\
  8: & return $\mathbf{T}^f$ \\
    \hline
  \end{tabular}
  \end{footnotesize}
\end{center}
\end{figure}

Before the DOs apply the first character encoding to each suffix in
their encoded suffix trees, each DO independently conducts a
frequency analysis on the existing encodings of the first characters
of each value in its database. As we describe 
in Sect.~\ref{sec:privacy}, if these existing encodings
of the first characters follow a uniform distribution in the
strings in the two databases that are to be matched, then the LU
will have no frequency information that it can exploit. In this case
our first character encoding technique is not required.

However, if this frequency analysis shows the encodings of the first
characters follow for example a Benford~\citep{Ben38} or Zipf
distribution~\citep{Zip49}, then the DOs would agree to apply the
secure first character encoding we describe next to each of their
suffixes. Algorithm~2 outlines the steps involved in our first
character encoding technique that will result in a new frequency
distribution of encodings that is closer to uniform and very
different from the original distribution, as we experimentally
validate in Sect.~\ref{sec:experiments}.

Prior to using Algo.~2, the DOs need to agree on $k > 1$, the
number of characters to use in the re-encoding of the first
character, and the secret salt $r$ known only to them. This salt
value can possibly be different from the one used in Algo.~1.
Further, the DOs need to agree on the hash function $h()$ to be
used in the encoding scheme, and the number of unique first
encodings to be generated, $n$, where $|\Sigma| \le n <
|\Sigma|^k$. We discuss the choice of $n$ in more detail in the
accuracy analysis in Sect.~\ref{sec:accuracy_analysis}.
%
%

In line 1 in Algo.~2, each DO initializes the list of first
character encoded suffix trees, $\mathbf{T}^f$. Next, in line 2,
the algorithm iterates over each encoded suffix tree
$\mathcal{T}^e$ generated using our basic encoding technique as
described in Sect.~\ref{sec:encrypted suffix tree}. In line 3, we
loop over each encoded suffix $E$ in $\mathcal{T}^e$ and get the
corresponding unencoded suffix $x$ of $E$ (line 4). 
In line 5, the function $\mathit{genEncFirstChar}()$ generates a new
encoding, $e'_1$, for the first character in $x$ using:
\begin{equation}\label{eqn:sha_mod_n}
e'_1 = h(x[1:k] + r) \Mod{n}.~~~~ 
\end{equation}

We show in Sect.~\ref{sec:analysis} how this secure first character
encoding approach improves privacy against frequency attacks by the
LU while keeping the accuracy of sequence comparisons. 
%
%
%
%
%
In line 6, the generated first character encoding, $e'_1$, is then
inserted into $\mathcal{T}^e$ by replacing the existing basic
encoding of the first character in an encoded suffix. The rest of
the encoded characters in the suffix stay unchanged.
We illustrate this first character encoding approach in the bottom
row of Fig.~\ref{fig:encoded-matches} for three string pairs with
matching beginning, middle, or ending, respectively.


\subsection{Privacy-Preserving String Matching}
\label{sec:matching}

In this section, we describe how the LU can compute the length of
the longest common sub-string across two encoded suffix trees.
Extensions of the functions to compute the longest common prefix
(beginning), the longest common suffix (ending), and the longest
common middle are also discussed.
These functions work both on normal suffix trees, ${\mathcal T}$,
where edges are sub-strings, as well as encoded suffix trees,
${\mathcal T}^e$, where edges are character based encodings as
discussed before. This is because individual encodings of characters
in an edge are the same if their original characters and their
prefixes were the same.


\smallskip

\noindent
\textbf{Longest Common Sub-string}:
Let $s_1$ and $s_2$ be two strings for which we want to compute
the length of their longest common sub-string, and
${\mathcal T}_{s_1}^e$ and ${\mathcal T}_{s_2}^e$ are their
respective encoded suffix trees. For convenience, we adopt the
usual functional-programming syntax to represent suffix trees. For
example, the suffix tree in Fig.~\ref{fig:suffix-tree} is written
as (where $\epsilon$ is the empty string):
\medskip

\indent({\it Tree} \; $\epsilon$ \; [({\it Node}\; 83321\$),
({\it Tree}\; 3
\;[({\it Node}\; 321\$), ({\it Node}\; 21\$)]), \\
\indent({\it Node}\; 21\$), ({\it Node}\; 1\$)]). 
\medskip

\noindent
We now define a recursive algorithm to compute the length of the
longest common sub-string, ${\it lcs(s_1,s_2)}$, given the suffix
tree representations of $s_1$ and $s_2$. In the following,
${\it size}(s)$ gives the length of string $s$, ${\it prefixes(s)}$
gives the set of all prefixes of string $s$,
${\it lprefix(s_1,s_2)}$ computes the length of the longest common
prefix of $s_1$ and $s_2$, and $s_1|s_2$ removes $s_2$ from the
beginning of $s_1$ (when it exists).
\medskip

{\it lcs} (({\it Node}\;$s_1$), ({\it Node}\;$s_2$)) = {\it
lprefix}($s_1$,$s_2$) 
\medskip

\indent{\it lcs} (({\it Tree} \;$s_1$\;[$i_1$,\ldots, $i_a$]), ({\it
Node}\;$s_2$)) = \\ 
\hspace*{2em}{\bf if} \;$s_2\in$ {\it prefixes($s_1$)}\;
  {\bf then}\; {\it size}($s_2$)\\
\hspace*{2em}{\bf else \;if}\;$s_1 \in$ {\it prefixes ($s_2$)}\;
  {\bf then}\; \\
\hspace*{2em}\indent {\it size} ($s_1$) + $\max$ \{
  {\it lcs}($s_2|s_1$, $i_1$),
$\ldots$, {\it lcs}($s_2|s_1$, $i_a$)  \} \\ 
\hspace*{2em}{\bf else} \; 0 
\medskip

{\it lcs} (({\it Node}\;$s_1$), ({\it Tree} \;$s_2$\;[$i_1$,
  $\ldots$, $i_a$])) = \\ 
  \hspace*{1em}\indent {\it lcs} (({\it Tree} \;$s_2$\;[$i_1$, $\ldots$,
  $i_a$]), ({\it Node}\;$s_1$)) \\ 
\indent{}{\it lcs} (({\it Tree} \;$s_1$\;[$i_1$, $\ldots$, $i_a$]),
  ({\it Tree} \;$s_2$\;[$j_1$, $\ldots$, $j_b$])) = \\ 
\hspace*{2em}{\bf if} \;$s_1 = s_2$ \; {\bf then}\; \\ 
\hspace*{2em}\indent {\it size}($s_1$) + $\max$\{ {\it lcs}($i_1$,
  $j_1$), {\it lcs}($i_1$, $j_2$), $\ldots$, \\
  \hspace*{4em}\indent {\it lcs}($i_a$, $j_{b-1}$),
  {\it lcs}($i_a$, $j_b$)\}\\ 
\hspace*{2em}{\bf else \;if}\;$s_1\in$ {\it prefixes($s_2$)}\;
  {\bf then} \;\\ 
\hspace*{2em}\indent {\it size}($s_1$)\;+ $\max$ \{ {\it lcs}
  (({\it Tree} \;$s_2|s_1$\;[$j_1$, $\ldots$, $j_b$]), $i_1$), $\ldots$,\\
  \hspace*{4em}\indent{\it lcs} (({\it Tree}
  \;$s_2|s_1$\;[$j_1$, $\ldots$, $j_b$]), $i_a$) \}\\ 
\hspace*{2em}{\bf else \;if}\;$s_2\in$ {\it prefixes($s_1$)}\;
  {\bf then} \;\\ 
\hspace*{2em}\indent {\it size}($s_2$)\;+ $\max$\{ {\it lcs}
  (({\it Tree} \;$s_1|s_2$\;[$i_1$, $\ldots$, $i_a$]), $j_1$),
  $\cdots$ ,\\
  \hspace*{4em}\indent{\it lcs} (({\it Tree} \;$s_1|s_2$\;[$i_1$, $\ldots$, $i_a$]), $j_b$) \}\\
\hspace*{2em}{\bf else} \; 0
\medskip

\noindent
\textbf{Longest Common Suffix}:
The problem of determining whether two strings represented by their
encoded suffix trees share a common suffix is straightforward to
compute. In fact, we can do better and compute the length of longest
common suffix of two strings, when one exists, via a simple
modification of the ${\it lcs}()$ function above by replacing the
base case by:
\begin{align*}
{\it lcs}&(({\it Node}\;s_1), ({\it Node}\;s_2)) =
  {\bf if}\;(s_1=s_2)\;{\bf
  then}\;v + {\it size}(s_1) \;{\bf else} \; 0
\end{align*}

\noindent
Here, $v$ is some arbitrary number that is larger than the longest
string in the database, such as $v=$ 999. A pair of encoded suffix
trees have a common suffix if the above modified function takes the
form $v + l$, where $l$ is the length of that longest common suffix.
In particular, if the returned value is less than $v$, then the two
strings do not share a suffix.
\smallskip


\noindent
\textbf{Longest Common Prefix}:
The longest common prefix of two strings represented by encoded
suffix trees can be computed by traversing the longest suffix
(path) in each tree and comparing them encoding by encoding to find
the longest match~\citep{Bab08}.
\smallskip

\noindent
\textbf{Longest Common Middle}:
The problem of finding the longest common middle of two strings
represented by encoded suffix trees can be computed easily using the
above algorithms: the ${\it lcs}()$ function must return a positive
value, and there cannot be a common prefix or a common suffix between
the two encoded suffix trees.
\smallskip

\noindent
\textbf{String Similarity}:
To calculate a similarity between two strings represented by their
encoded suffix trees, we use ${\it lcs}()$ as
described above, and then calculate a normalized similarity,
$sim_{lcs}$, as:
\begin{equation}
sim_{lcs}(s_1,s_2) = \frac{lcs(s_1,s_2)}{max(l_1, l_2)},
\label{eqn:lcs}
\end{equation}
where $l_1 = |s_1|$ and $l_2 = |s_2|$ are the lengths of strings
$s_1$ and $s_2$, respectively. The LU can calculate $l_1$ and $l_2$
from the longest suffixes of the corresponding encoded trees,
$\mathcal{T}_{s_1}^e$ and $\mathcal{T}_{s_2}^e$, respectively.
The similarity is normalized such that $0 \le sim_{lcs} \le 1$,
where $sim_{lcs}=0$ means two strings have no sub-string of at
least length $m$ in common, $sim_{lcs}=1$ means two strings are the
same, and a value of $sim_{lcs}$ means they have a sub-string of at
least $m$ characters in common. Also, it is important to note that 
in the event of using the secure first character encoding scheme upon 
suffix trees we can only calculate $lcs()$ of a certain minimum 
length $k$, where $m \ge k$.


\section{Analysis of Our Protocol}
\label{sec:analysis}

We now analyze our protocol in terms of complexity, accuracy,
and privacy. We assume each database owner (DO) has a database 
$\mathbf{D}$ containing $|\mathbf{D}|$ records each consisting 
of a string $s$, where we assume the average length of these
strings is $l$. We also assume all parties participating in the
protocol are directly connected to each other through a secure
communication channel.


\subsection{Complexity Analysis} 

We calculate the computational complexities for each step of our
protocol shown in Fig.~\ref{fig:pp-sub-string-protocol}.
%
%
As described in Sect.~\ref{sec:encrypted suffix tree}, we use
Ukkonen's algorithm~\citep{Ukk93} to construct the suffix tree for
each string value $s \in \mathbf{D}$ which is of linear complexity
in the length $l = |s|$ of $s$. Hence the generation of a suffix
tree for all string values in $\mathbf{D}$ is of $O(|\mathbf{D}|
\cdot l)$ complexity.
Assuming $l$ suffixes can be generated for each string $s$, there
can be at most $2l-1$ edges in a suffix tree which are (1) the
number of paths leading to the $l$ leaves, plus (2) the number of
edges leading to internal nodes ($\le l-1$). The worst case occurs
when each character of a string is different, such as `12345',
leading to $l$ suffixes, one each of length 1 to $l$, and a total
of $l(l+1)/2$ characters to be encoded.

By assuming each hash operation on a character of $s$ is of $O(1)$
complexity, then the encoding of all paths in a suffix tree is of
$O(l^2)$ worst-case complexity. Hence the overall complexity of
encoding all suffix trees in $\mathbf{D}$ (as well as sending them
to the LU), each with $l$ suffixes, is $O(|\mathbf{D}|\cdot l^2)$.
As detailed in Sect.~\ref{sec:first_encoding}, the first character
encoding is applied on each suffix in all encoded suffix trees
which is of $O(|\mathbf{D}| \cdot l)$ complexity for all strings
in $\mathbf{D}$.

For the matching operations performed by the LU, in
Sect.~\ref{sec:matching} we have provided recursive functions for
computing ${\it lcs()}$ and other related operations. In practice,
these recursive functions can be implemented either as a
breadth-first or a depth-first search algorithm, whichever is more
efficient~\citep{Ukk93}.

The comparisons of encodings (hash values) instead of sub-strings
will add a constant time to their time complexities. Let us assume
two encoded suffix trees $\mathcal{T}^e_{s_1}$ and
$\mathcal{T}^e_{s_2}$ of strings $s_1$ and $s_2$, and each
containing $l$ suffixes, respectively. To check if any of the
suffixes of $s_1$ matches with any suffixes in $s_2$, a naive
approach requires a traversal through each path (suffix) in
$\mathcal{T}^e_{s_1}$ for each path in $\mathcal{T}^e_{s_2}$,
resulting in a complexity of $O(l^2)$.

However, work by Babenko and Starikovskaya~\citep{Bab08} has shown
that the longest common sub-string between two strings can be
calculated in linear time, $O(l)$, when sorted suffix arrays are
used (assuming $O(l \times \log~l)$ for sorting) to efficiently
obtain the longest common prefixes. In our implementation, evaluated
experimentally in Sect.~\ref{sec:experiments}, we employ this
efficient matching approach.


\subsection{Accuracy Analysis}
\label{sec:accuracy_analysis}

We first show that running the $\mathit{lcs}()$ function defined in
Sect.~\ref{sec:matching} on basic encoded suffix trees as
described in Sect.~\ref{sec:encrypted suffix tree} gives the same
result as running $\mathit{lcs}()$ on regular suffix trees with high
probability. To see this, note that all we are doing is replacing
operations like $s_1 = s_2$ and $s_1 \in \mathit{prefixes}(s_2)$ in
$\mathit{lcs}()$ with the corresponding operations on the encoded
characters.
\smallskip

\noindent\textbf{Basic Chained Hash Encoding:}~
We can only get errors in the longest common sub-string algorithm
if there are hash collisions that map different characters to the
same encoded value. In the case when the hash function $h()$ is
SHA256~\citep{Sch96}, for example, the probability of a hash
collision in a set of $w$ strings is approximately $\frac{1}{2}(w /
2^{128})^2$ \citep{upfal2005probability}. The probability of an
incorrect longest common sub-string of length $l$ is thus
upper-bounded by:
\[ 2^{-l} \prod_{i=1}^l \biggl( \frac{|\Sigma|^l}{2^{128}} \biggr)
^2, \]
which decreases rapidly to zero with increasing $l$.
\smallskip

\noindent\textbf{First Character Encoding:}~
Consider next the setting of running the $\mathit{lcs}()$ function
on encoded suffix trees with the first character encoding as 
described in Sect.~\ref{sec:first_encoding}. As before, we can get 
errors in the longest common sub-string computation if there are
hash collisions in the encoded characters. Note that in the
encoding scheme from Sect.~\ref{sec:first_encoding} only the first
character of each suffix is changed while the remaining characters
continue to be encoded in the chained hash approach described in
Sect.~\ref{sec:encrypted suffix tree}. Consider two suffixes 
$x_1x_2\ldots x_{l_1}$ and $y_1y_2 \ldots y_{l_2}$ where $x_1
\neq y_1$. Let us also assume that we use $k=2$ in calling
Algo.~2. For small $n$ (we discuss the choice of $n$ in more detail
below), there is a good chance that when using
Eqn.~(\ref{eqn:sha_mod_n}) it holds:
\[ h(x_1x_2 + r) ~\equiv~ h(y_1y_2 + r) \Mod{n}, \]
resulting in an incorrect match of the encodings of $x_1$ and $y_1$.
There are now
two cases to consider: $x_2 = y_2$ and $x_2 \neq y_2$. In both cases,
the basic encoding of $x_2$ and $y_2$ given by $h(x_2 + h(x_1 + r) +
r)$ and $h(y_2 + h(y_1 + r) + r)$ will not match with high
probability when $h()$ is SHA256, since $x_1 \neq y_1$ in the first
case and $x_2 \neq y_2$ in the second case. The argument holds more
generally for arbitrary $k>1$, which means the computation of the
longest common sub-string of length at least $k$ would be correct
with high probability, with the error (collision) probability
upper-bounded by:
\[ \mathit{hc}(n,|\Sigma|,k) \cdot 2^{-(k-1)} \prod_{i=2}^k
   \biggl( \frac{|\Sigma|^k}{2^{128}} \biggr) ^2, \]
where $\mathit{hc}(n,|\Sigma|,k)$ is the probability of collision
when hashing $|\Sigma|^k$ possible suffixes into $n$ possible
values using Eqn.~(\ref{eqn:sha_mod_n}). For most practical values
of $|\Sigma|$, $k$, and $n$, we have $\mathit{hc}(n,|\Sigma|,k) = 1$.
Nevertheless, the error probability decreases rapidly to zero with
increasing $k$.

How to select the values of $k$ and $n$ used in
Eqn.~(\ref{eqn:sha_mod_n}) depends upon the size of the alphabet,
$|\Sigma|$, from where strings are being generated.

First of all, $k$ must be larger than 1. To see why, assume $k=1$
and consider two cases: $n < |\Sigma|$ and $n \ge |\Sigma|$. In the
first case, multiple input characters will be mapped to the same
first character encoding. This can result in false matches of
encoded suffixes leading to inaccurate similarity results. In the
second case, $n \ge |\Sigma|$, the first character encoding will
generate one hash encoding per input character in $\Sigma$
(assuming no hash collision). The frequency distribution of the
original first characters is therefore preserved in the frequency
distribution of the first character encodings computed using
Eqn.~(\ref{eqn:sha_mod_n}). This will allow the LU to conduct a
frequency attack (as we discuss in more detail below) by mapping
encodings back to characters if the distribution of these
characters follow for example Benford’s Law~\citep{Ben38}. Therefore
setting $k=1$ results in either inaccurate ${\it lcs}()$
calculations or insecure character encodings.

We have thus established the need for $1 < k \le m$, where $m$ is
the minimum length of ${\it lcs}()$ we want to calculate. For any
such $k$, the value of $n$ does not have an effect on the accuracy
of our approach. To see why, consider two strings $s_1$ and $s_2$.
If they agree on the first $k$ characters, then the encoding of the
first $k$ characters for $s_1$ and $s_2$ will be the same
regardless of what $n$ is. If $s_1$ and $s_2$ do not have the
same first $k$ characters, then their hash encodings will disagree
at the first position where $s_1$ and $s_2$ disagree or earlier,
again regardless of what $n$ is. 

Given the choice of $n$ does not affect the accuracy of our
approach, should we simply set $n=1$? The answer is no, and the
reason relates to privacy rather than accuracy. Note that the LU
is not given the value of $k$ in our protocol. If $n$ is too small
compared to $|\Sigma|$, it becomes easy for the LU to guess what
$k$ is, and leakage of that information opens a (small but)
possible door for the LU to employ frequency attacks on the encoded
suffix trees it receives from the DOs. If $n \ge |\Sigma|^k$, the
frequency distribution of the original first $k$ characters are
preserved in the distribution of the first character encodings
computed using Eqn.~(\ref{eqn:sha_mod_n}), again opening a door to
frequency attacks by the LU.
From the above, we can conclude that we should have $|\Sigma| \le n
< |\Sigma|^k$. In practice, we set $n = |\Sigma|$, which we show
empirically to work well in Sect.~\ref{sec:experiments} for a range
of data sets.


\subsection{Privacy Analysis}
\label{sec:privacy}

%

We assume the DOs and the LU follow the honest-but-curious (HBC)
adversary model without any collusion~\citep{Lin09}. The HBC model
is commonly used in other PPRL and private string comparison
protocols~\citep{Vat17} because of its applicability to real
scenarios. In the HBC model each party in a protocol tries to
learn as much as possible about other parties' data based on what
it receives from other parties,  while following the protocol
steps. We next analyze the privacy of our approach in terms of
security against privacy attacks by a DO and the LU.
\smallskip

\noindent\textbf{Frequency Attacks by a DO:}~
We assume the DOs do not collude with the LU. Though each DO agrees
upon the same hash function and secret salt value $r$ in the suffix
tree encoding in Sect.~\ref{sec:encrypted suffix tree}, and the
number of first characters, $k$, in the first character encoding in
Sect.~\ref{sec:first_encoding}, neither of the DOs will learn the
set of plain-text strings of the other DOs. This is because the
encoded suffix trees are not shared between the DOs but only sent
to the LU for comparisons. Hence, a frequency attack by a DO upon
the database of another DO is impossible. 
\smallskip

\noindent\textbf{Dictionary and Frequency Attacks by the LU:}~
Once the DOs send their encoded suffix trees to the LU, the LU
compares pairs of trees to identify possible matching sub-strings
encoded in these trees. The LU can identify the character patterns
based on the encodings in the trees. This includes the number of
hash values that match between two trees and their positions.
However, as described in Sect.~\ref{sec:encrypted suffix tree},
each character in a suffix is encoded individually based on the
previous character's hash value concatenated with the secret salt
$r$. This chained hashing provides strong privacy against
dictionary attacks because the LU cannot attack the encoded suffix
trees by generating its own encoded trees based on a database of
plain-text values without knowing the secret salt $r$ as used
by the DOs to encoding their string databases.

However, when only the basic chained hash encoding described in
Sect.~\ref{sec:encrypted suffix tree} is applied on each suffix,
from the set of all encoded suffix trees it receives the LU can
conduct a frequency analysis on the hash encodings that occur at
certain positions in the suffixes of the encoded trees. From
these learned frequency distributions the LU can try to
re-identify which hash encoding could correspond to a certain
character in the alphabet $\Sigma$, assuming the LU knows the
type of strings encoded in the suffix trees. The success of such
an attack by the LU depends on the frequency distribution of 
characters and the availability of a similar plain-text database
to the LU~\citep{Chr18a}.

From the longest suffixes in all trees the LU can learn the length
distribution of all encoded strings, and therefore guess what type
of information is encoded in these trees. For example, if all trees
encode strings of length 16 then these are likely credit card
numbers, while strings of length 9 could be UK mobile phone
numbers. One way to overcome this leakage of information is for the
DOs to pad their strings with characters that are not part of
the alphabet $\Sigma$ before they are processed, where they need to
make sure each DO has their own set of extra characters to prevent
accidental matches of such added extra characters.

One important aspect of re-identification is however that the LU
needs to be able to identify every character in an encoded string,
because partial identifications might not provide useful information.
A partially identified telephone number of the form `?1??1??2??',
where `?' means the digit is unknown, will unlikely help the
attacking LU to re-identify an individual. This is different from
attacks on names and addresses as conducted on
PPRL~\citep{Chr18a,Nie14}, where even a few identified q-grams can
help re-identify a person (identity disclosure). For example, if an
attacker learns that a name string contains three identified
q-grams, and only one rare name in a database contains these three
q-grams, then the attacker learns both the name and the individual
with that name~\citep{Chr18a}. This is because of the smaller domain
of names and addresses (even in large population databases there
are commonly only a few hundred thousand unique names~\citep{Chr14c})
compared to the much larger domains for example of credit cards
which is in the order of $10^{16}$.


Assuming the LU does have access to a plain-text database with a
highly similar frequency distribution of string values, it can
mount a frequency attack whereby it concentrates on the first
character encoding in a suffix, because these encodings are all
based on the same secret salt value $r$ (lines 11 and 13 in Algo.~1).
If there are distinct frequency patterns in a database of
plain-text strings then these will be reflected in a corresponding
frequency distribution of encodings and potentially allow the
attacker to re-identify certain individual characters in the
encoded trees. 
We discuss the success of such an attack under three scenarios:
\smallskip

\textit{1. Uniformly distributed characters}:
If we assume every character at every position is selected uniformly
random from the alphabet $\Sigma$ with probability $1/|\Sigma|$,
then the LU has no frequency information that can be exploited.
This is because each encoding at the beginning of each suffix of
the encoded suffix trees will occur with the same frequency. In
such an ideal situation our chained hash encoding approach will be
secure from any frequency based attack.
\smallskip

\textit{2. Value distribution follows a specific law}:
For a given encoded suffix tree, the LU can identify the
longest suffix and then the first character in this suffix. The
encoding of this first character in a suffix only depends on its
value and the secret salt value $r$ (unknown to the LU). 
However, in real scenarios the distribution of the first 
character in values usually follows a specific distribution law, 
such as Benford's law~\citep{Ben38} for telephone numbers or
Zipf's law~\citep{Zip49} for surnames. For example, by assuming the
input strings contain digits only then it is possible that the
first digits in these strings follow Benford's law, which
states that in many naturally occurring collections of numerical
values, the leading first digit is likely to be small (i.e.\ 1
occurs more often than 2, 2 more often than 3, and so on).

The LU can perform a frequency analysis of the hash encodings that
correspond to the first position of a string across all encoded
suffix trees. This potentially allows the LU to learn the first
digit in each string. Additionally, each repeat of the first
digit later in a string (which means the digit is again encoded in
the top level of a suffix tree with the secret salt value $r$) will
be the same hash encoding. Therefore, the LU can learn all positions
in a string where the first digit occurs. Further, due to the basic
chained hash encoding approach, if there is a correlation between
occurrences of the second character based on the first character
in a string, the LU will be able to identify the second character
in suffixes using a frequency analysis.
\smallskip

\textit{3. Specific patterns at beginning of strings}:
Apart from the distribution of the first character, certain prefixes 
in string values can occur frequently in a database leading to
distinct patterns in strings. For example, in international
telephone numbers certain country codes might be more frequent than
others (`+44' for the UK likely occurs more often than `+354' for
Iceland). A similar frequency analysis as discussed above can be
applied on the encoded suffix trees, where the LU will be able to
identify those sequences at the beginning of strings that occur
more often than others. This will however only provide the LU with
information about frequent sub-strings at the beginning of strings,
which by themselves will neither allow the identification of all
characters in a string nor the actual re-identification of
individuals.
\smallskip

As we discussed in the first scenario above, if the characters of
the strings that are encoded in suffix trees follows a uniform 
distribution it is highly unlikely for the LU to be able to identify 
all characters (or digits) in a string with high accuracy. As we 
discussed in Sect.~\ref{sec:first_encoding}, if such a uniform 
distribution occurs in the databases to be matched then the DOs do
not need to perform the extra first character encoding outlined
in Algo.~2. 

However, the first character encoding technique described in
Sect.~\ref{sec:first_encoding} provides privacy of string values
encoded in suffix trees against a frequency attack by the LU under
the second and third scenarios discussed above. As we outlined in 
Algo.~2, the DOs need to agree on the number of characters, $k > 1$,
to be used for the re-hashing of the first character. In the
first character encoding process, a higher value for $k$ results 
in more distinct hash values generated, as we discussed in
Sect~\ref{sec:accuracy_analysis} above. Further, the modulo
operation ensures the resulting encodings are uniformly distributed
within the range of $n$. If we set $n = |\Sigma|$ then
$|\Sigma|^k > n$ if $k>1$.
Further, we add a secret salt value $r$ in the first character
encoding scheme. The use of $r$ provides strong privacy against
dictionary attacks on first digits encodings. This is because the
LU is not capable of identifying the correct encoding that has
been applied on different first characters without knowing $r$
that is used by the DOs.

As we show in our experiments below, each hash encoding of the
first characters of the encoded suffix trees will occur with nearly
the same frequency, especially with larger values of $k$, even if
the unencoded first characters follow a certain distribution, for
example Benford's law. This assures that the LU will not be able to
exploit any frequency information about the first characters in
strings and therefore cannot directly map hash encodings to their
corresponding plain-text values. This makes our approach secure
from any frequency based attacks. In Sect.~\ref{sec:experiments} we
experimentally evaluate how frequency distributions of the first
characters of strings of different data types change with different
$k$.


%
\smallskip

\noindent\textbf{Similarity graph attack by the LU:}~
%
%
As we described in Sect.~\ref{sec:matching}, the LU calculates the 
length of the longest common suffix between each pair of encoded
suffix trees. Once all encoded suffix tree pairs are compared the
LU can construct a similarity graph where each encoded suffix tree
becomes a vertex while the edges between these vertices represent
the length of the longest common suffix between a pair of encoded
suffix  trees. 

Once such a graph is generated, the LU can construct a similar graph 
based on a publicly available plain-text database that has similar 
characteristics as the encoded databases. Then the  LU can conduct
a sub-graph matching~\citep{Hei18} between the two graphs to identify
possible plain-text values that correspond to the encoded suffix
trees. One possible way of carrying out such matching would be to
identify any sub-graphs that are unique and can obviously be
identified based on the vertices that have a unique set of
edges in the sub-graph. If such unique sub-graphs can be found then
the plain-text values that can be mapped to vertices in the encoded
suffix tree graph can be identified with high probability. 

Such an attack by the LU requires the accessibility to a plain-text 
database that has a highly similar distribution of characters in 
string values as those in the encoded database. Though such attacks
are limited in the literature~\citep{Cul17,vidanage2020graph}, there are several
counter-measures that the DOs can apply on their databases before
encoding and sending them to the LU, including applying
blocking~\citep{Chr12} and block-specific salt values, adding faked
values into their databases, or employing several LUs for the
comparison of encoded suffix trees. We aim to investigate such
counter-measures as future work.

\begin{figure}[th!]
	\centering
	\includegraphics[width=0.22\textwidth]
	{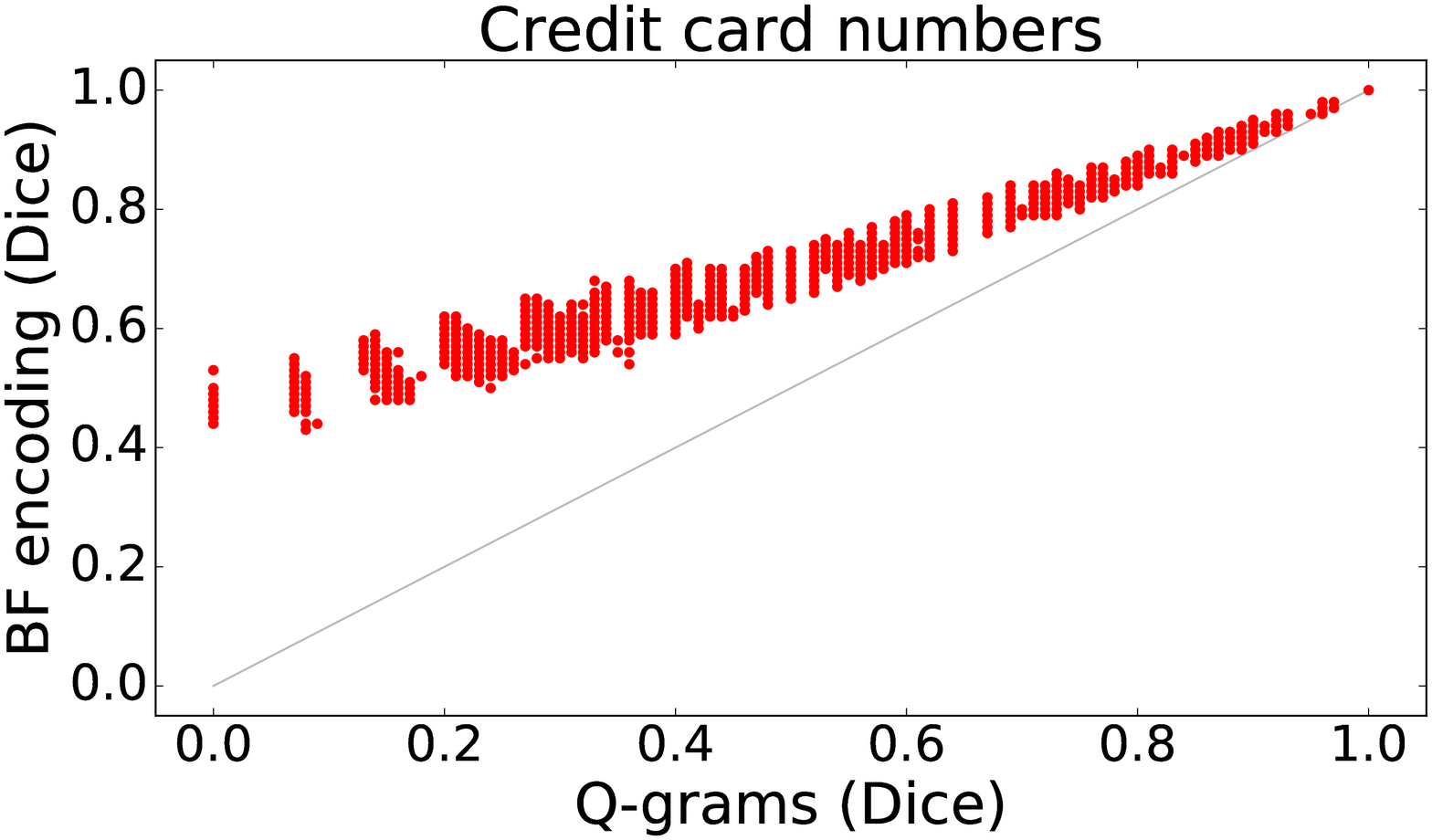}
	\hspace{2mm}
	\includegraphics[width=0.22\textwidth]
	{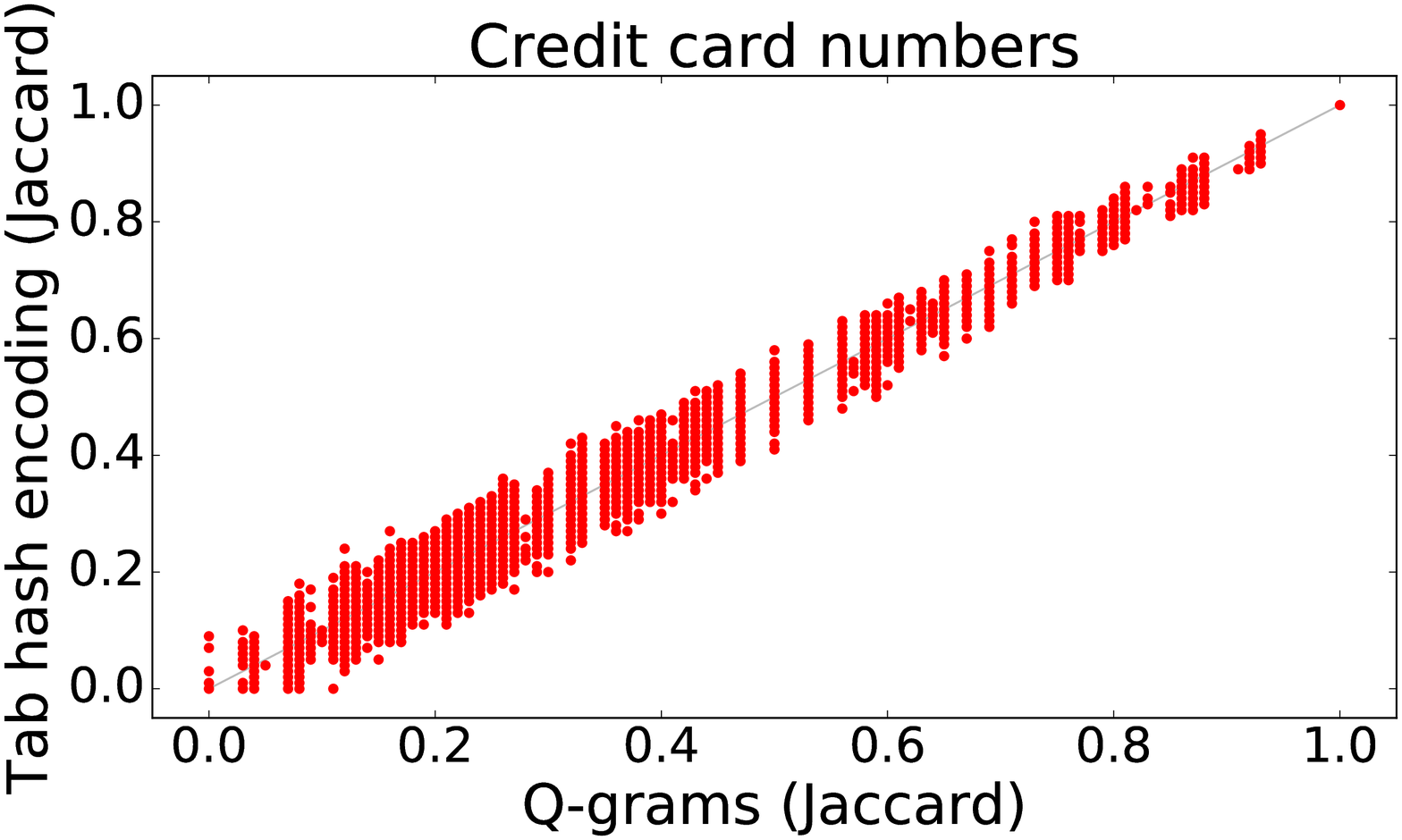}
	\hspace{2mm}
	\includegraphics[width=0.22\textwidth]
	{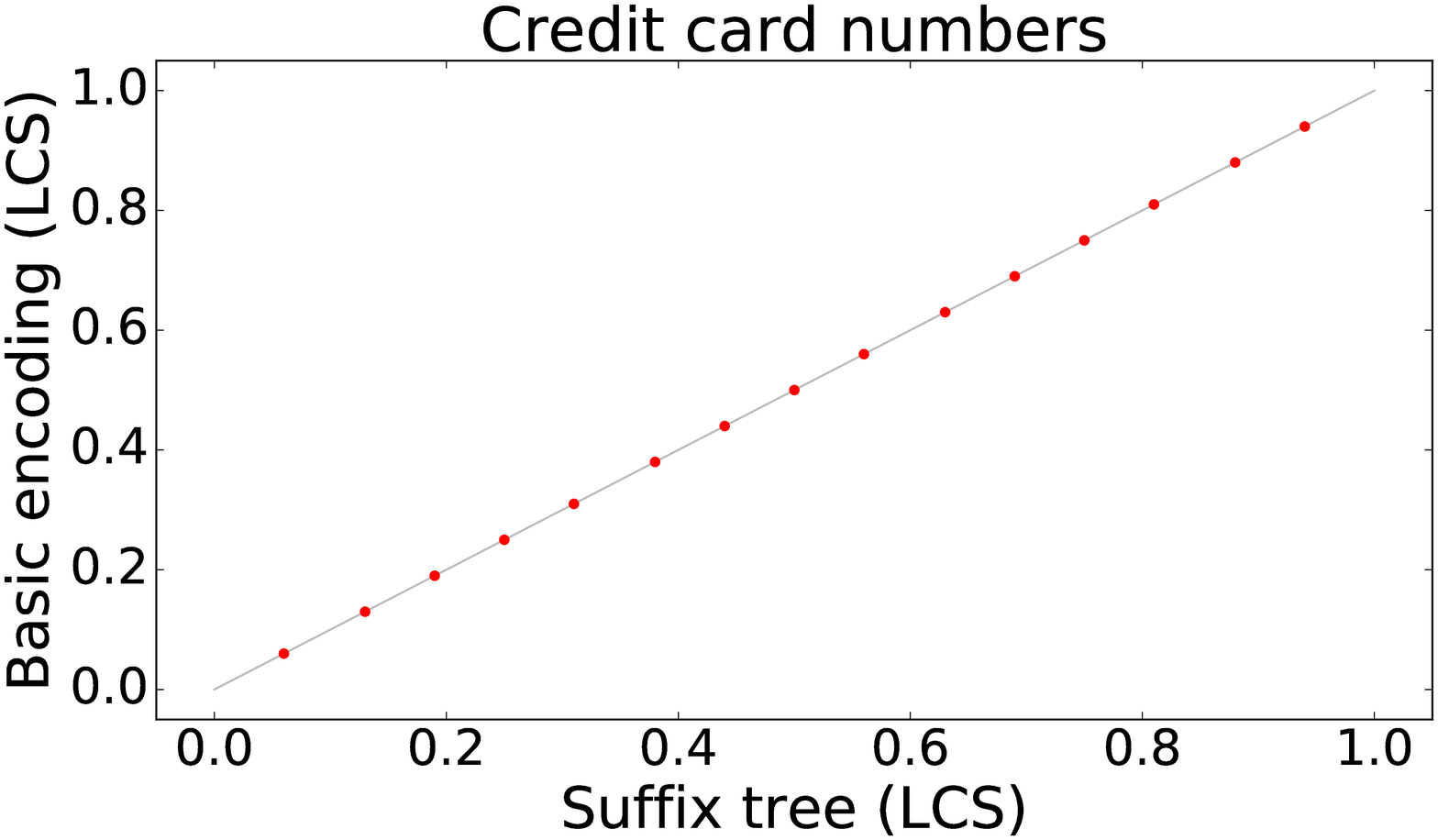}
	\hspace{2mm}
	\includegraphics[width=0.22\textwidth]
	{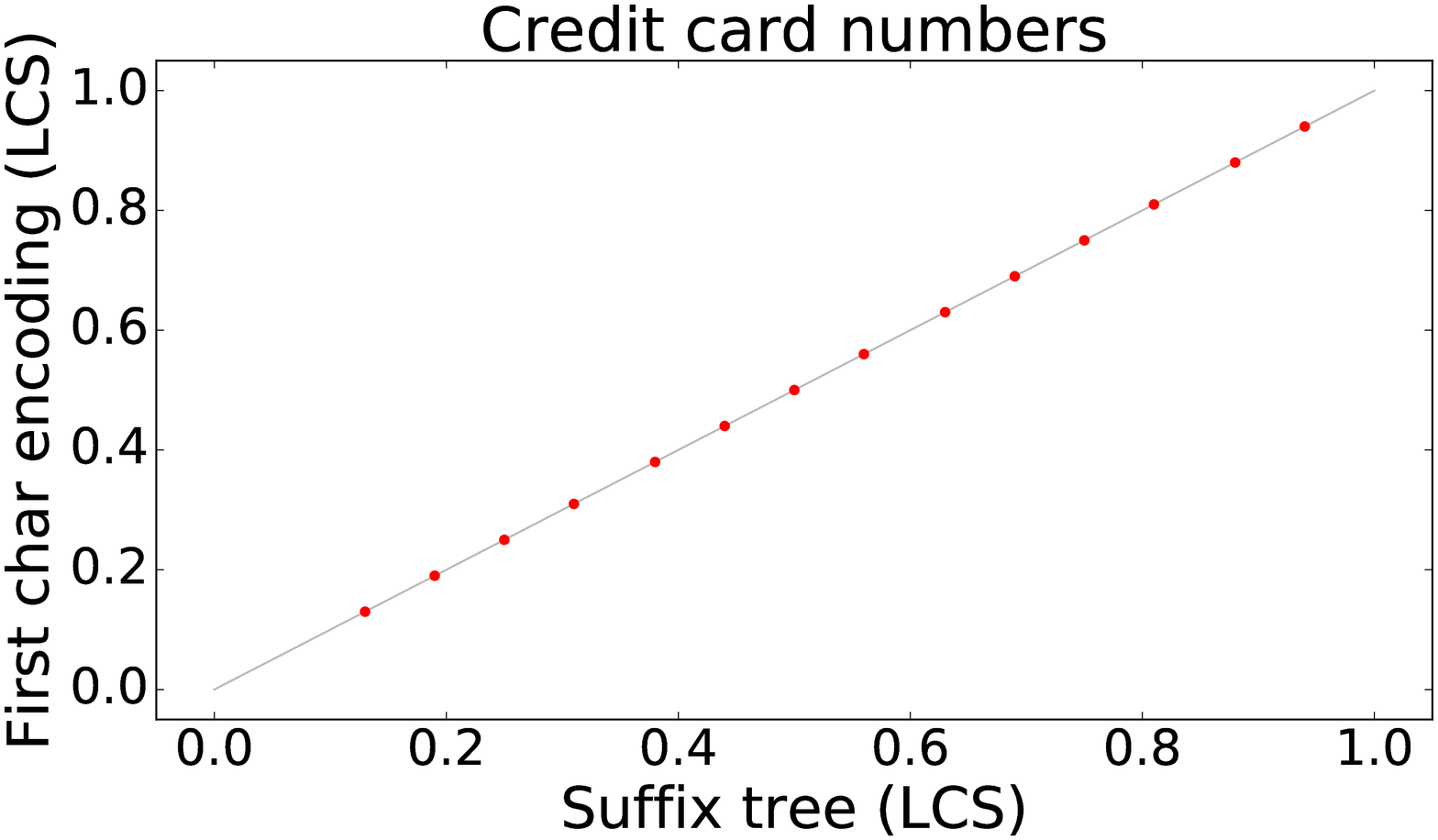}
	~ \\[2mm]
	\includegraphics[width=0.22\textwidth]
	{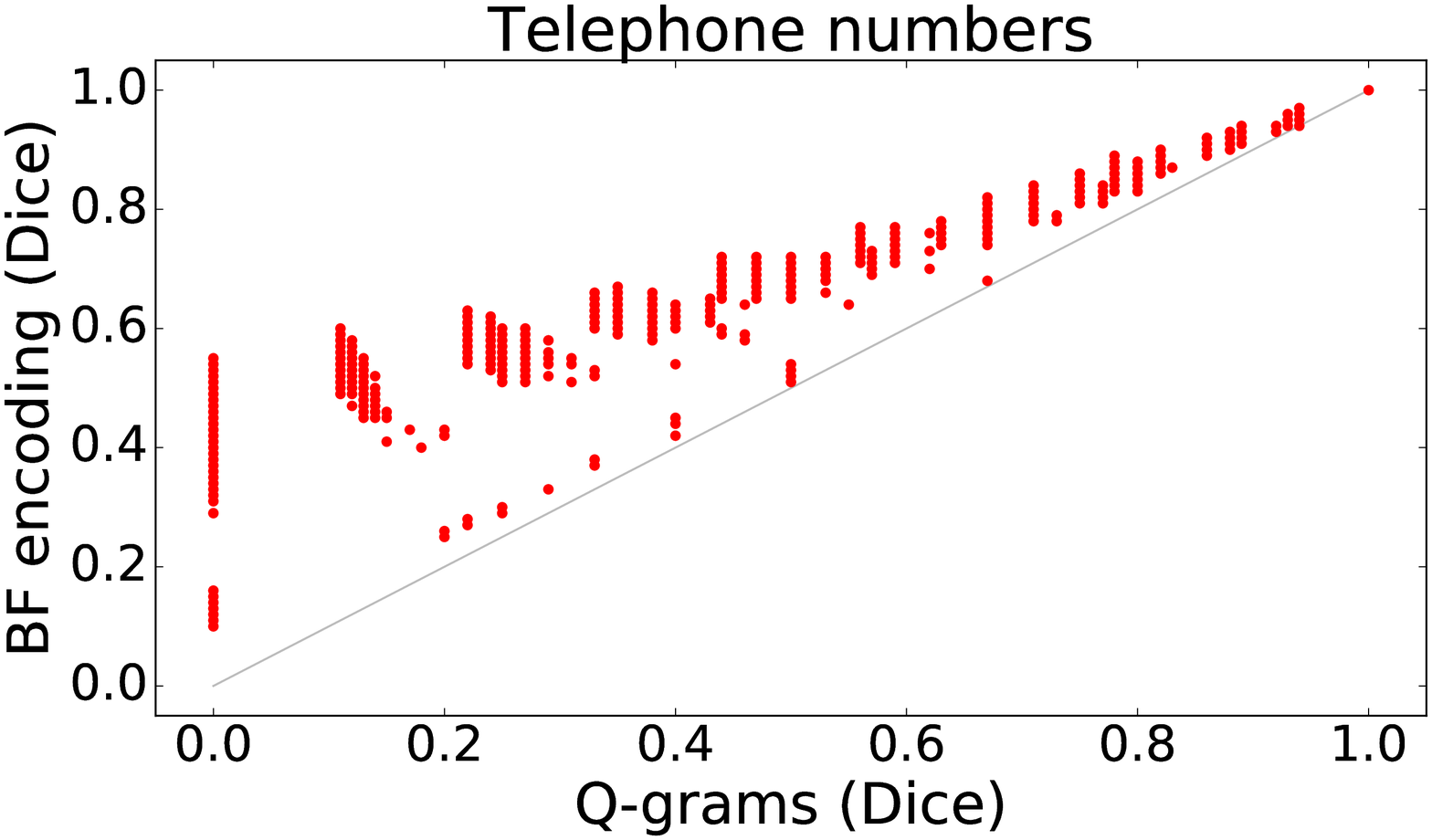}
	\hspace{2mm}
	\includegraphics[width=0.22\textwidth]
	{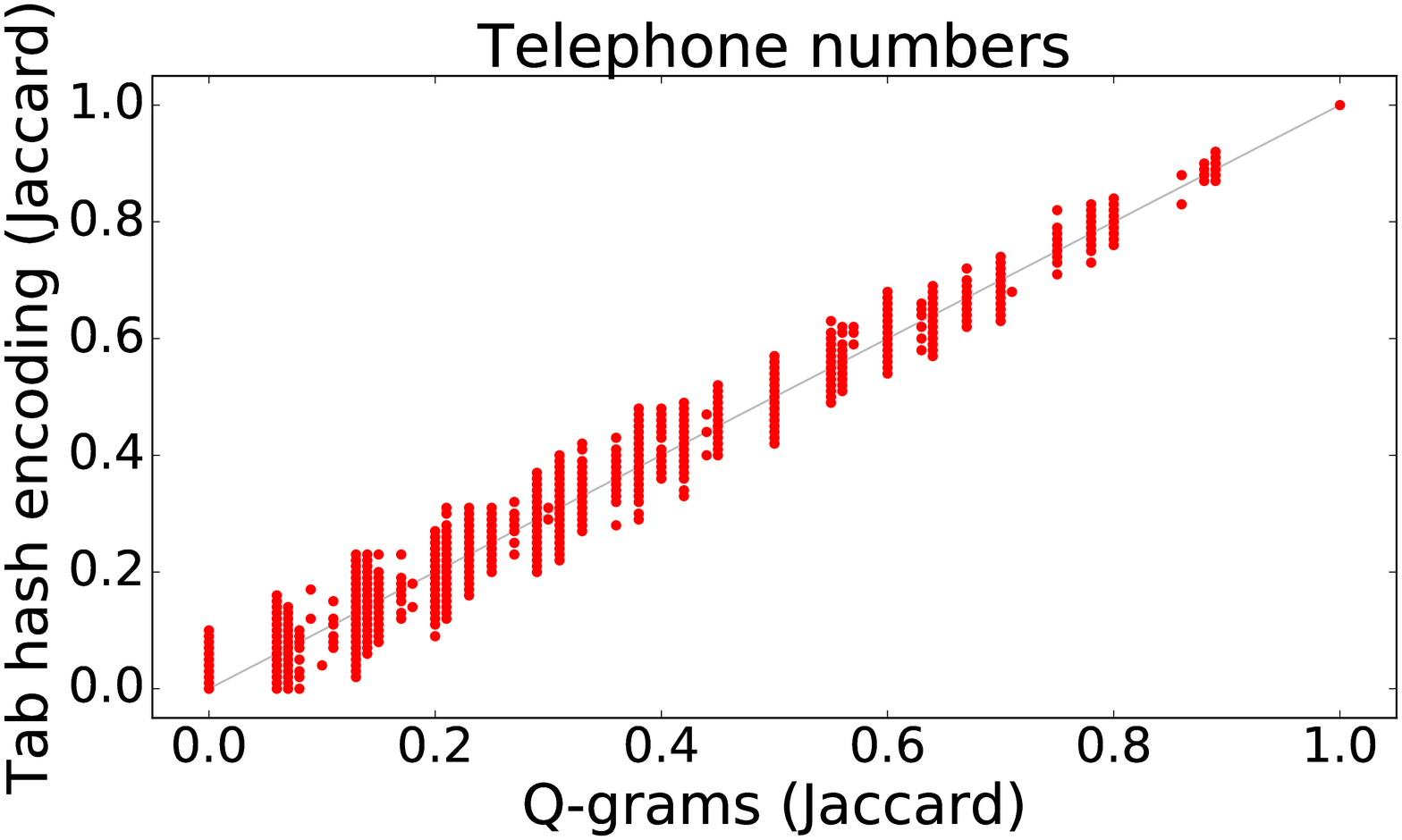}
	\hspace{2mm}
	\includegraphics[width=0.22\textwidth]
	{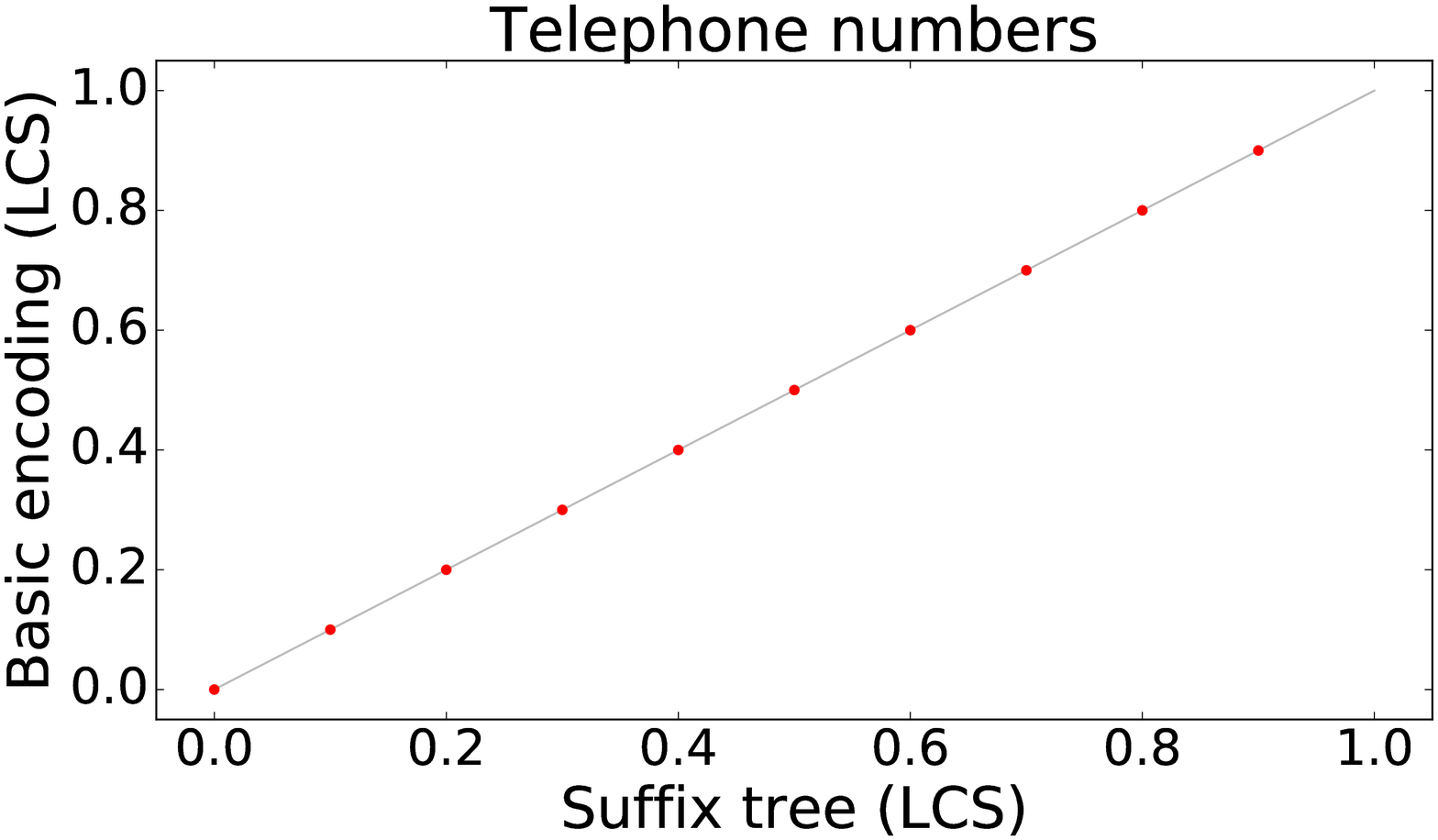}
	\hspace{2mm}
	\includegraphics[width=0.22\textwidth]
	{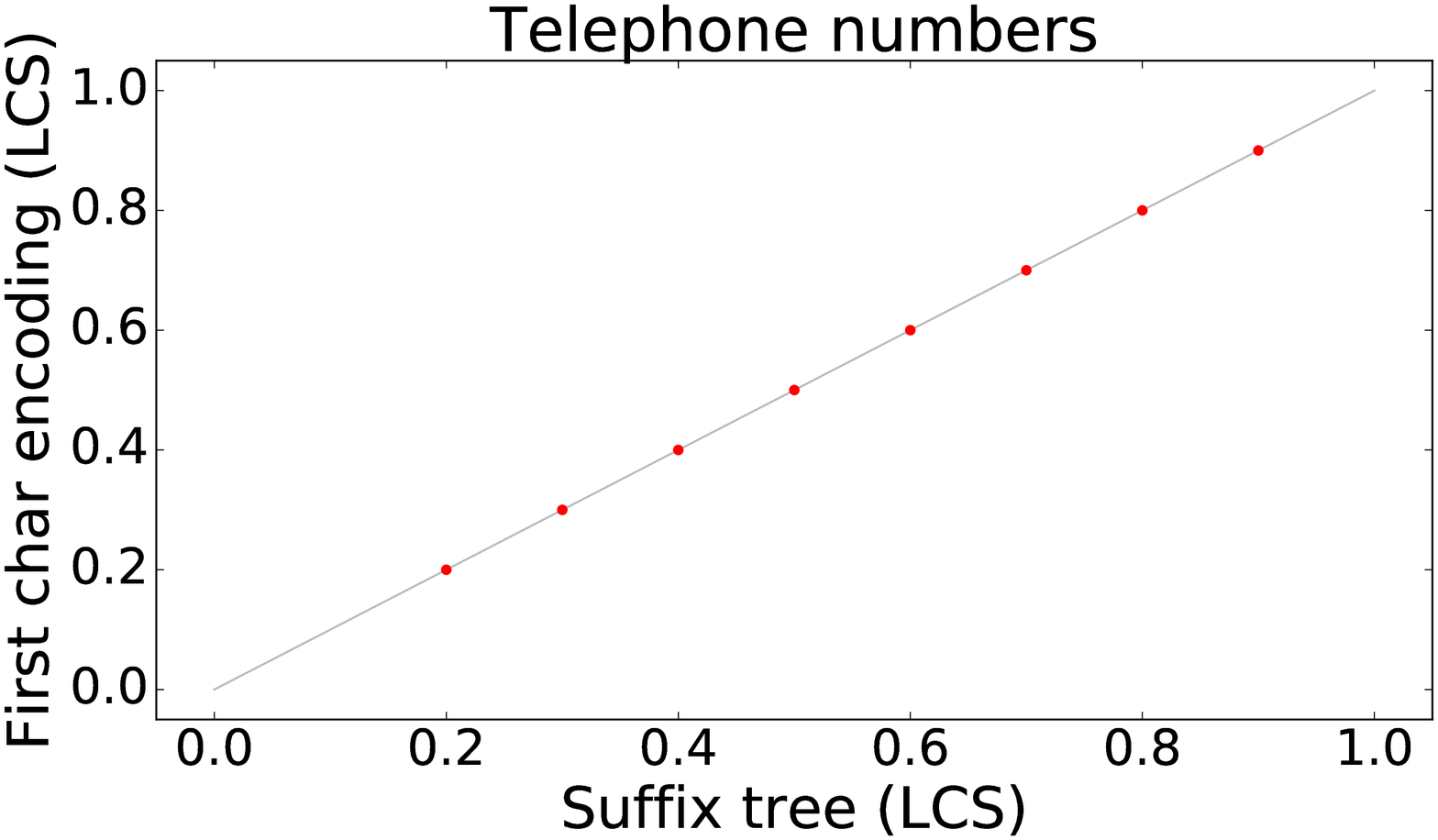}
	~ \\[2mm]
	\includegraphics[width=0.22\textwidth]
	{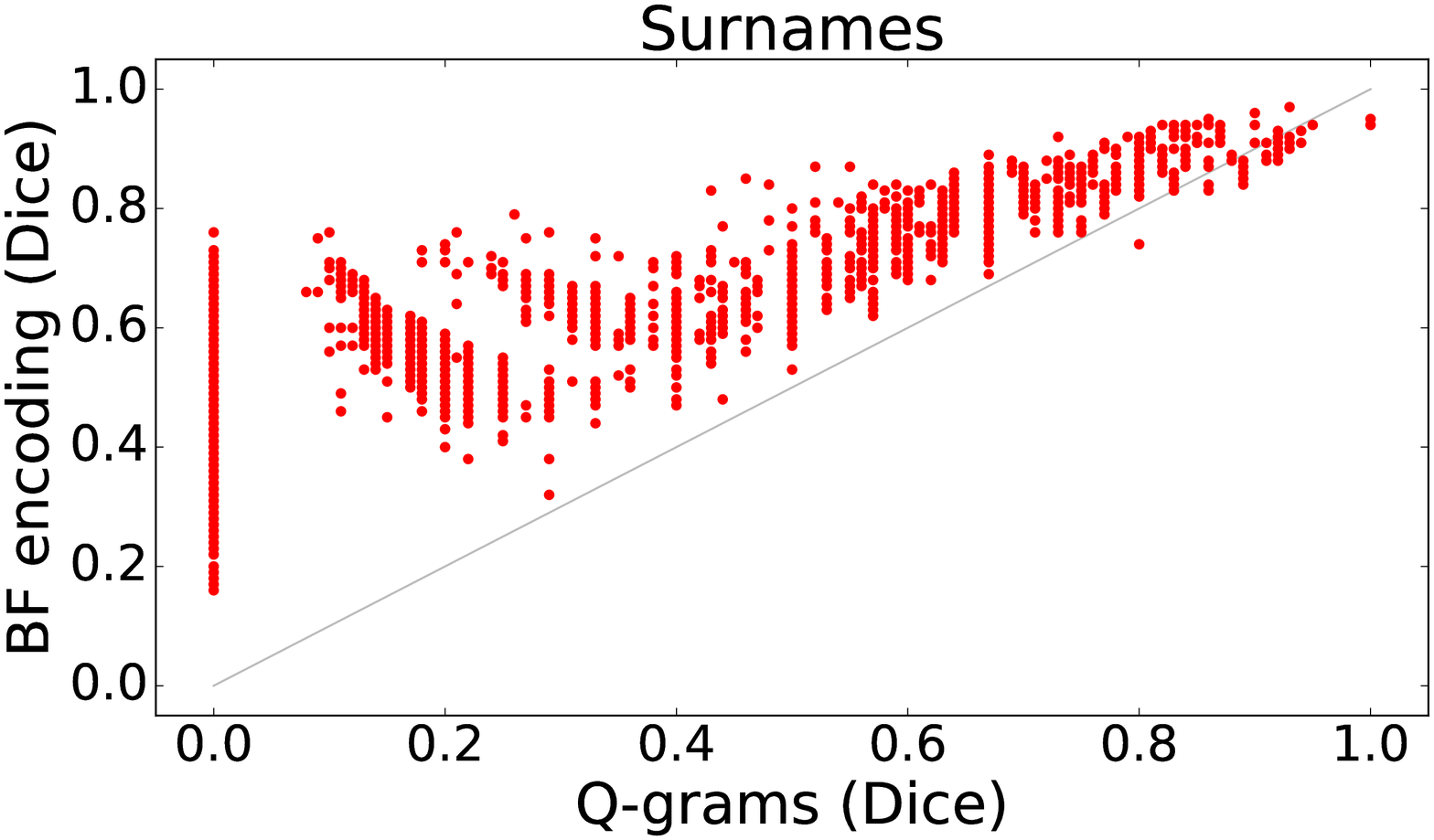}
	\hspace{2mm}
	\includegraphics[width=0.22\textwidth]
	{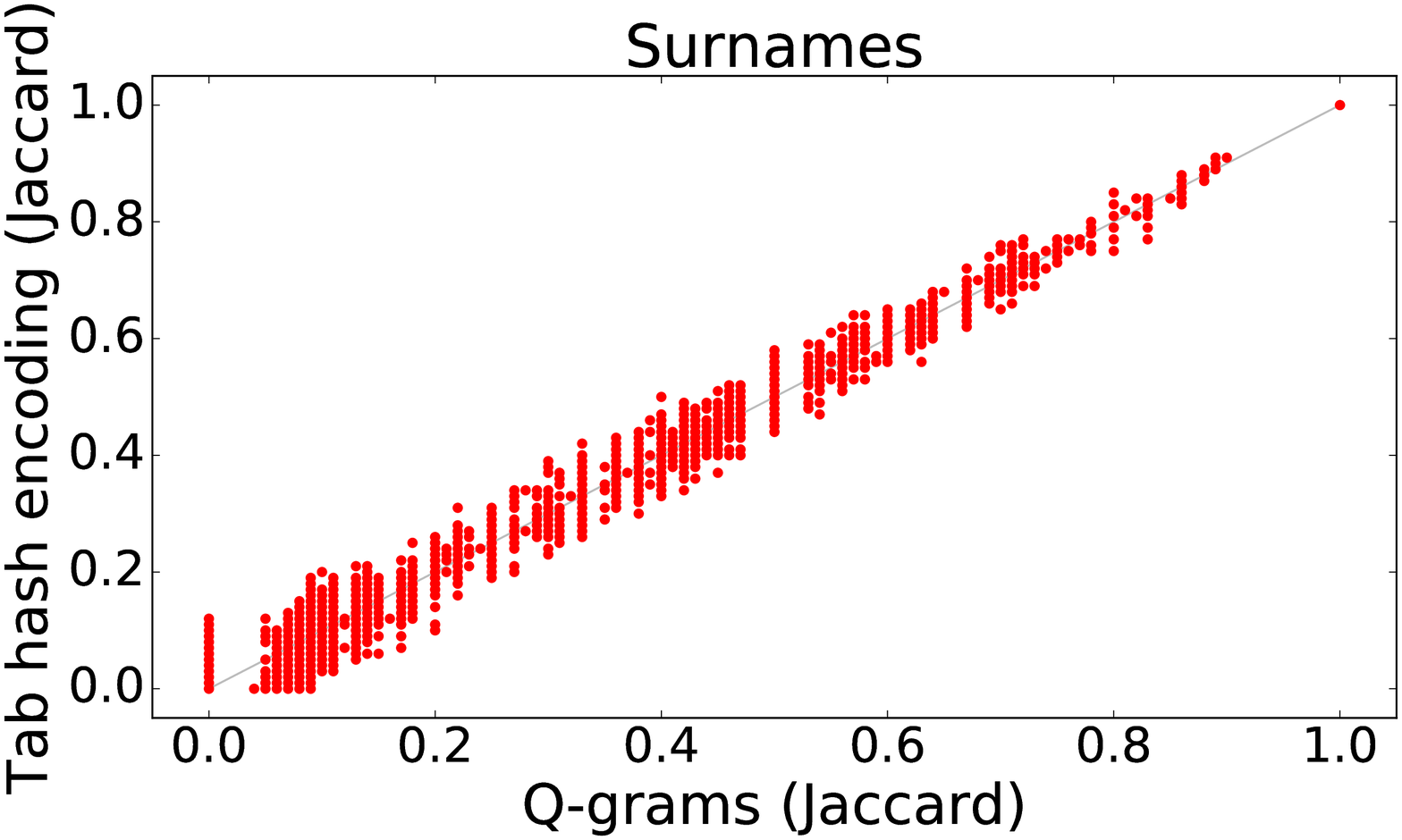}
	\hspace{2mm}
	\includegraphics[width=0.22\textwidth]
	{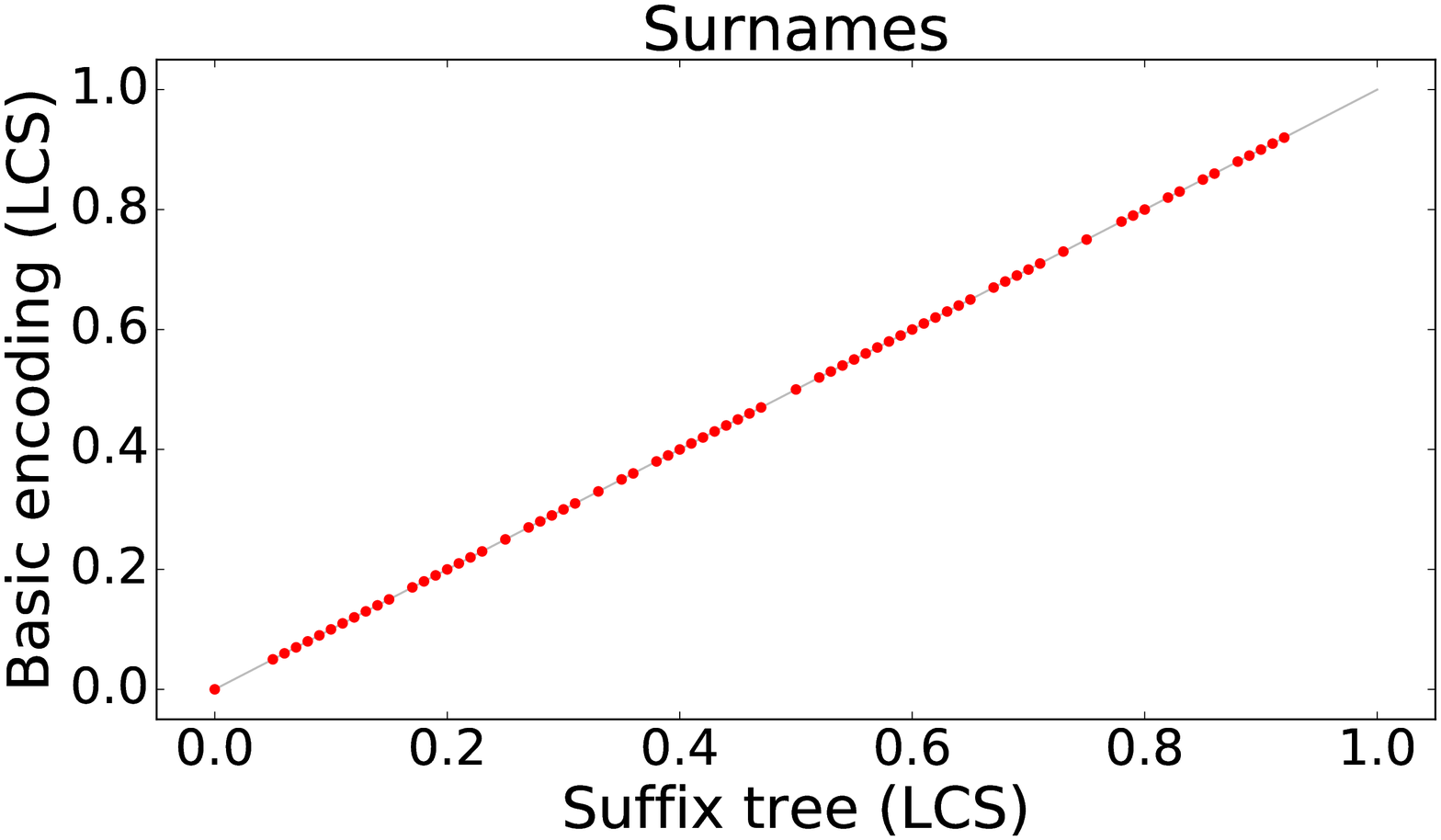}
	\hspace{2mm}
	\includegraphics[width=0.22\textwidth]
	{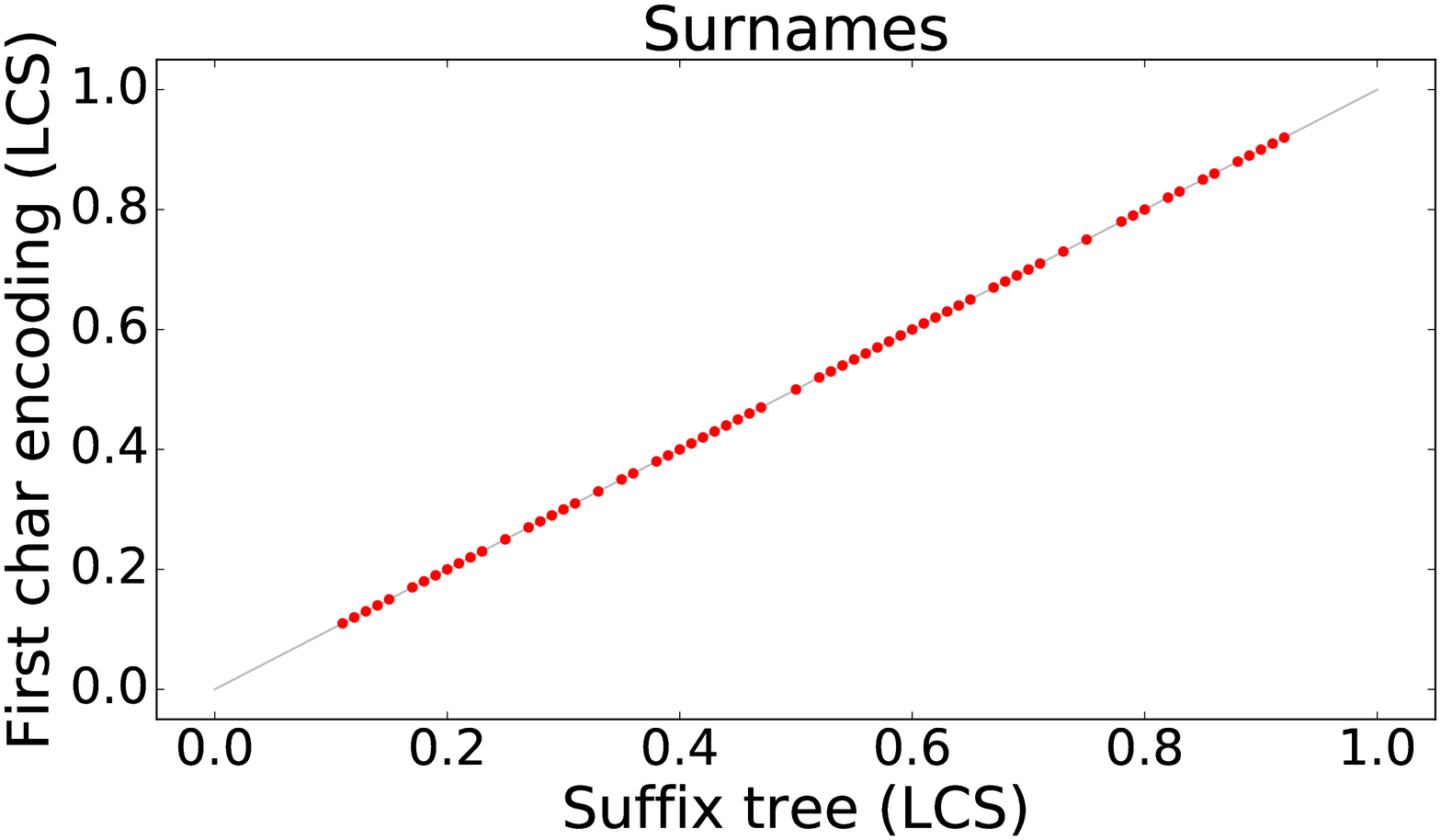}
	~ \\[2mm]
	\includegraphics[width=0.22\textwidth]
	{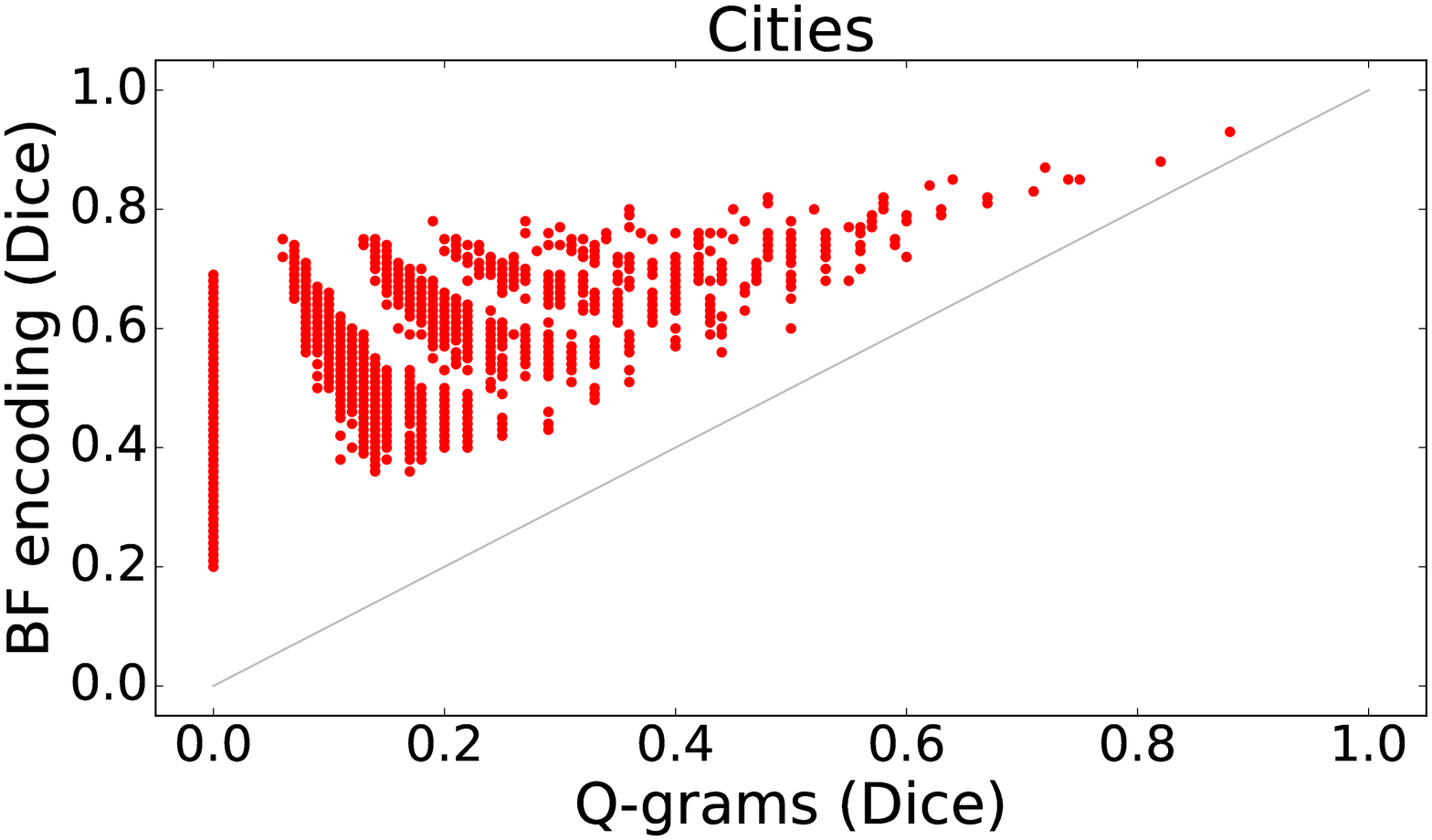}
	\hspace{2mm}
	\includegraphics[width=0.22\textwidth]
	{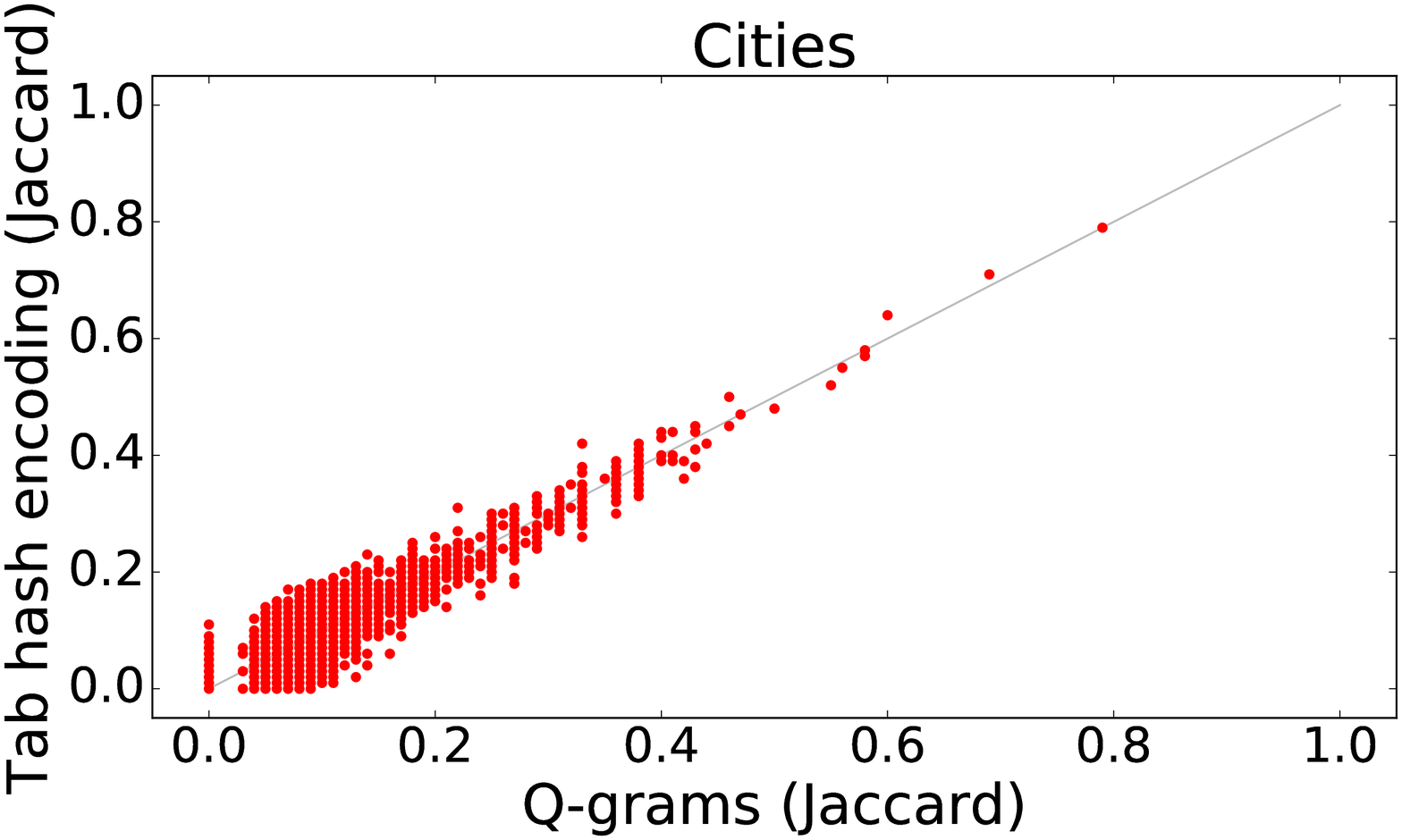}
	\hspace{2mm}
	\includegraphics[width=0.22\textwidth]
	{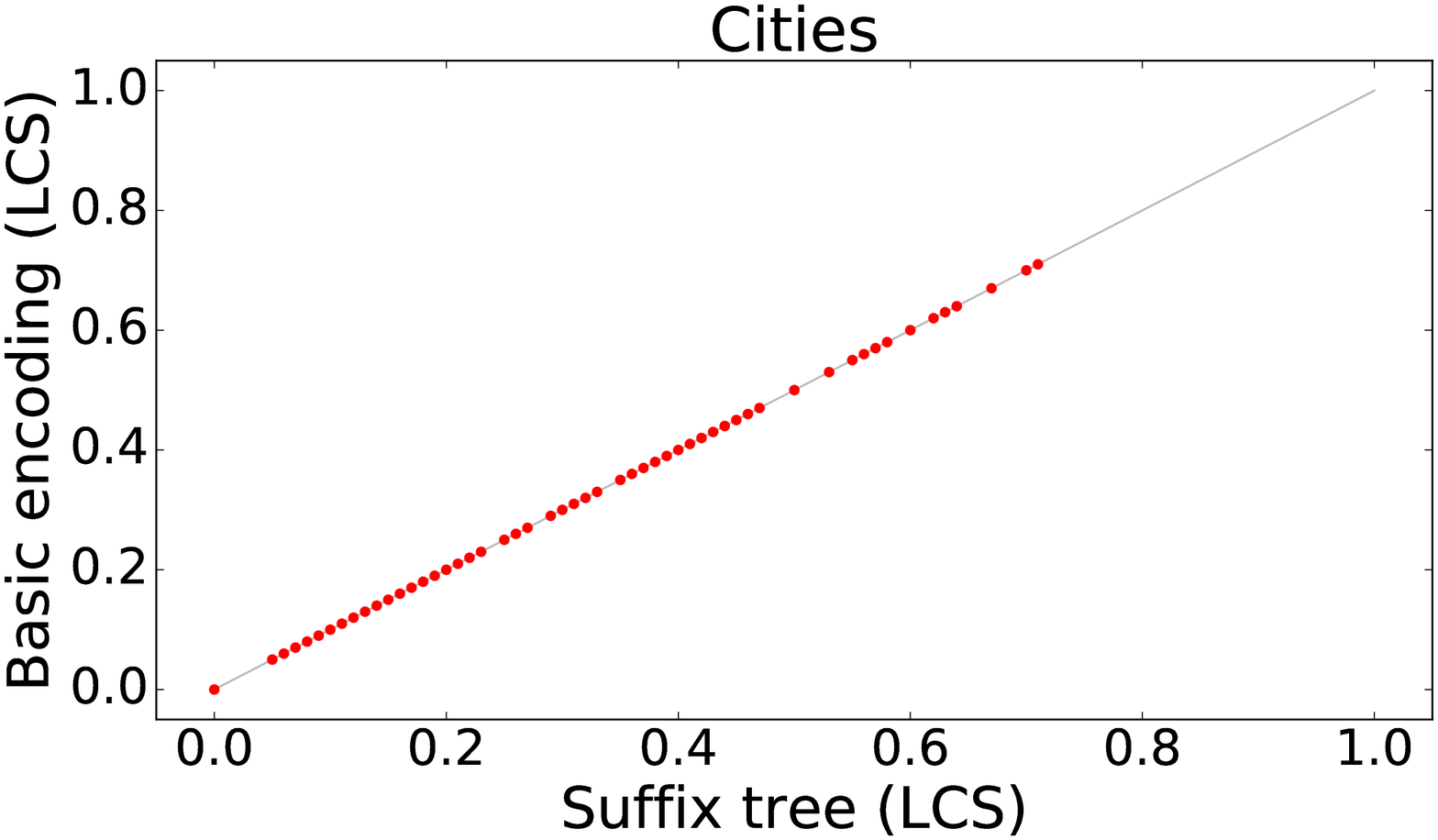}
	\hspace{2mm}
	\includegraphics[width=0.22\textwidth]
	{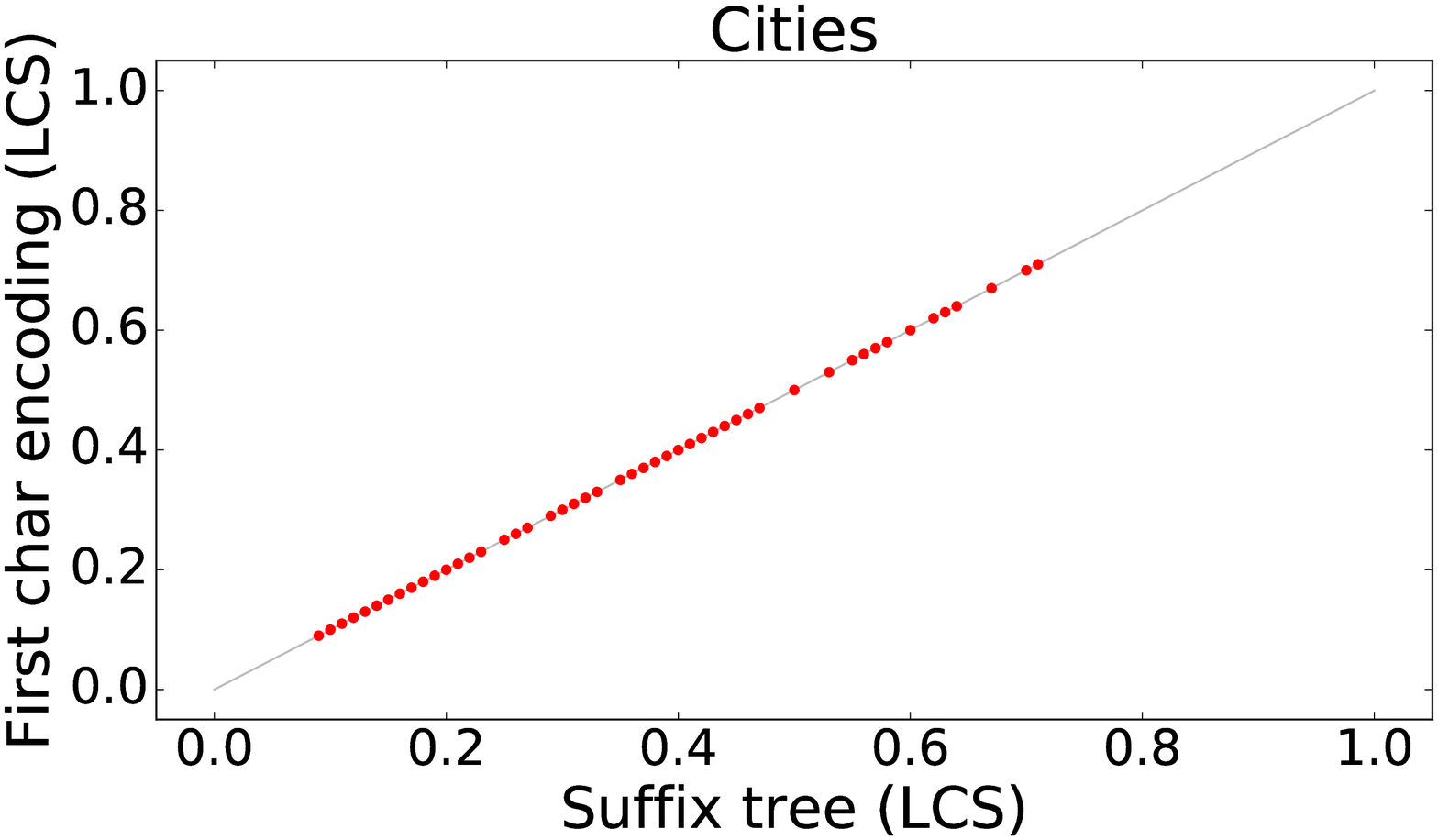}
	~ \\[2mm]
	\includegraphics[width=0.22\textwidth]
	{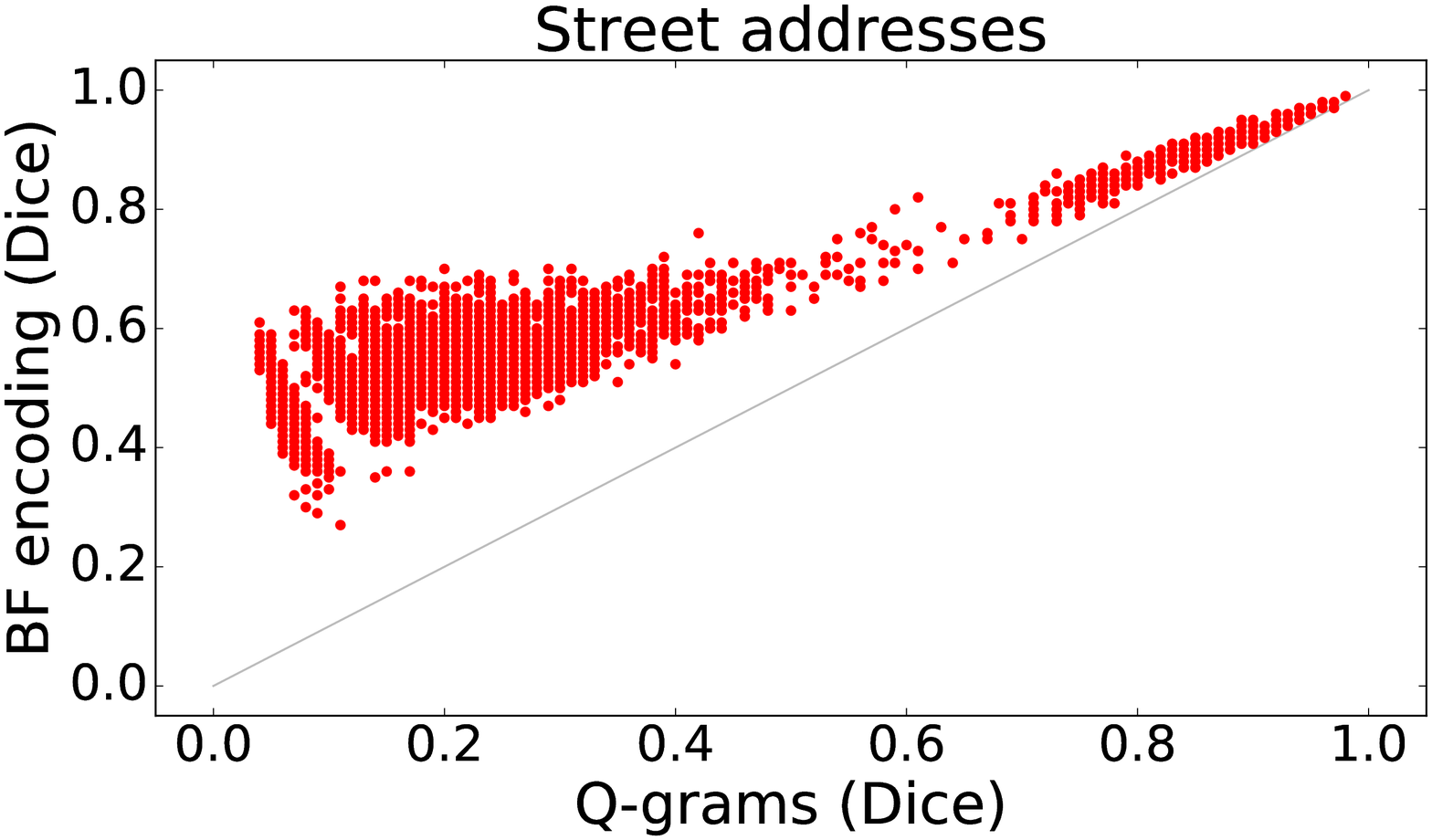}
	\hspace{2mm}
	\includegraphics[width=0.22\textwidth]
	{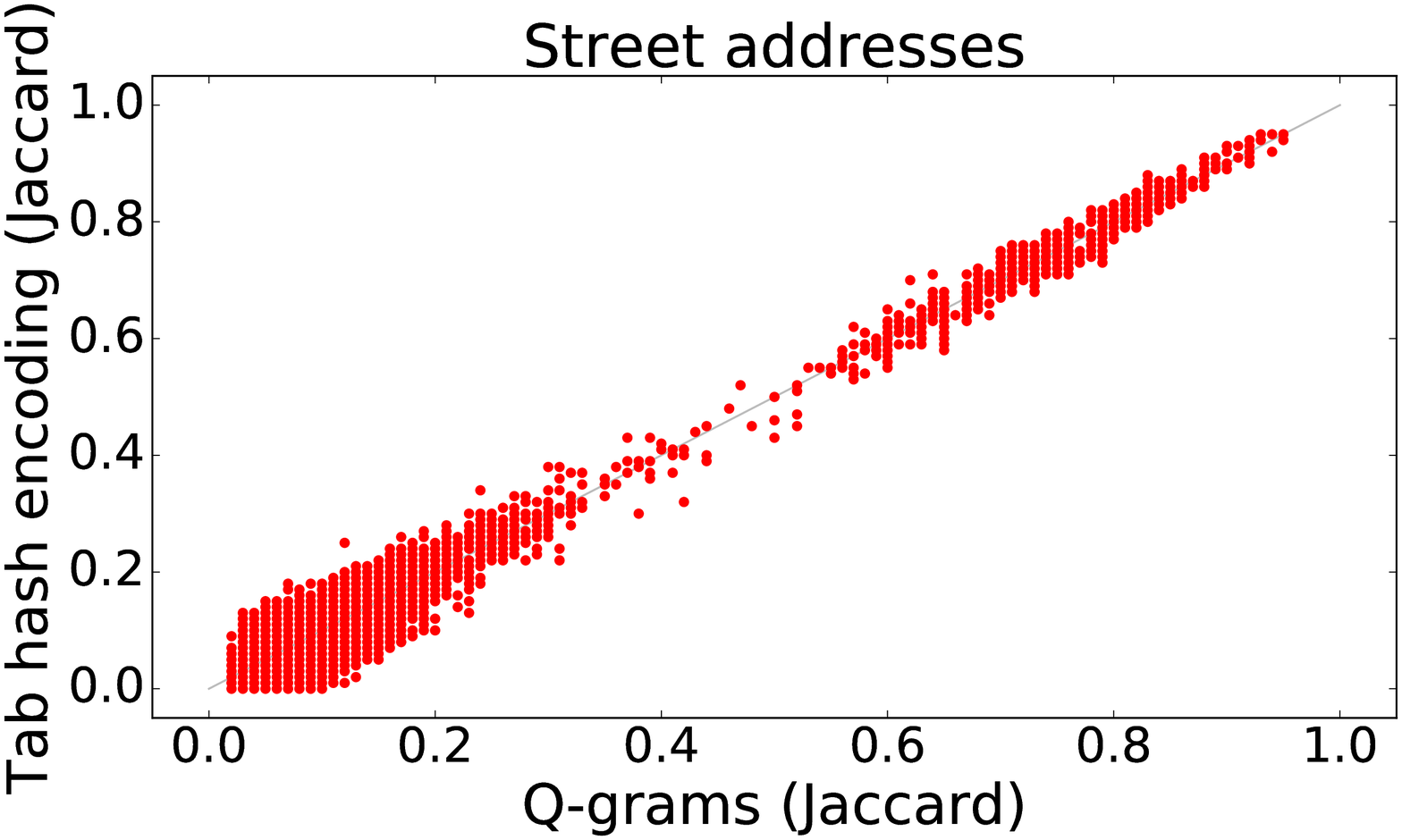}
	\hspace{2mm}
	\includegraphics[width=0.22\textwidth]
	{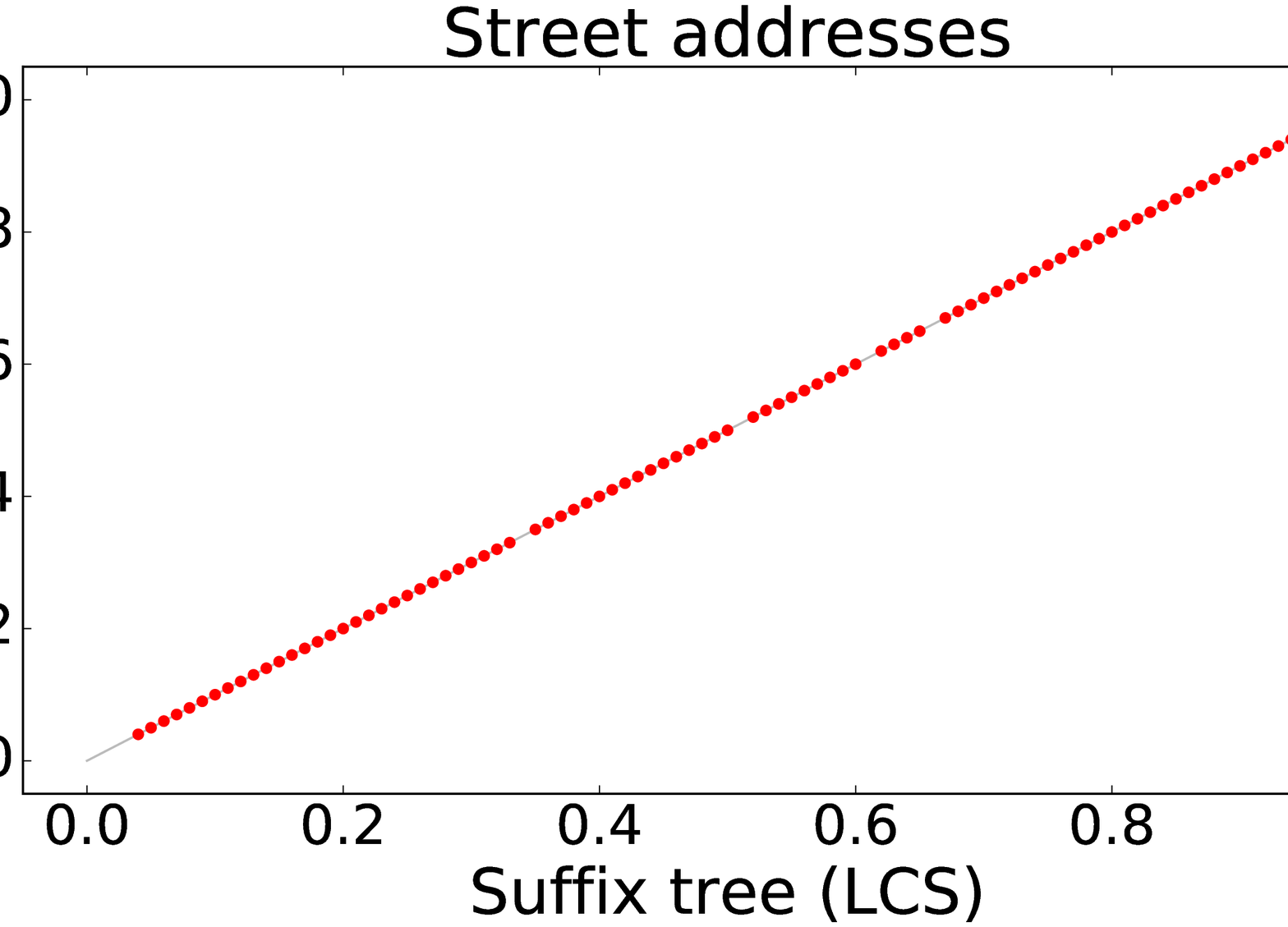}
	\hspace{2mm}
	\includegraphics[width=0.22\textwidth]
	{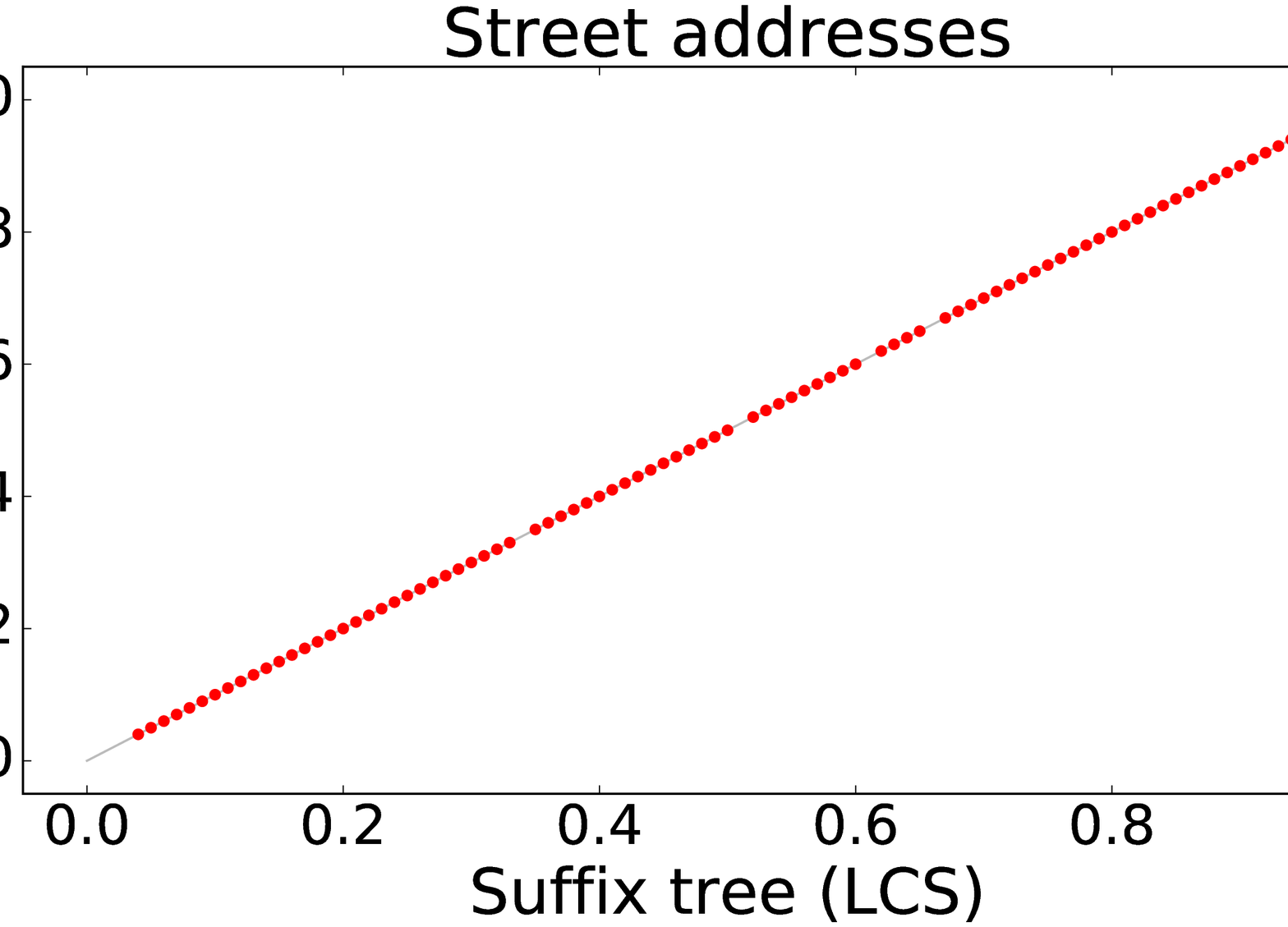}
	~ \\[2mm]
	\includegraphics[width=0.22\textwidth]
	{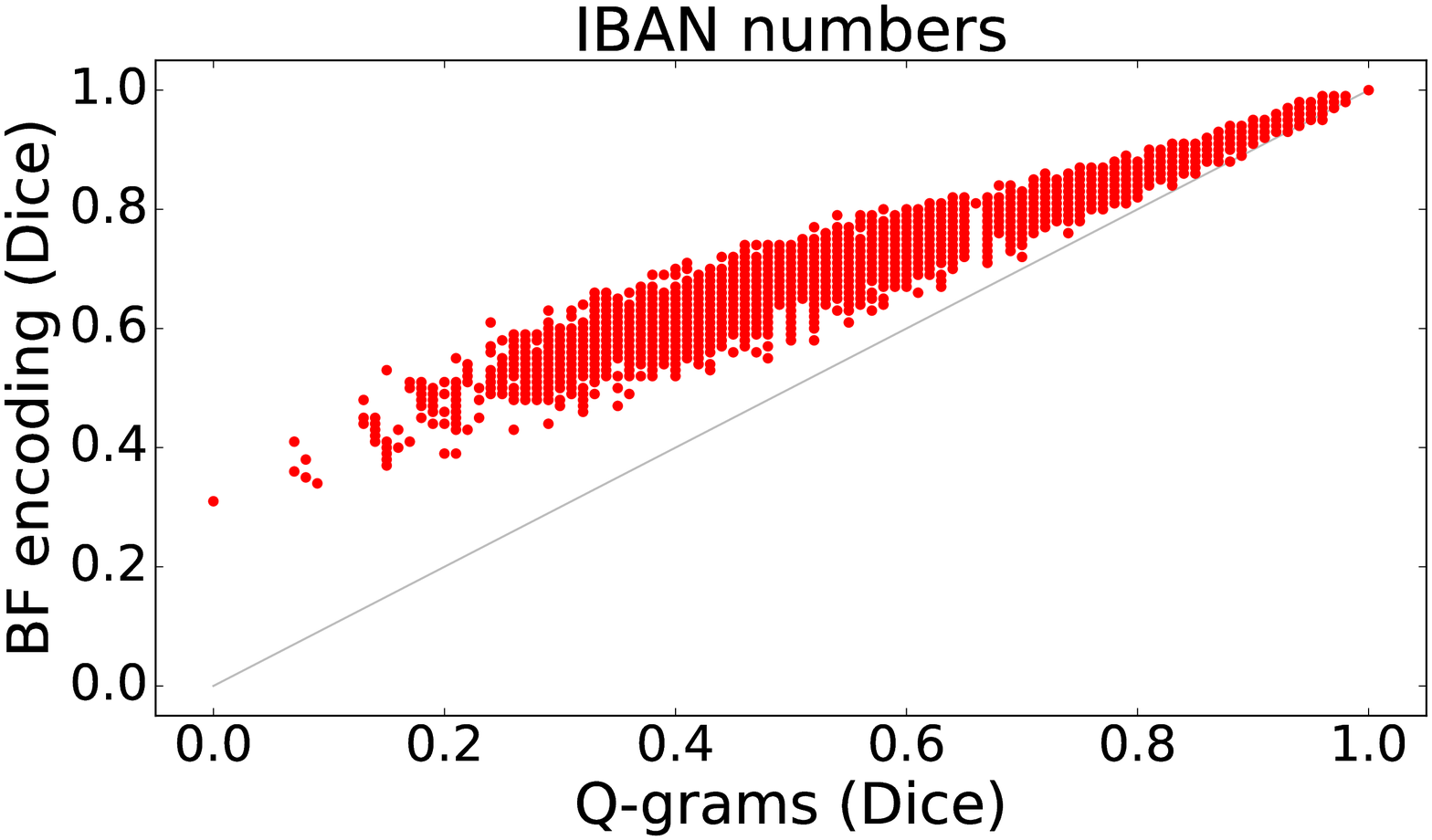}
	\hspace{2mm}
	\includegraphics[width=0.22\textwidth]
	{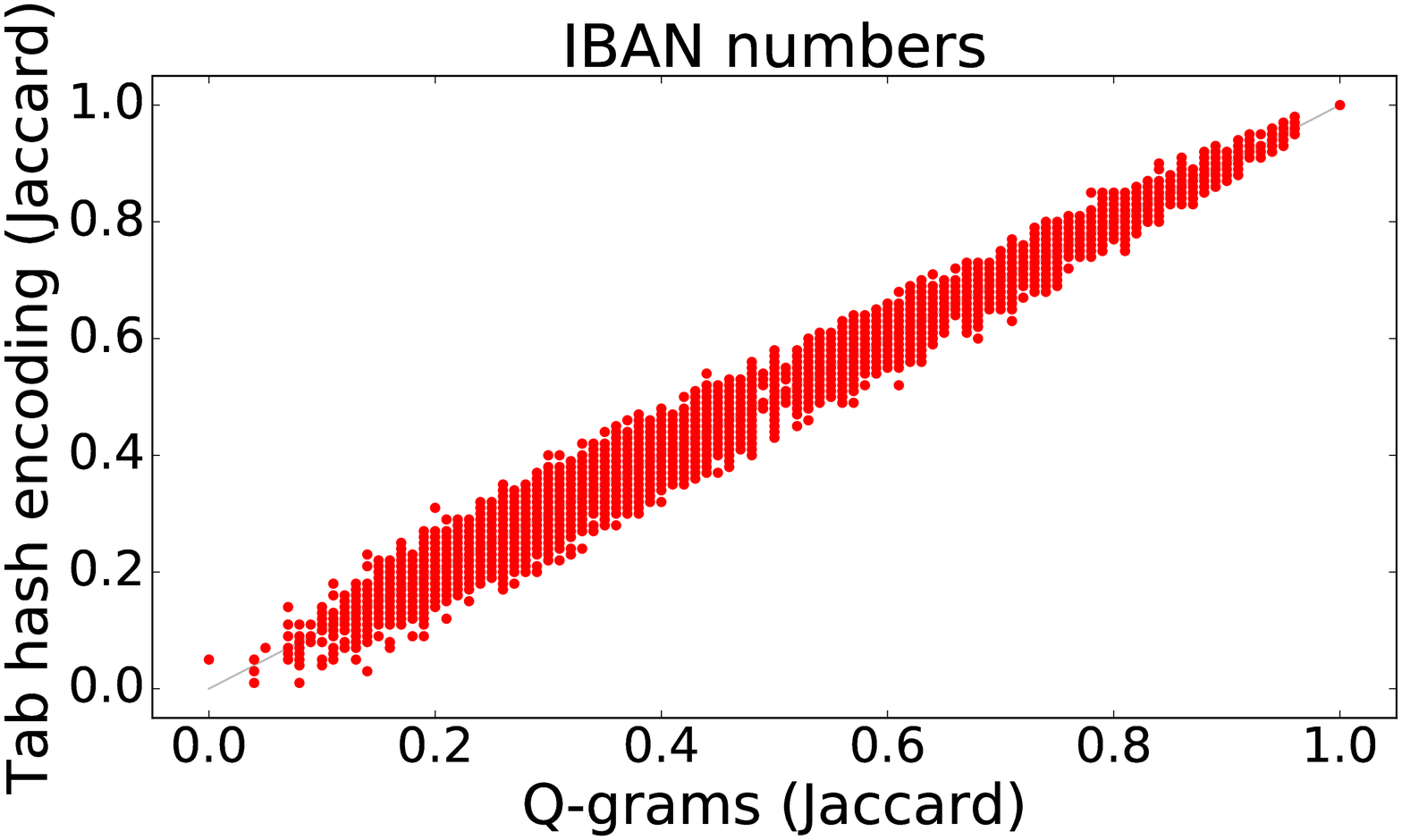}
	\hspace{2mm}
	\includegraphics[width=0.22\textwidth]
	{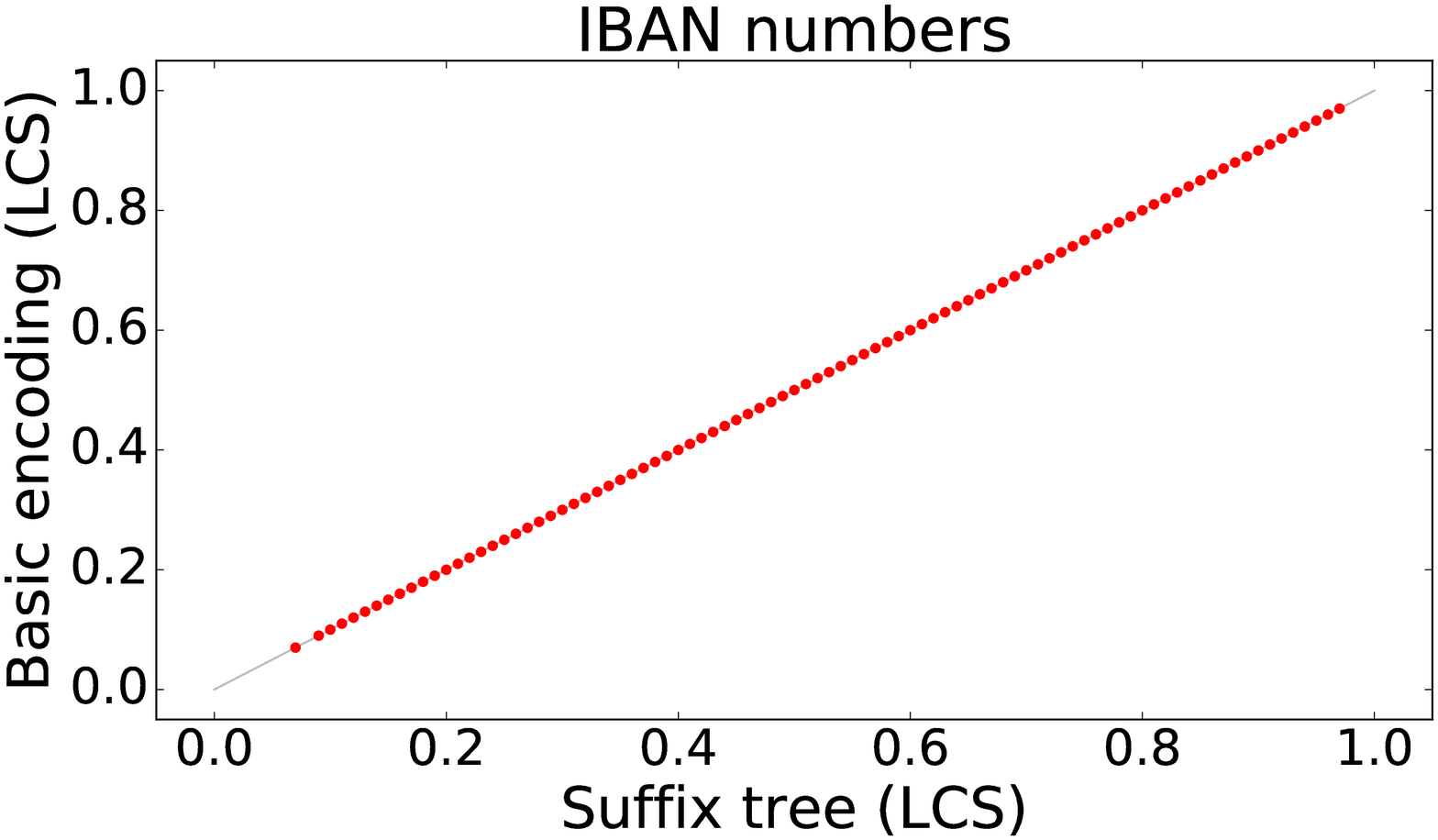}
	\hspace{2mm}
	\includegraphics[width=0.22\textwidth]
	{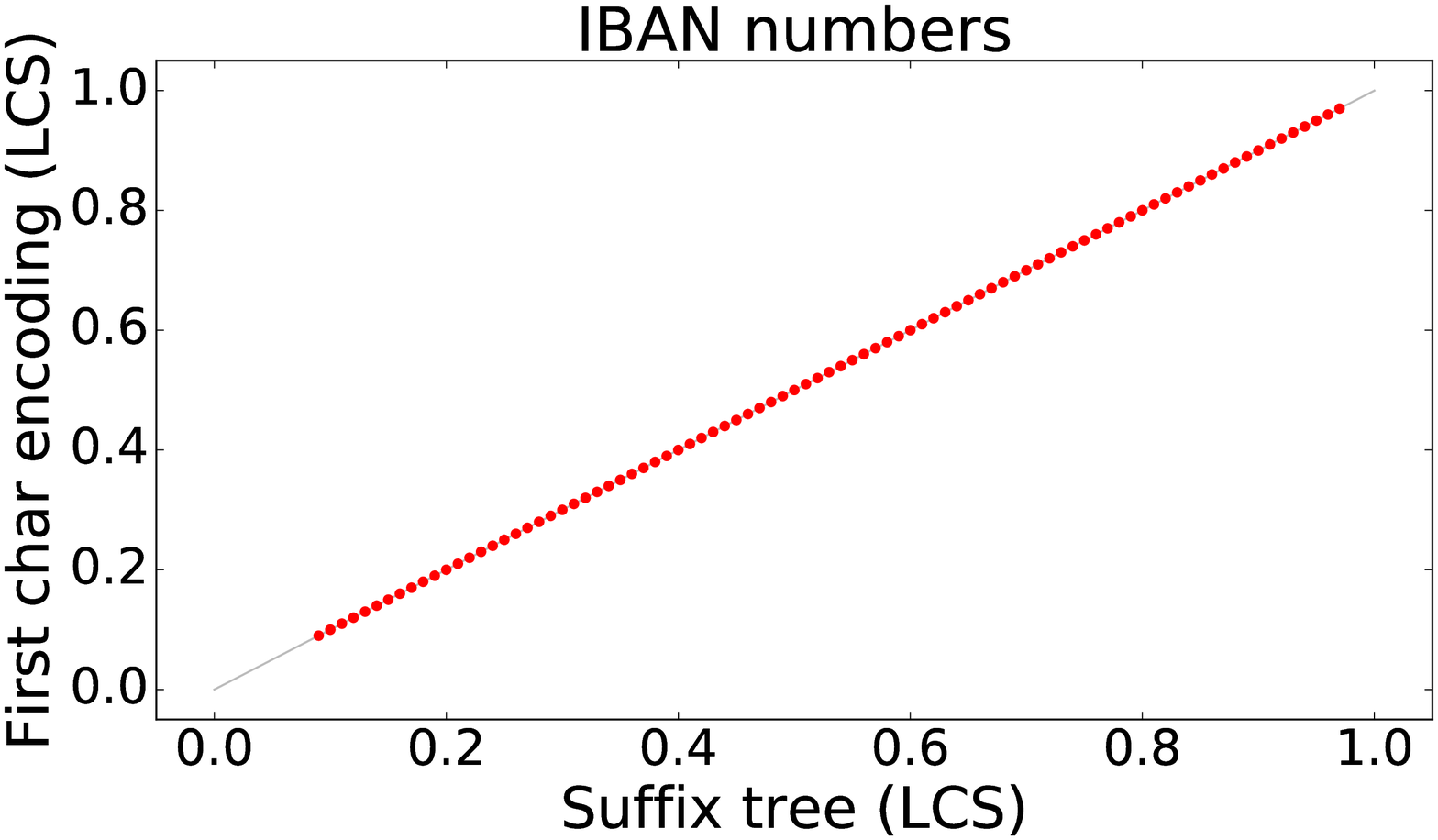}
  \caption{Similarity plots of 
    Bloom filter (BF) encoding~\citep{Sch09} (left) and tabulation
    based hashing (TabHash)~\citep{Smi17} encoding (second left),
    basic encoded suffix trees (second right), and secure first
    character encoded suffix trees with \emph{m\,=\,}2,
    \emph{k\,=\,}2, and \emph{n\,=\,}$|\Sigma|$ (right). As
    can be seen, both our suffix tree based encoding methods provide
    accurate similarity calculations, while BF and TabHash
    encoding can lead to substantially changed similarities even
    between very similar strings. 
	\label{fig:sim-plots-k}}
\end{figure}

\section{Experimental Evaluation}
\label{sec:experiments}


We used both synthetic as well as real data of different types to
evaluate our novel privacy-preserving string matching approach.
We used the Mockaroo synthetic data generator
(see: \url{https://www.mockaroo.com}) to create 10,000 strings with
unique credit card and IBAN (International Bank Account Number)
numbers. From these strings we then generated corrupted versions by
randomly replacing between 1 and 10 characters from the same
alphabet (digits only for credit card, and digits and letters for
IBAN), resulting in 10,000 pairs of credit card and IBAN numbers.

We extracted two different data sets with telephone numbers,
surnames, city names, and street addresses from the North Carolina
Voter Registration (NCVR) database (see: \url{https://dl.ncsbe.gov}),
where the first data sets were from a
snapshot of NCVR from 2015 and the second data sets from a snapshot
of NCVR from 2019. We paired records from these two data sets based
on the corresponding voter identifiers, ensuring we only had pairs
where the strings were not the same. We then selected 10,000 pairs
of strings for each of the four attribute types.

Overall, our data sets consist of strings of different types
(digits only, letters only, or mixed) and of different lengths.
They reflect the types of data commonly used in applications such
PPRL where sensitive databases are to be linked across
organizations \citep{Christen2020lsd}.

We implemented our approach using Python 2.7 and ran experiments on
a server with 128 GBytes of memory and 2.4 GHz CPUs running Ubuntu
16.04.

\begin{figure}[t!]
	\centering
	\includegraphics[width=0.22\textwidth]
	{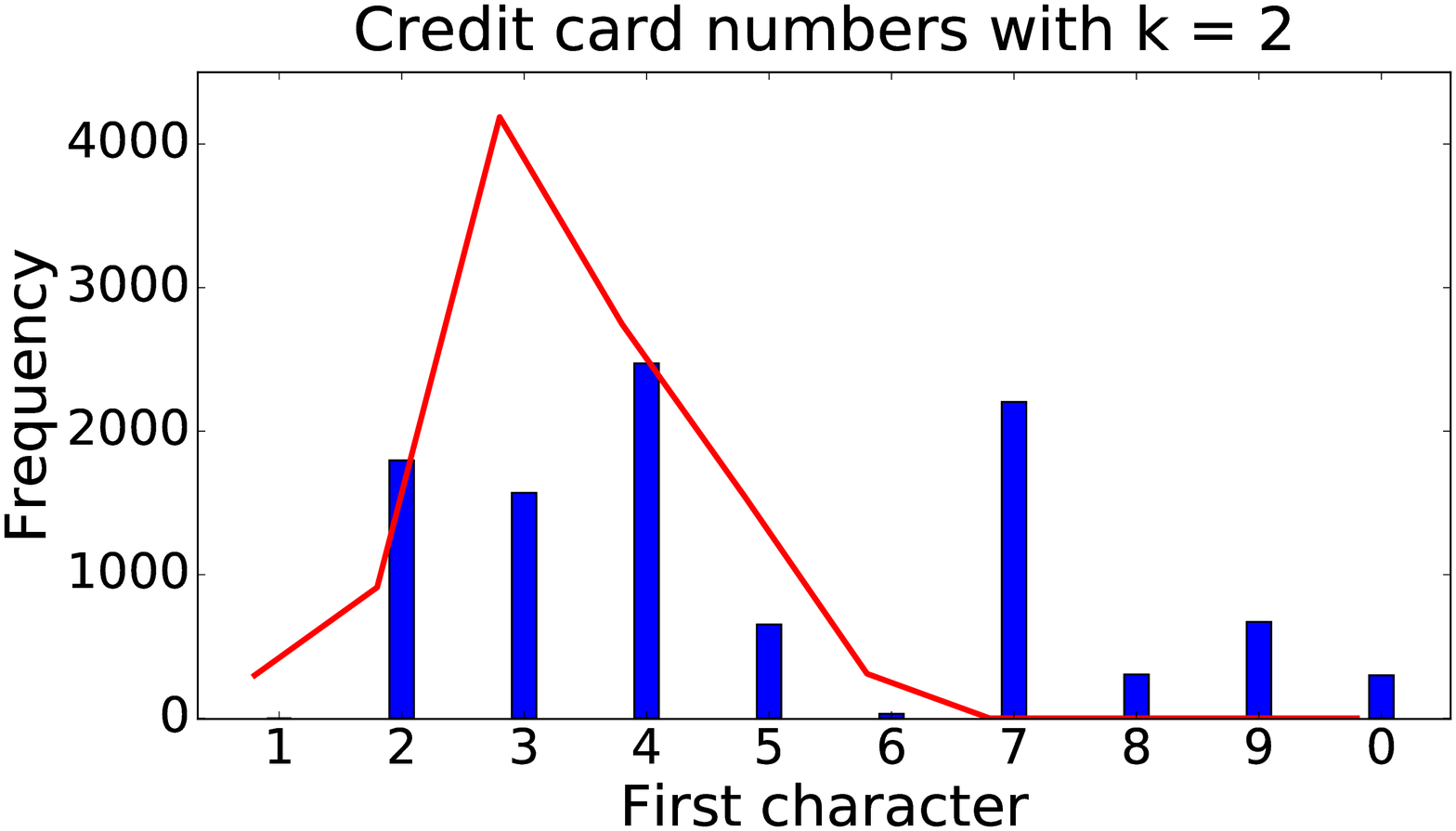}
	\hspace{2mm}
	\includegraphics[width=0.22\textwidth]
	{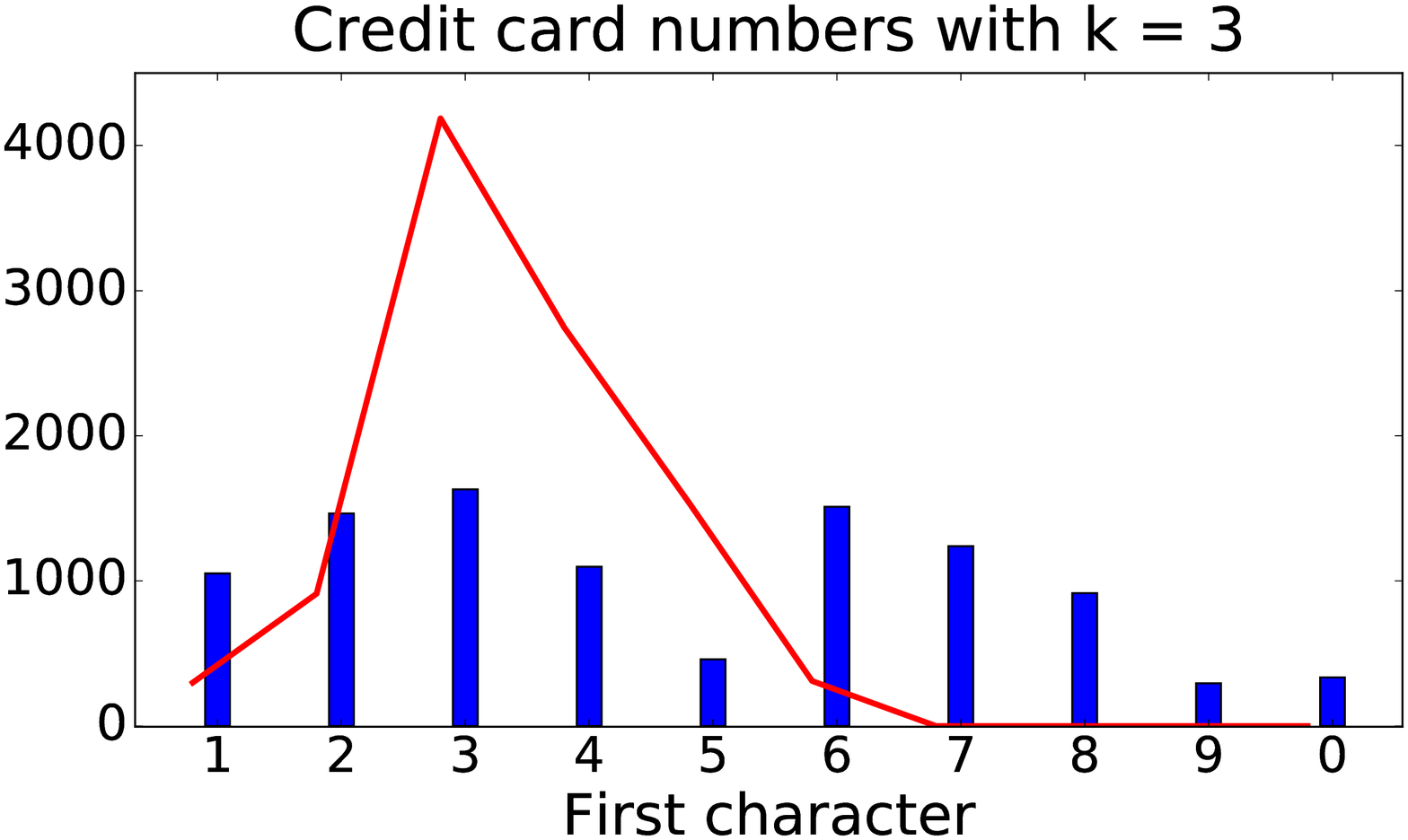}
	\hspace{2mm}
	\includegraphics[width=0.22\textwidth]
	{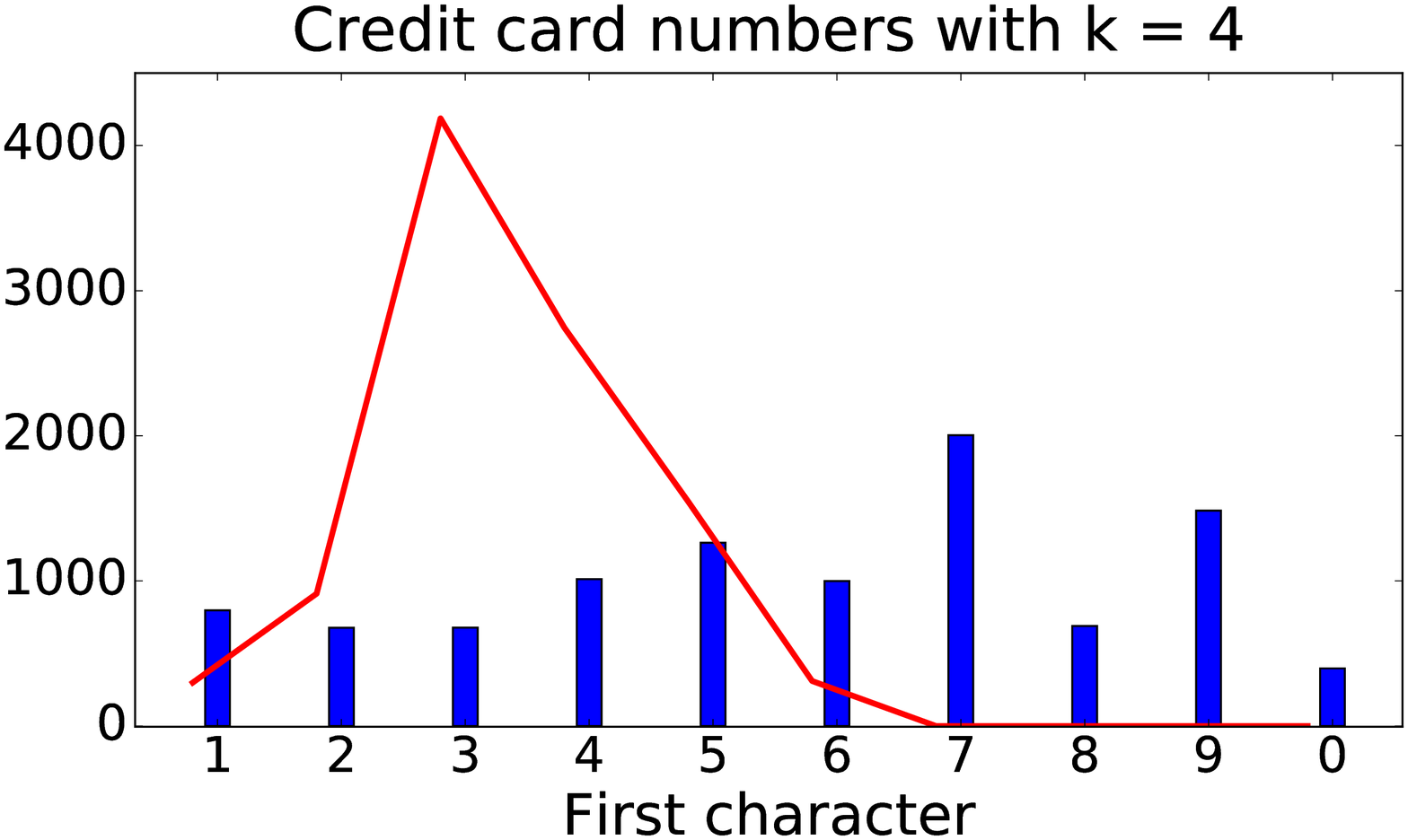}
	\hspace{2mm}
	\includegraphics[width=0.22\textwidth]
	{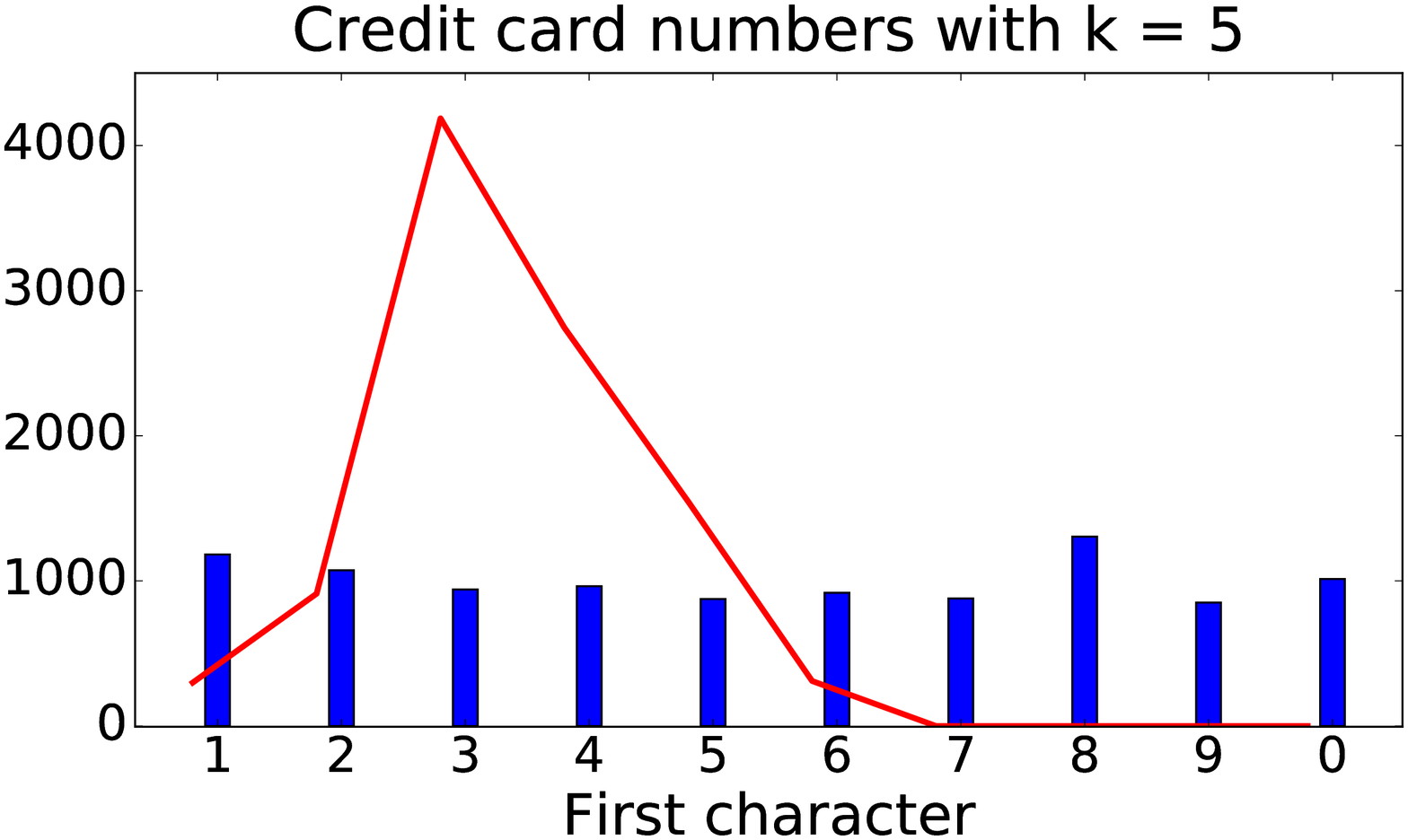}
	~ \\[2mm]
	\includegraphics[width=0.22\textwidth]
	{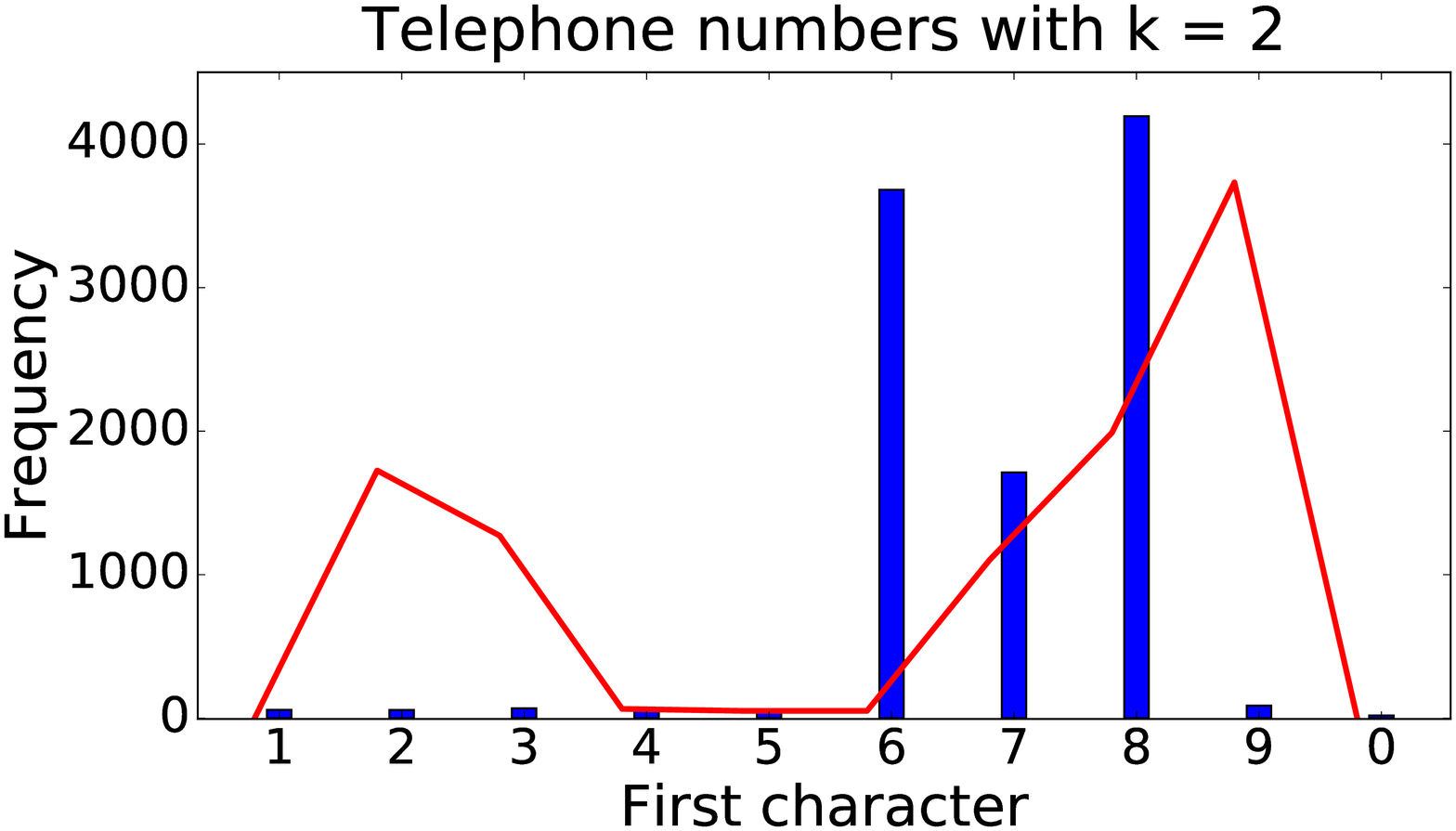}
	\hspace{2mm}
	\includegraphics[width=0.22\textwidth]
	{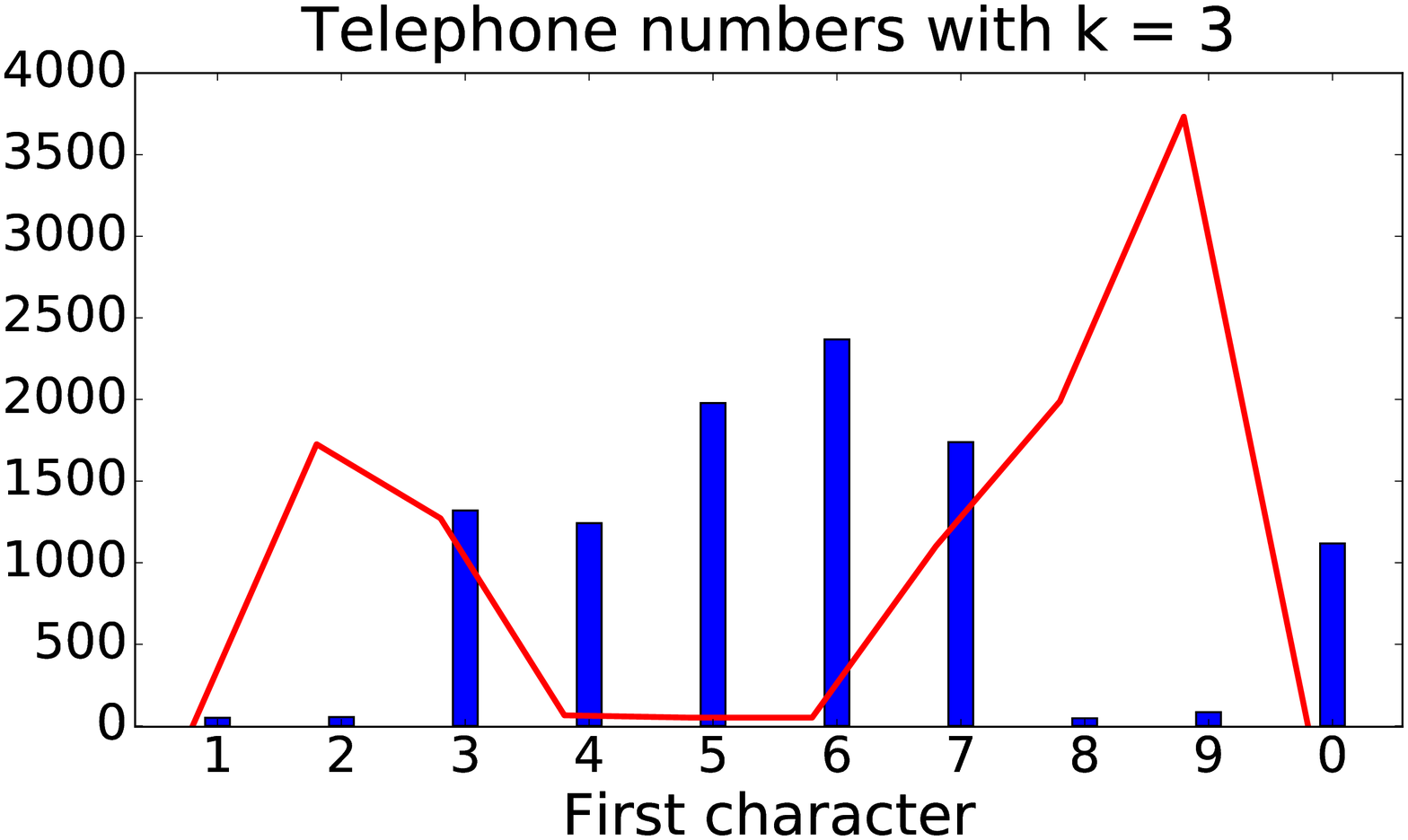}
	\hspace{2mm}
	\includegraphics[width=0.22\textwidth]
	{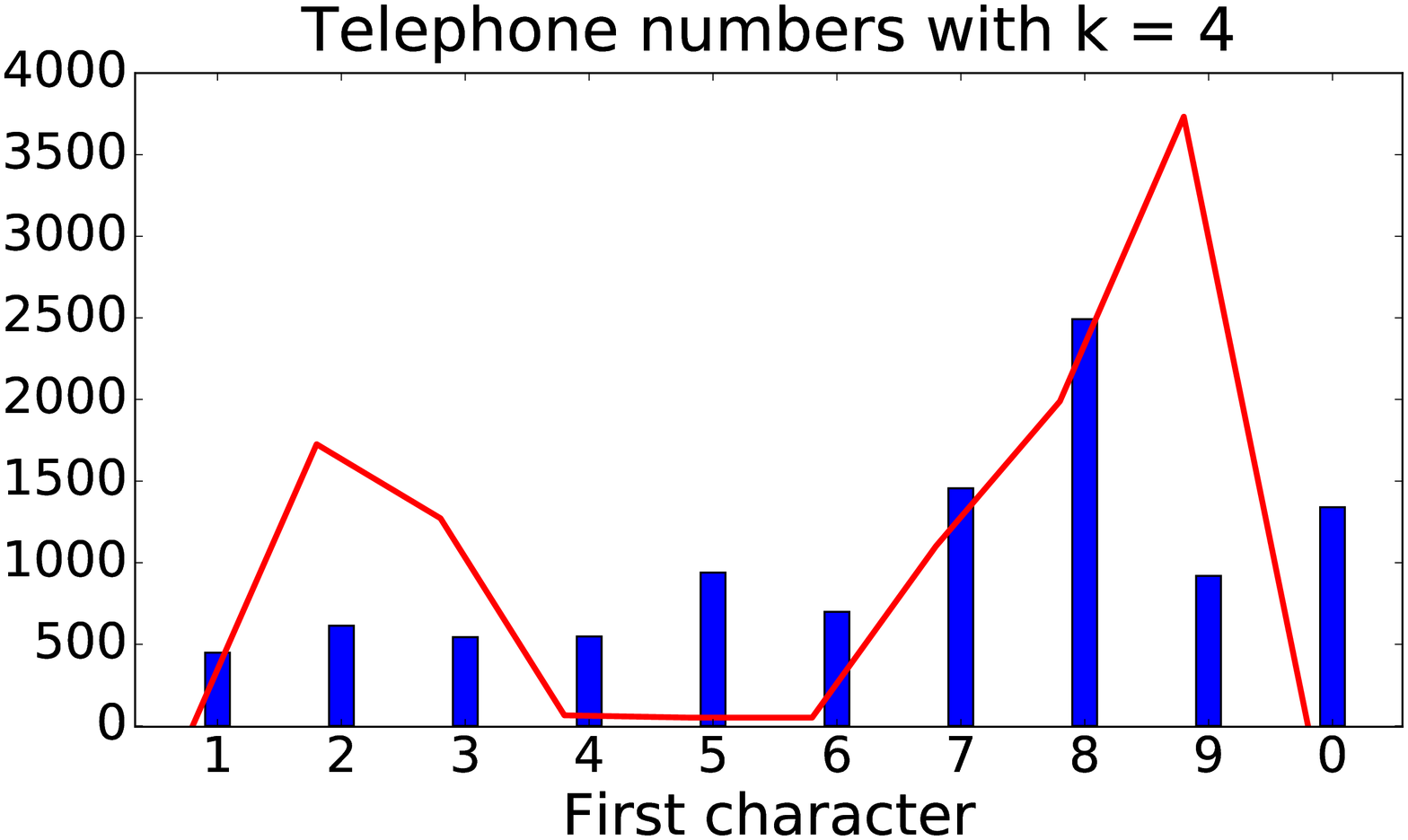}
	\hspace{2mm}
	\includegraphics[width=0.22\textwidth]
	{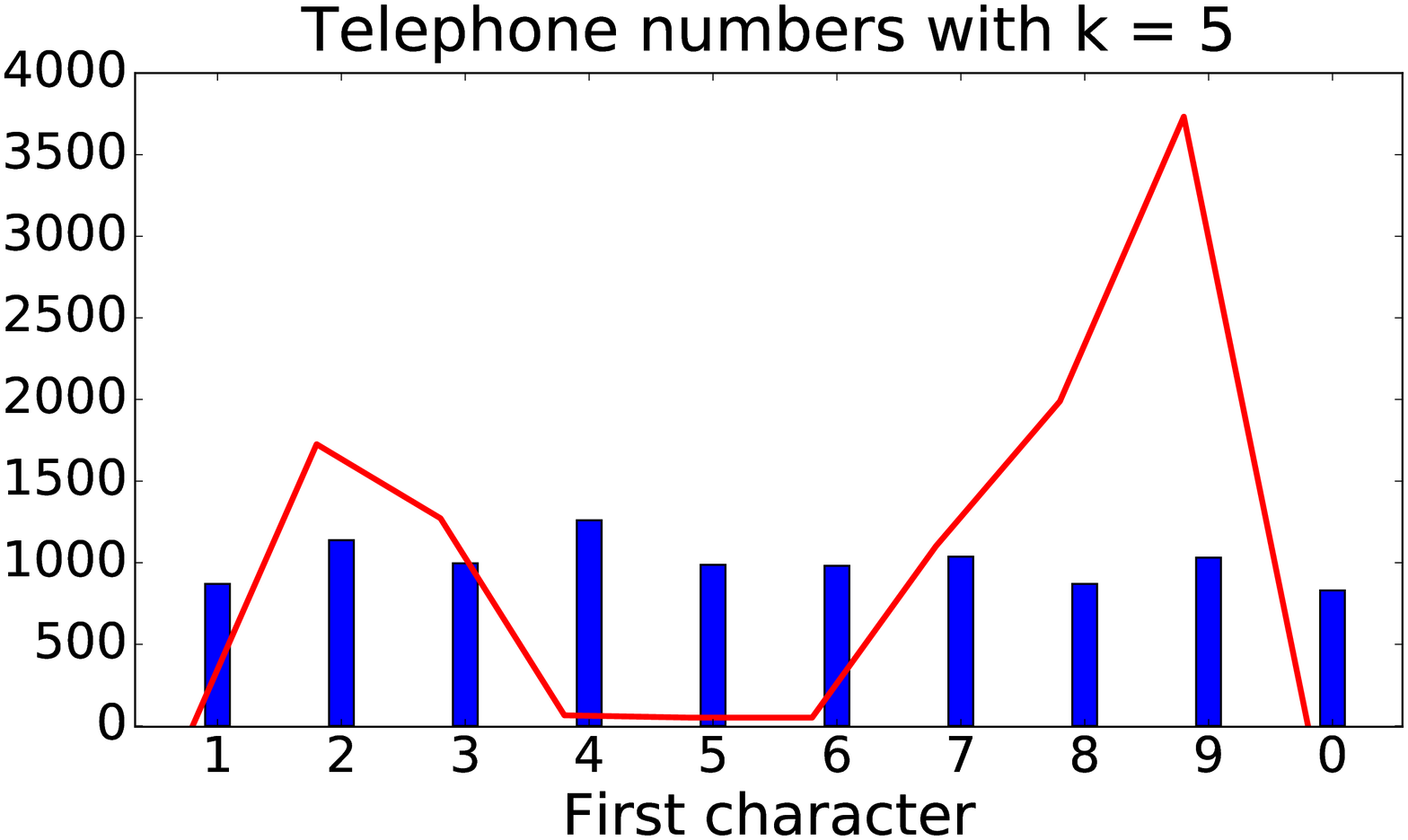}
	~ \\[2mm]
	\includegraphics[width=0.22\textwidth]
	{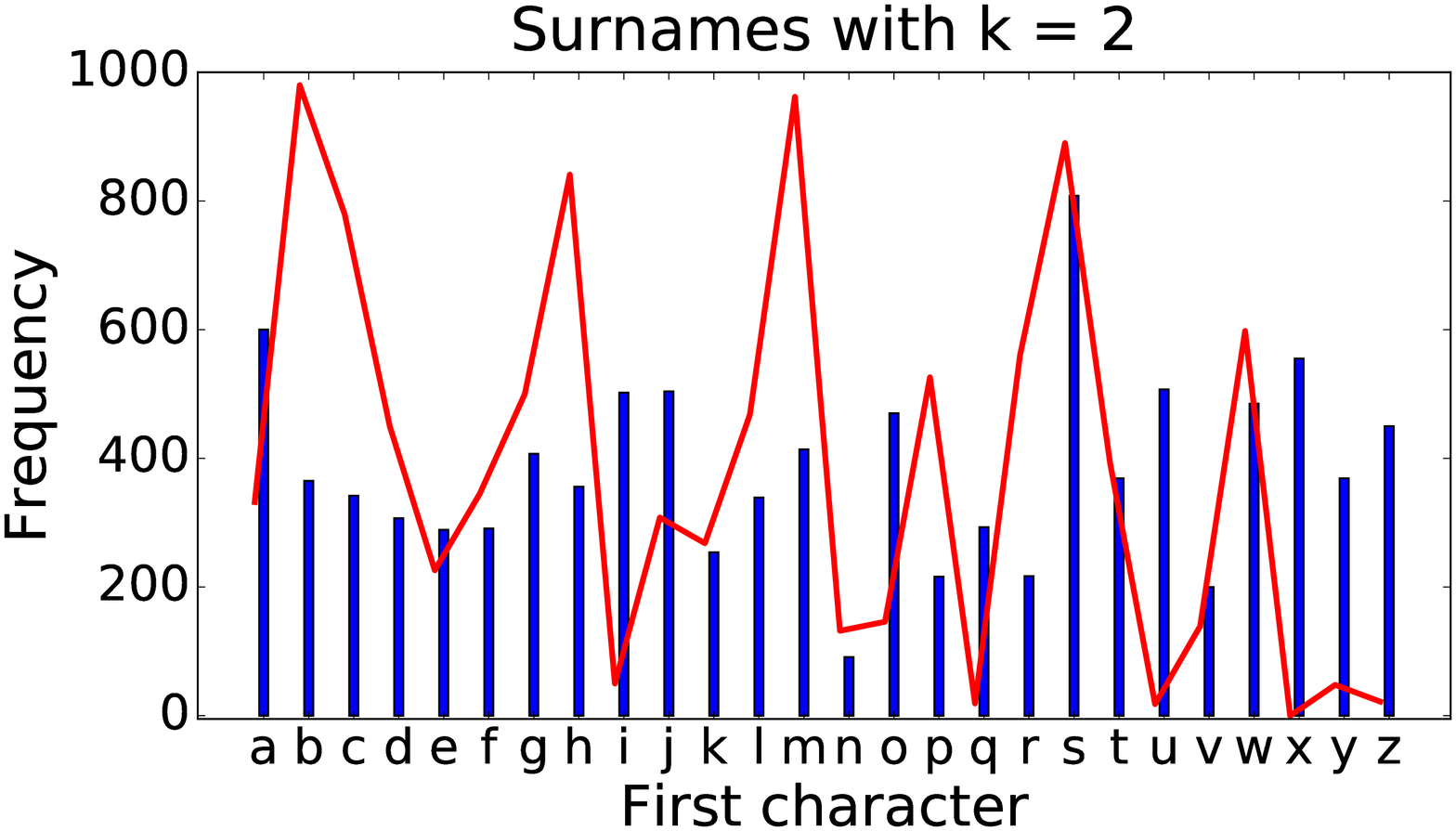}
	\hspace{2mm}
	\includegraphics[width=0.22\textwidth]
	{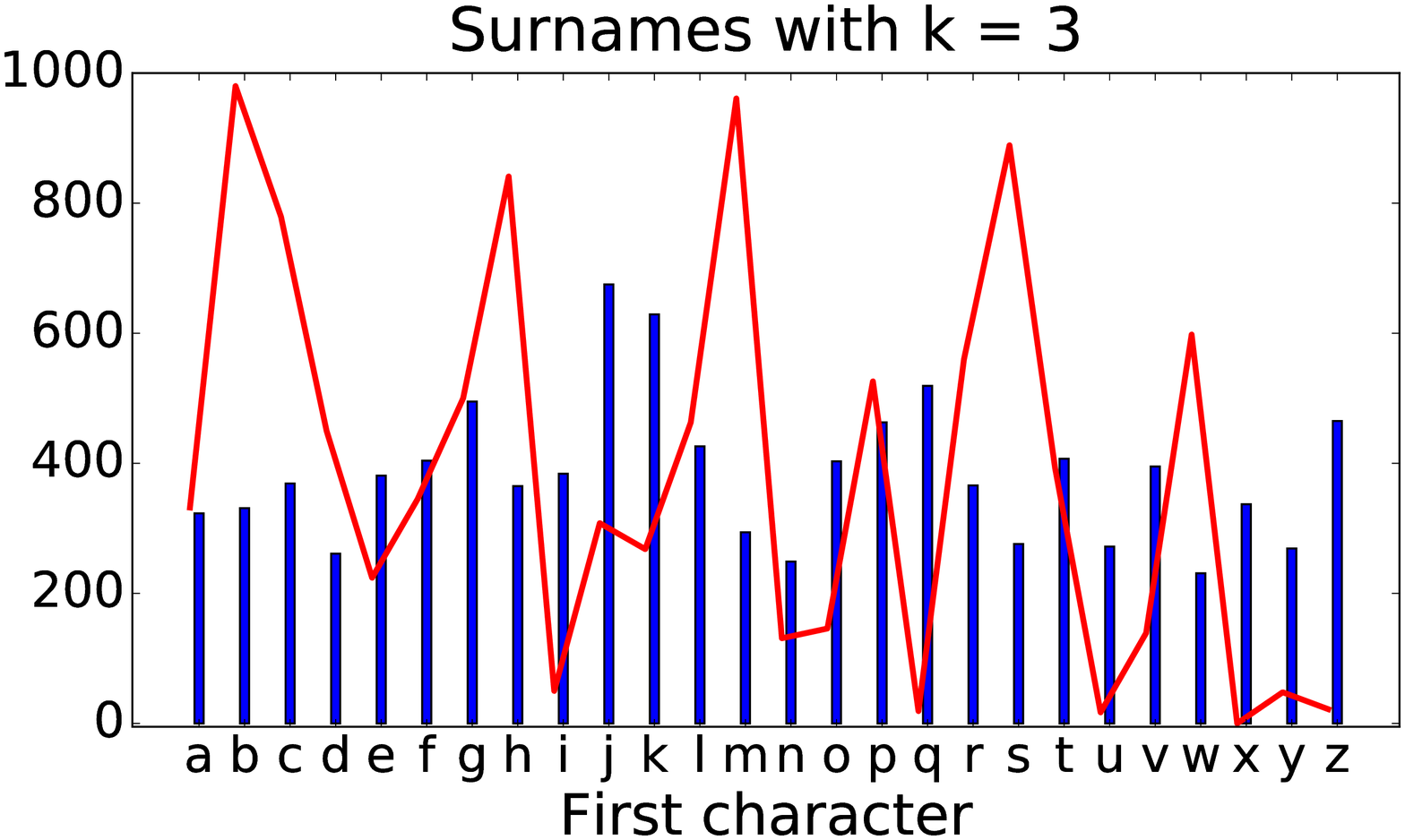}
	\hspace{2mm}
	\includegraphics[width=0.22\textwidth]
	{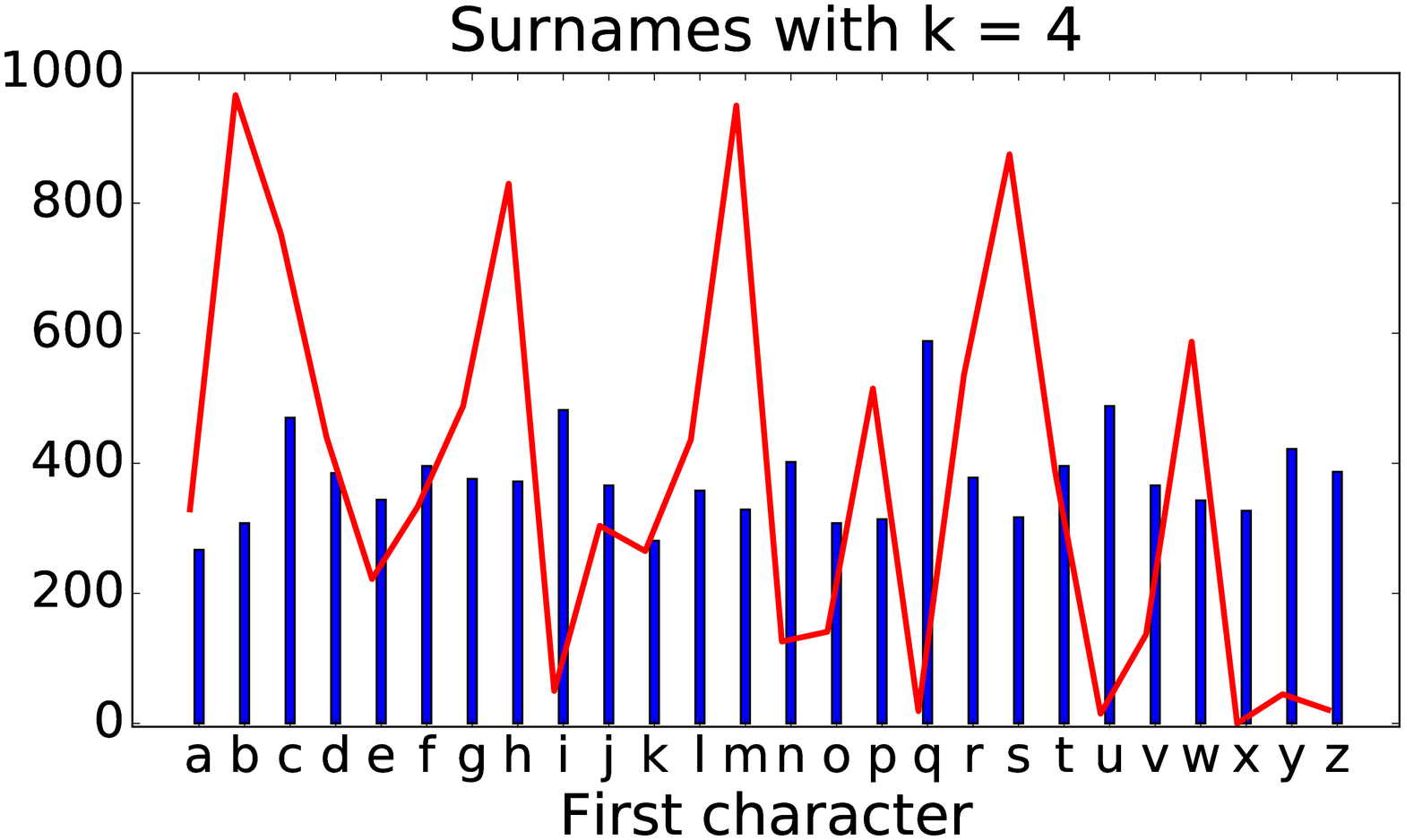}
	\hspace{2mm}
	\includegraphics[width=0.22\textwidth]
	{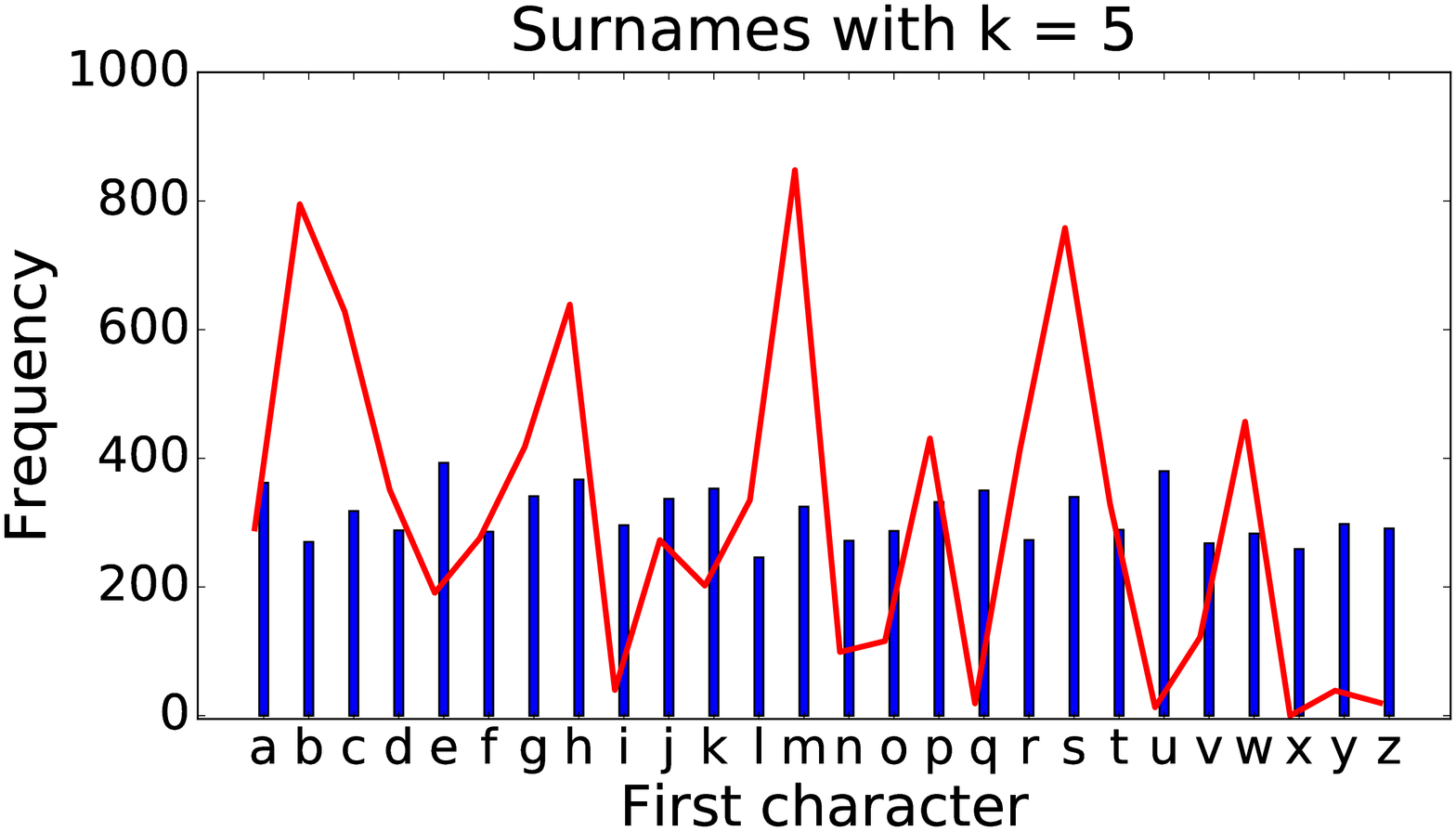}
	~ \\[2mm]
	\includegraphics[width=0.22\textwidth]
	{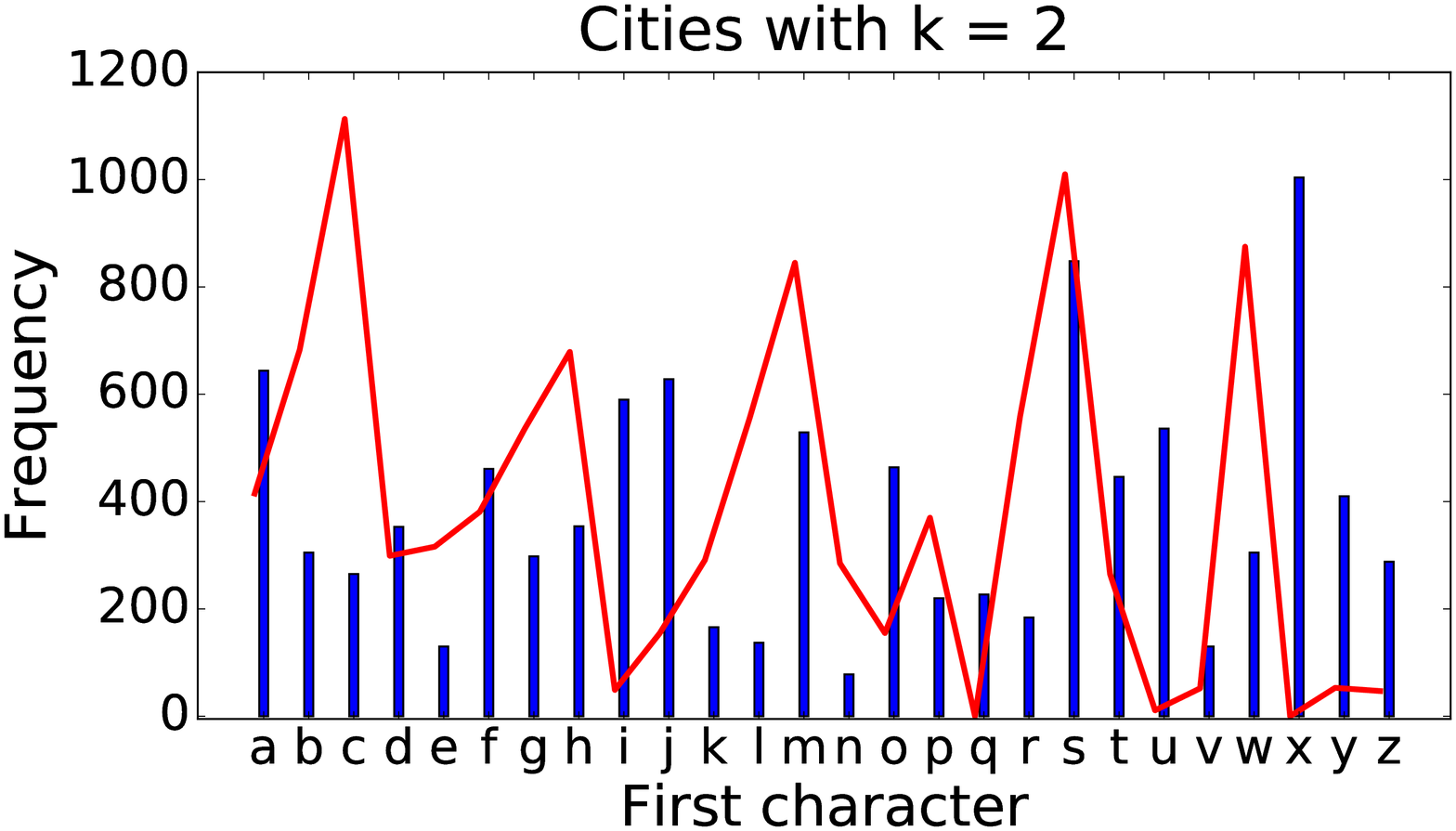}
	\hspace{2mm}
	\includegraphics[width=0.22\textwidth]
	{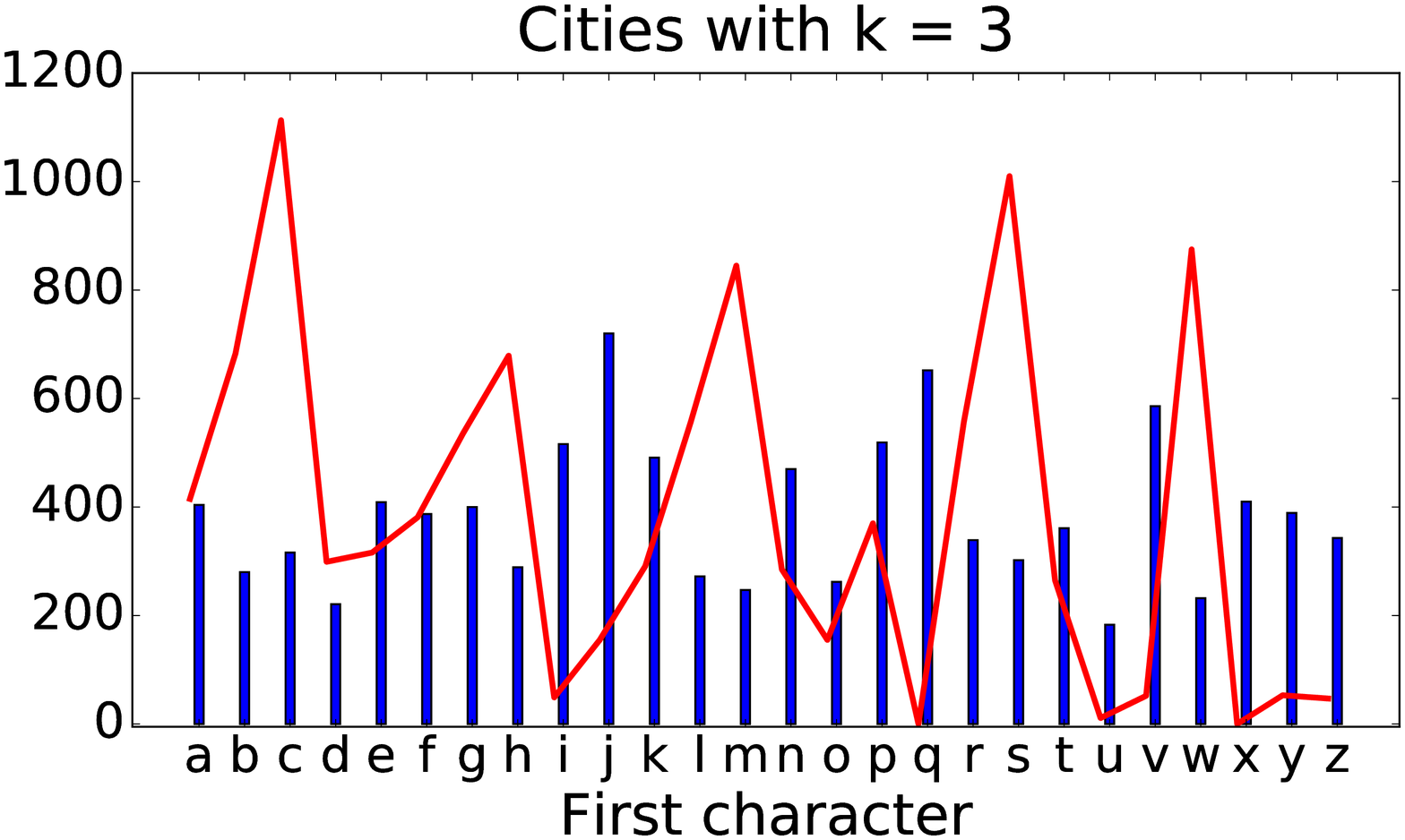}
	\hspace{2mm}
	\includegraphics[width=0.22\textwidth]
	{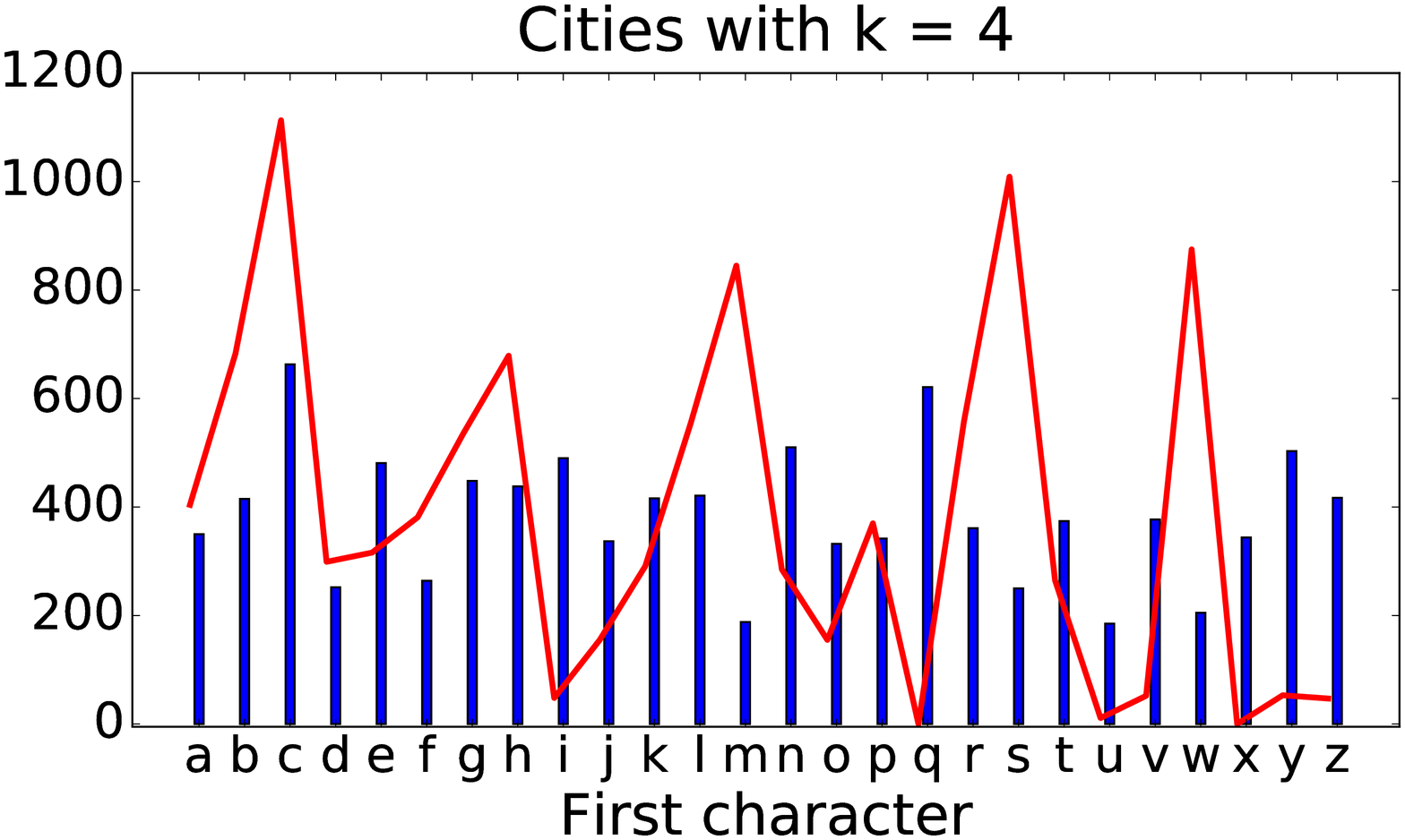}
	\hspace{2mm}
	\includegraphics[width=0.22\textwidth]
	{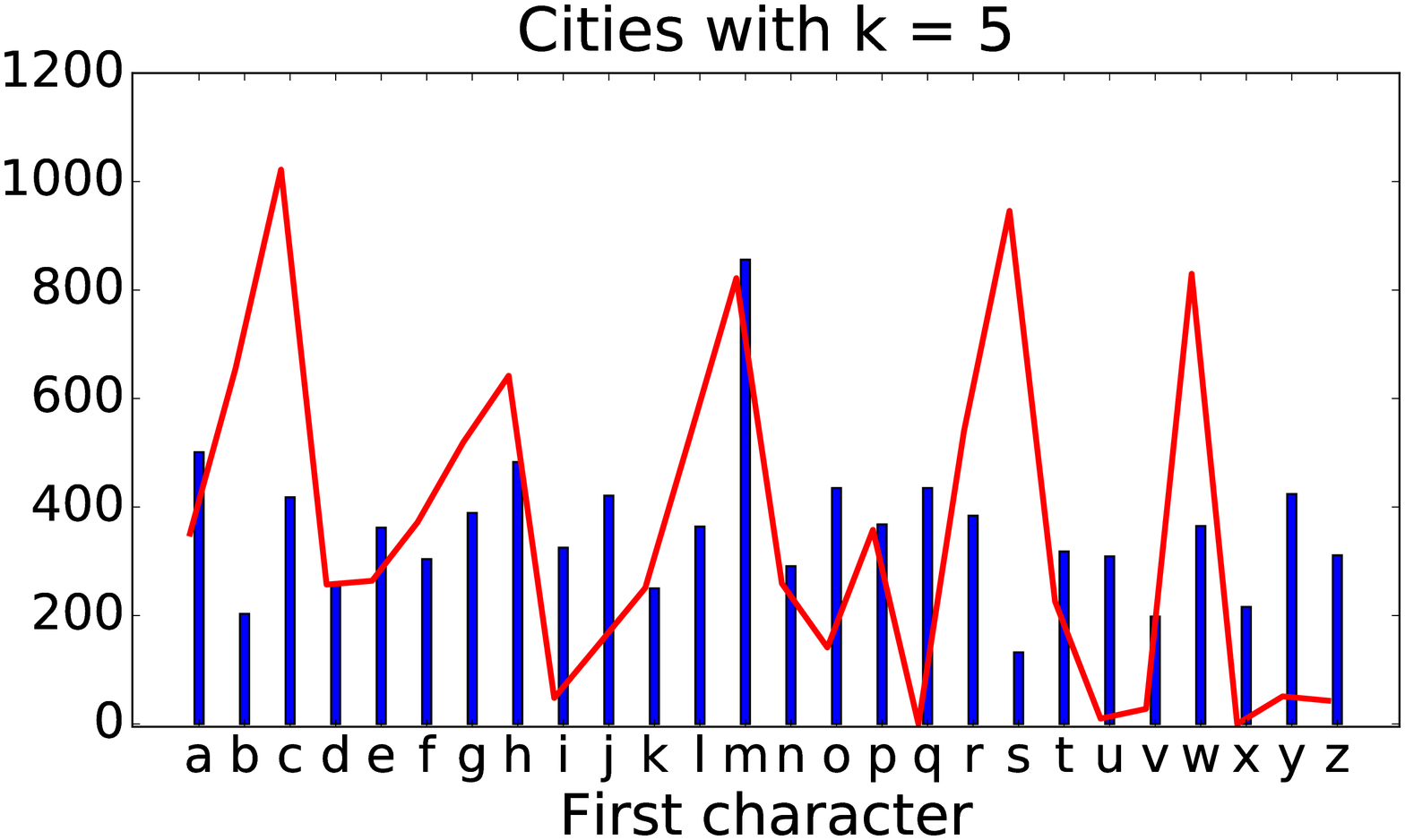}
	~ \\[2mm]
	\includegraphics[width=0.22\textwidth]
	{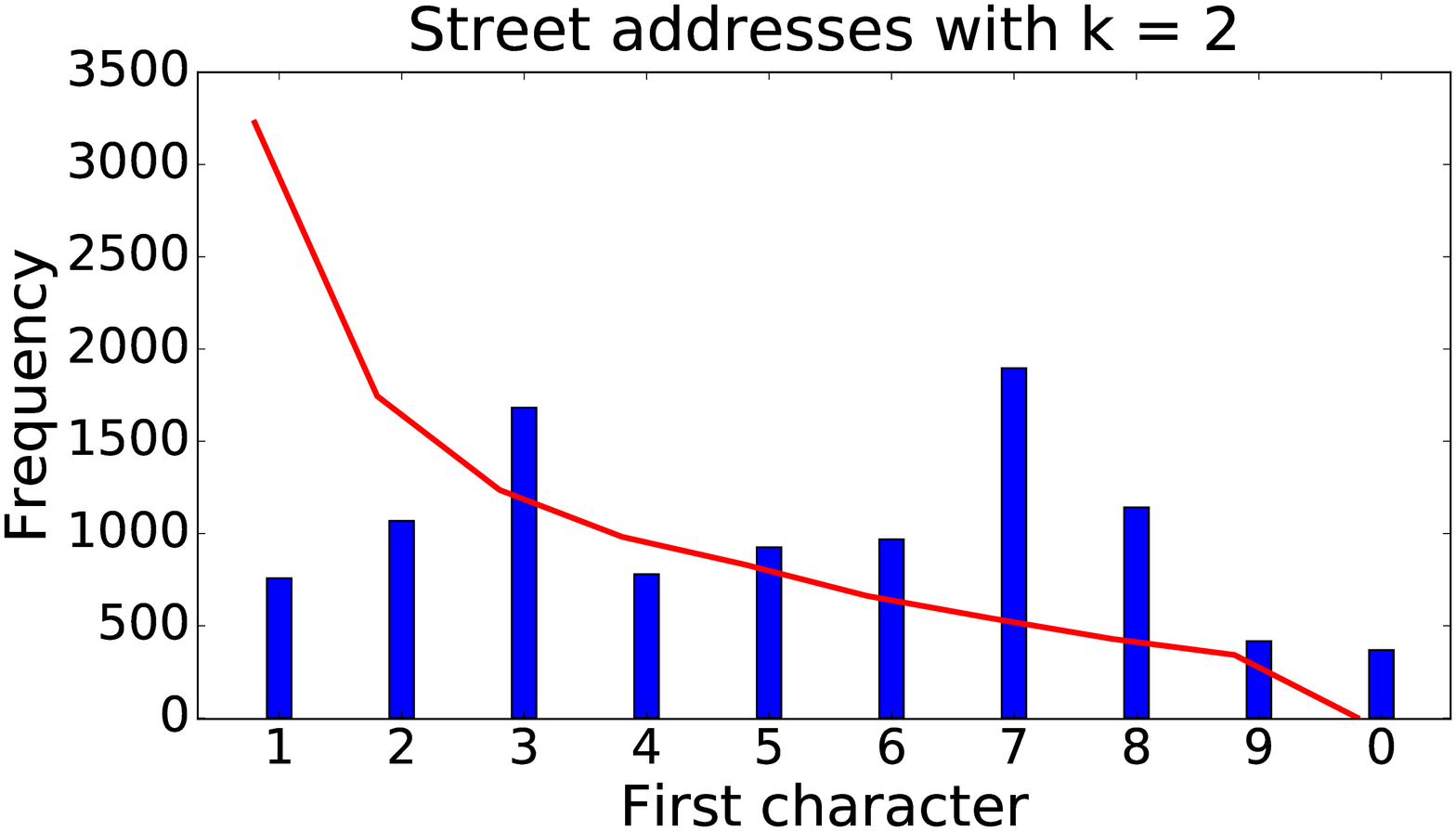}
	\hspace{2mm}
	\includegraphics[width=0.22\textwidth]
	{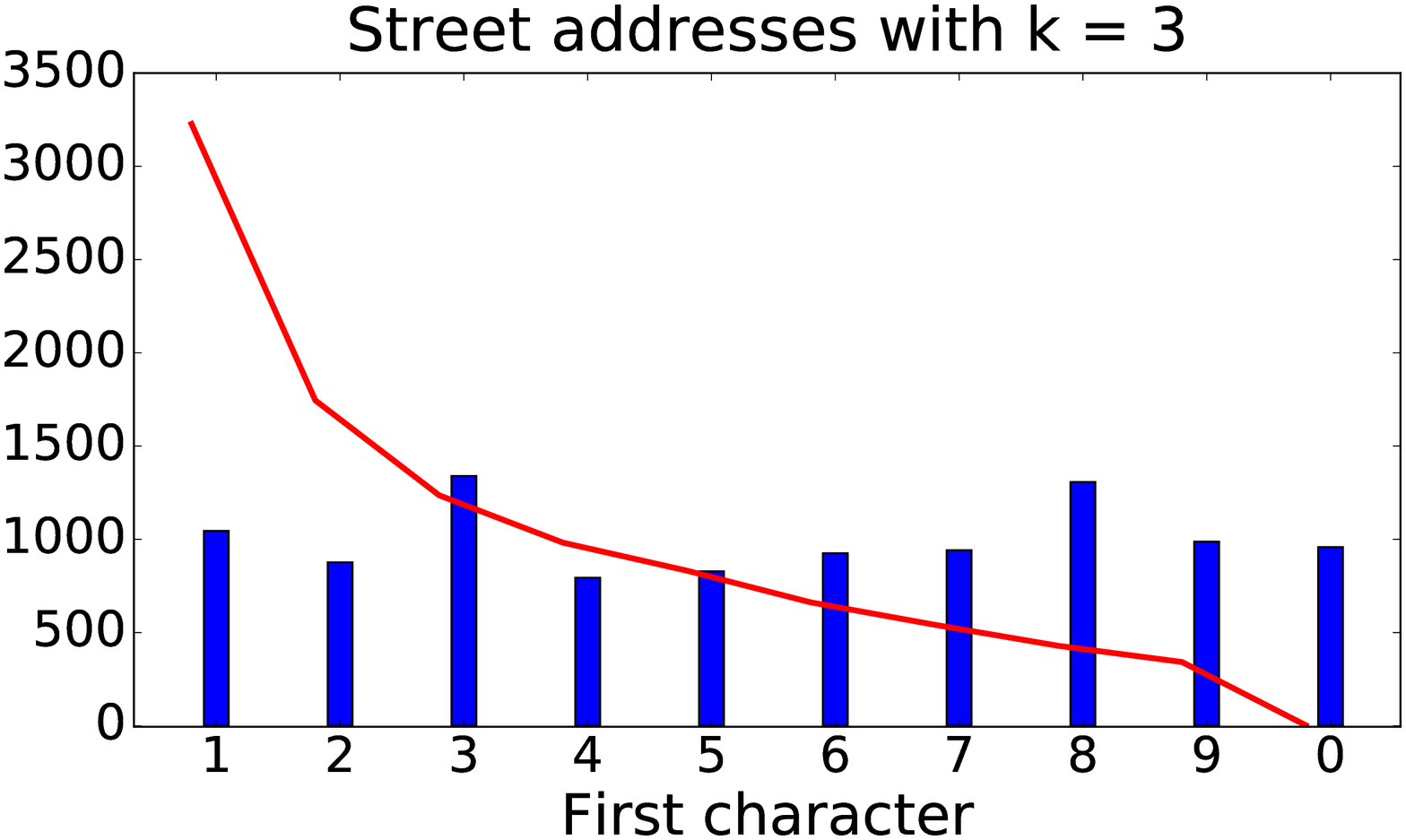}
	\hspace{2mm}
	\includegraphics[width=0.22\textwidth]
	{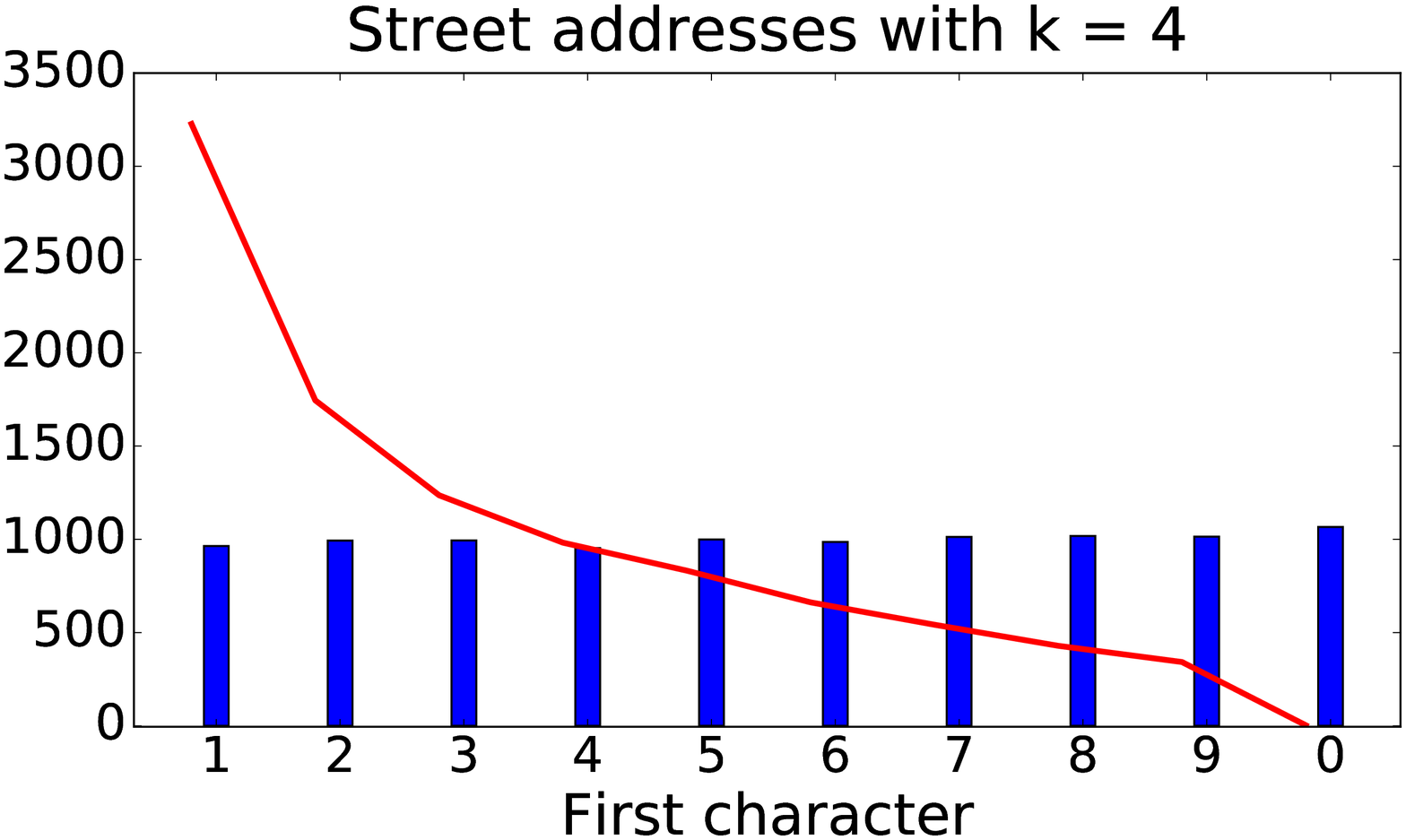}
	\hspace{2mm}
	\includegraphics[width=0.22\textwidth]
	{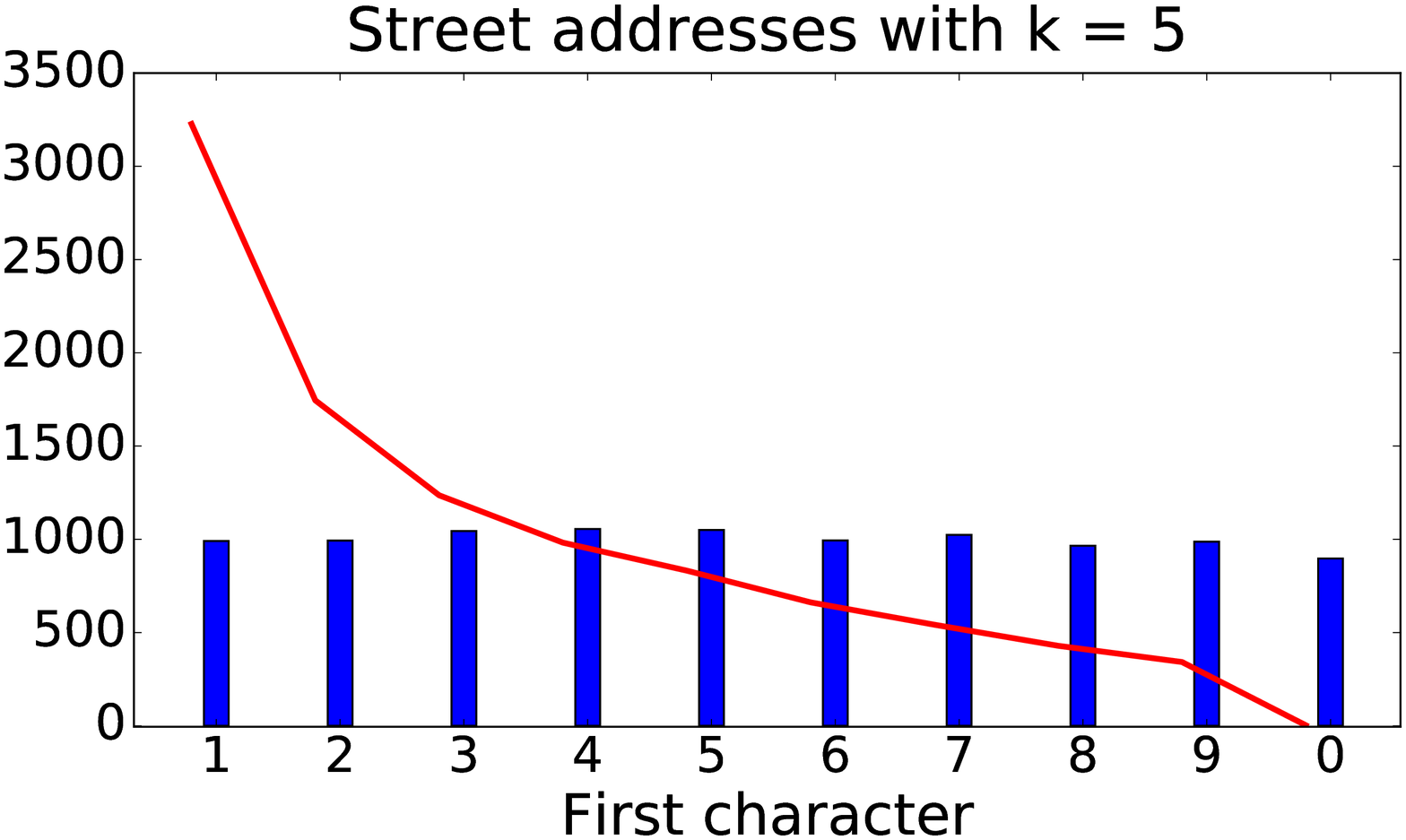}
	~ \\[2mm]
	\includegraphics[width=0.22\textwidth]
	{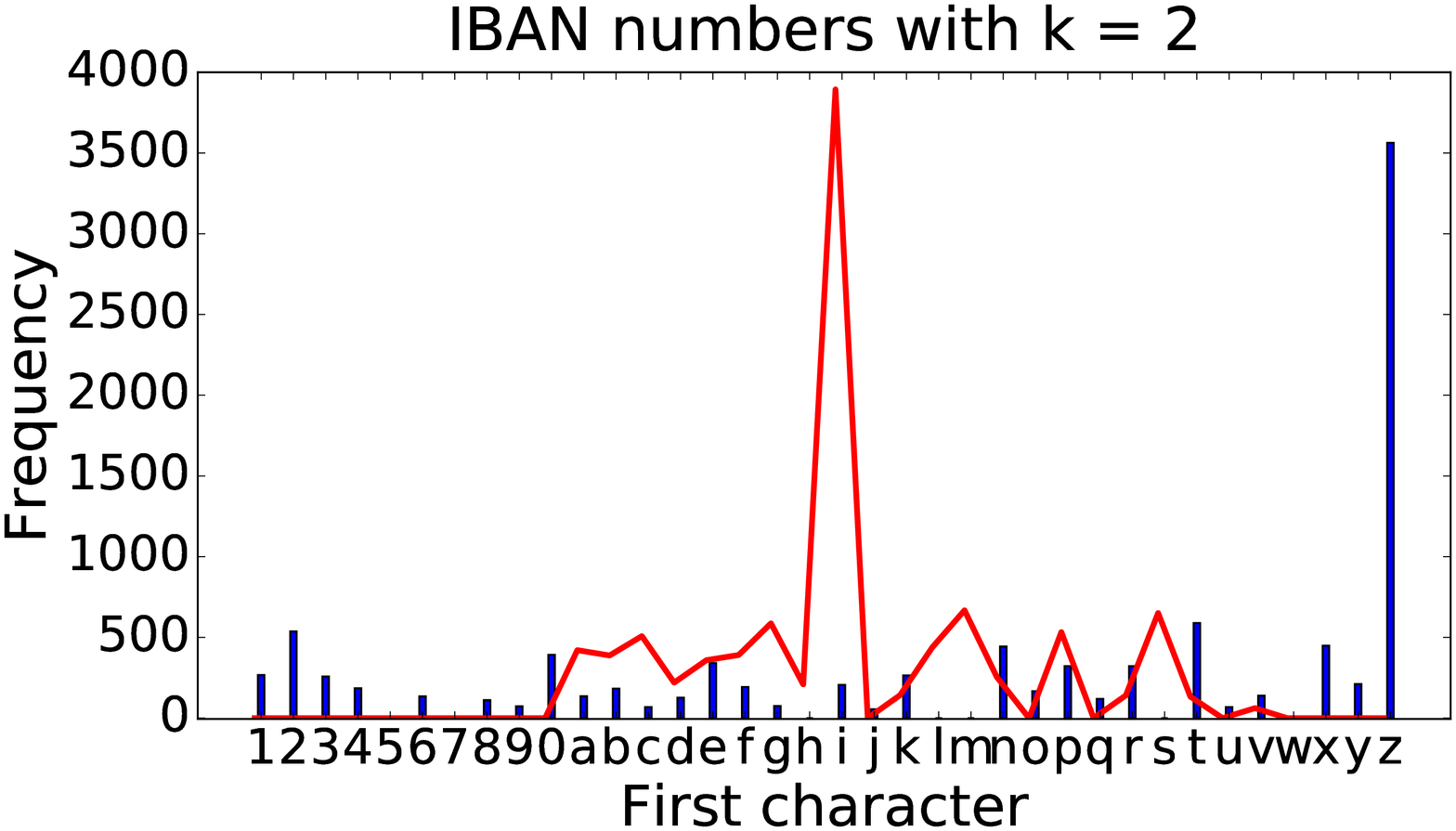}
	\hspace{2mm}
	\includegraphics[width=0.22\textwidth]
	{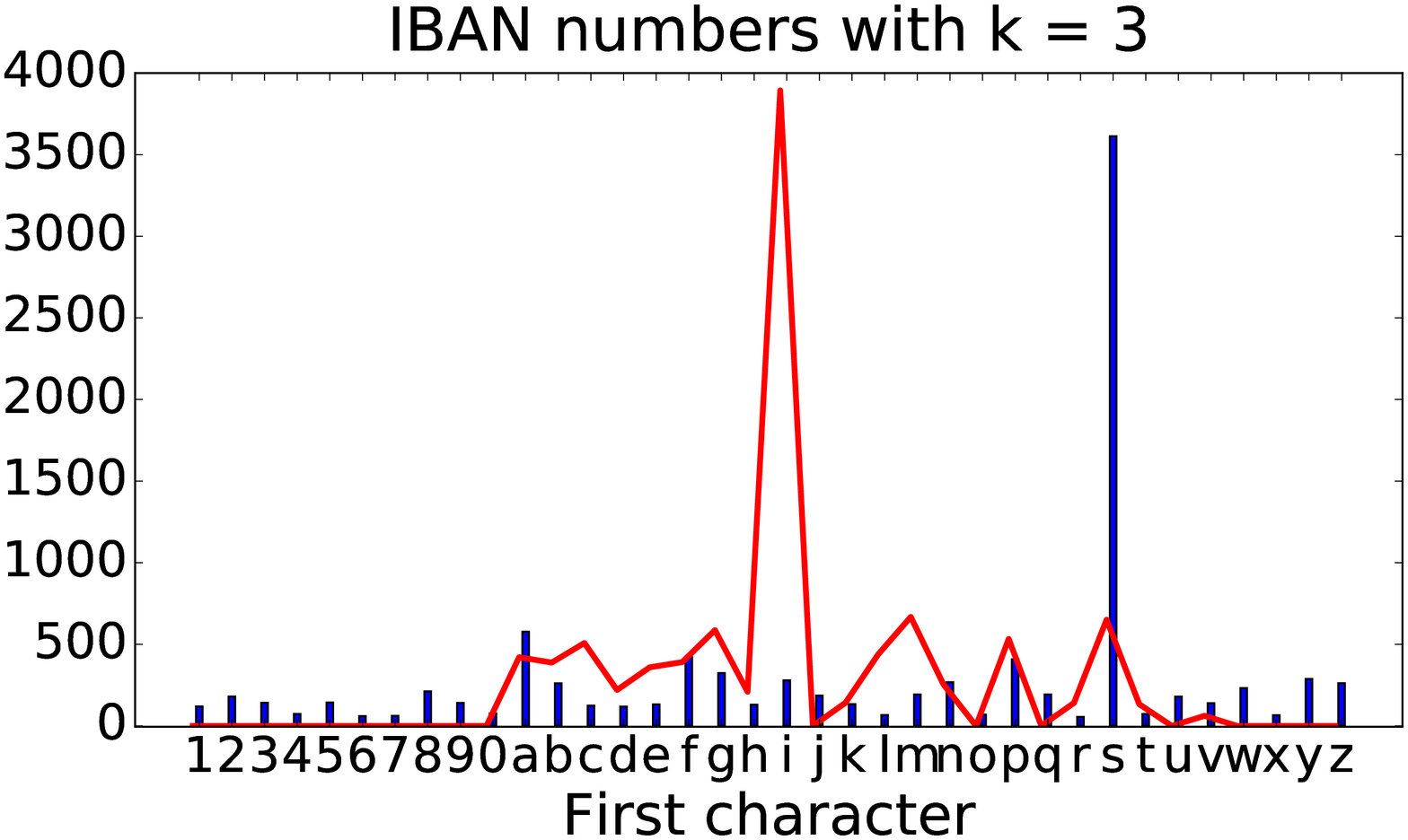}
	\hspace{2mm}
	\includegraphics[width=0.22\textwidth]
	{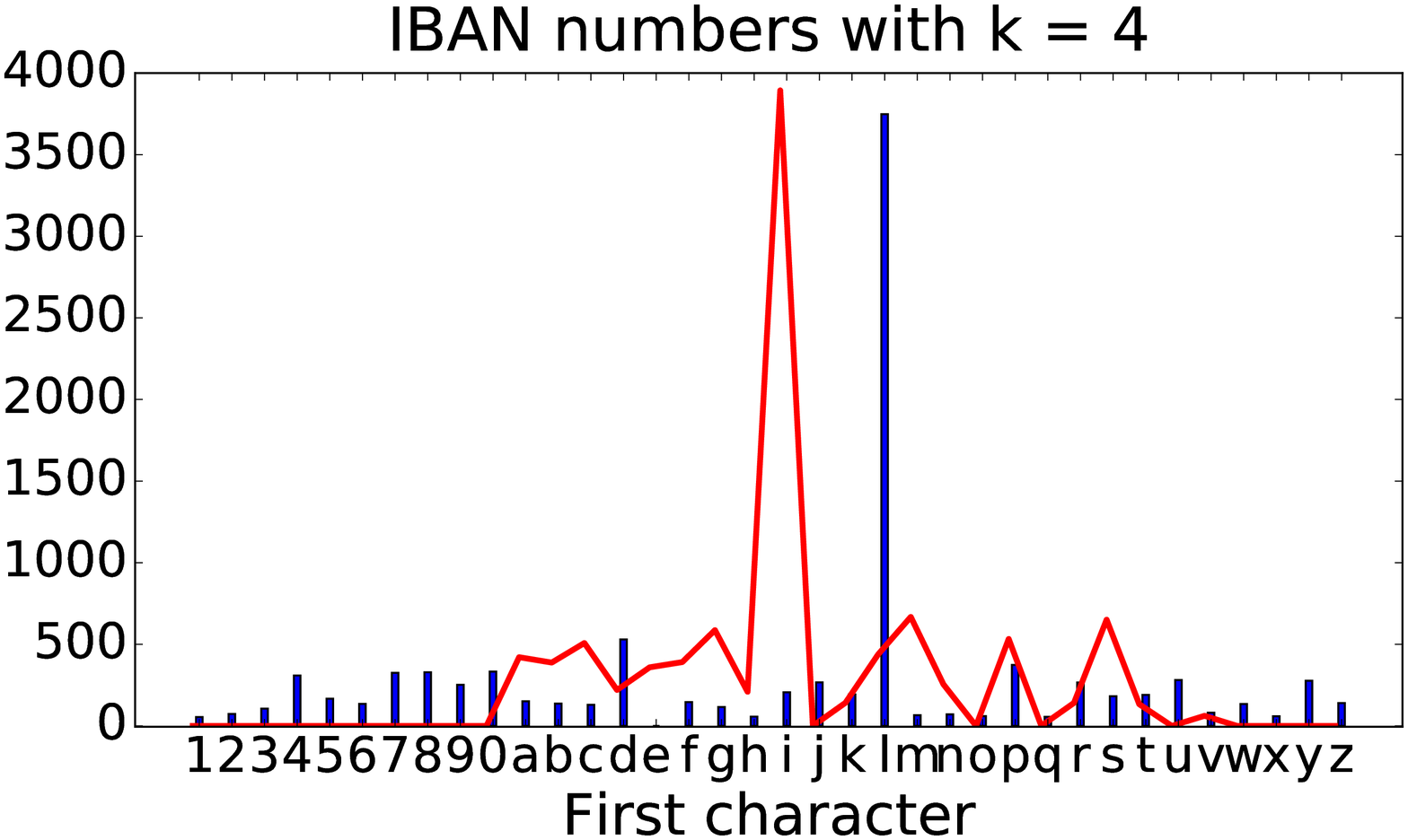}
	\hspace{2mm}
	\includegraphics[width=0.22\textwidth]
	{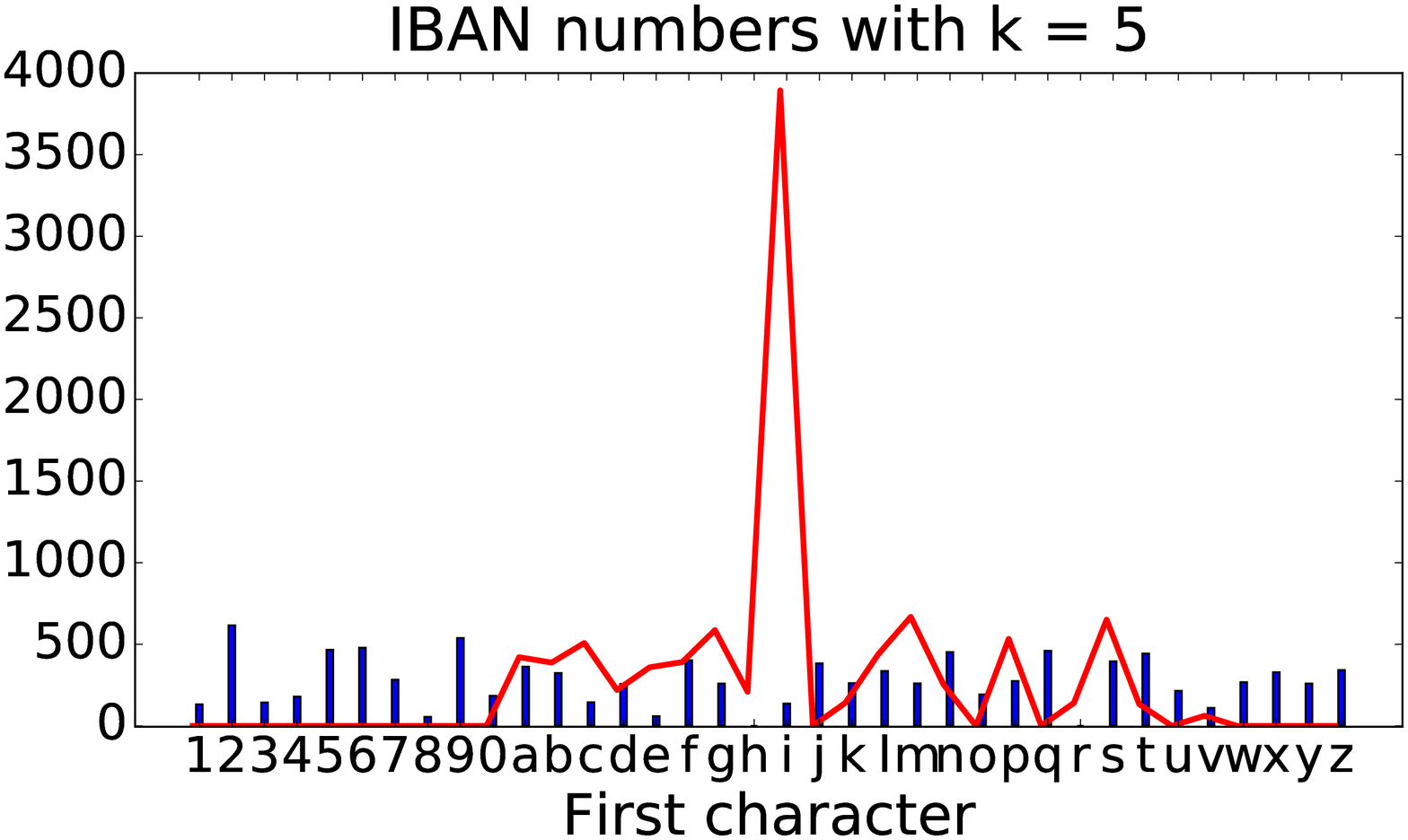}
	\caption{Frequency distributions of the first characters in
	  strings with different \emph{k} from 2 (left) to 5 (right).
	  The red lines show the original first character distributions
	  while the blue bars show the distributions of the first
	  character encodings, with \emph{n\,=\,}$|\Sigma|$.
	  As can be seen, as \emph{k} increases the distributions of
	  the first character encodings become more uniform.
	  \label{fig:freq-dist}}
\end{figure}

We compared our approach with Bloom filter (BF) encoding as commonly
used in PPRL~\citep{Sch09,Vat17}. The BFs were generated by
converting each string into a q-gram set with $q=2$, and by then
hashing each q-gram set into one BF of length 1,000 bits (a commonly
used BF length for PPRL~\citep{Sch09}) per q-gram set. We used
different optimal numbers of hash functions $k_{opt}$ that lead to
the smallest number of false positives~\citep{Vat16b}: 46 for credit
card and 30 for IBAN numbers, 116 for surnames, 87 for city names,
36 for street addresses, and 77 for telephone numbers.

As a second baseline method we used a tabulation hashing based
approach for PPRL recently proposed by Smith~\citep{Smi17}, where
again q-gram sets are hashed into bit arrays using a tabulation
approach which provides min-hashing properties~\citep{Pat11}. This
approach was shown to calculate more accurate similarities. We
used 8 tabulation keys each of 64 bits length to generate one bit
array of length 1,000 bits to encode one string.

There are many different techniques to calculate similarities
between strings~\citep{Chr12}. Because of the encodings used in the
three methods we compare, we need to employ different such string
matching techniques. We are however not interested in the absolute
similarities calculated between two strings; rather we want to know
if for the same string pair the same similarity method applied on
the unencoded and the encoded strings gives the same similarity
value or not.
For our suffix tree based approach, as described in
Sect.~\ref{sec:overview}, we calculated the longest common
sub-string similarity using Eqn.~(\ref{eqn:lcs}) on both unencoded
and encoded suffix trees (both the basic and first character
encoding described in Sects.~\ref{sec:matching}
and~\ref{sec:first_encoding}, respectively). For BF encoding we
calculated the Dice coefficient similarity on the q-gram sets and
on BFs~\citep{Sch09}, while for tabulation based hashing we calculated
the Jaccard similarity on q-gram sets and on the bit arrays generated
by this encoding technique~\citep{Smi17}.

In Fig.~\ref{fig:sim-plots-k} we show scatter plots where the
horizontal axis shows unencoded similarities and the vertical axis
shows the corresponding encoded similarities. A pair of strings
where both the unencoded and the encoded similarities are the same
will generate a point in a scatter plot that is shown on the
diagonal, while any point off the diagonal shows differences in the
calculated similarities between unencoded and encoded strings. An
accurate (exact) privacy-preserving string similarity measure
should only result in pairs of similarities that are the same and
are therefore located on the diagonal.

As can be seen from Fig.~\ref{fig:sim-plots-k}, the similarities
calculated on suffix trees from both our encoding approaches always
result in the same similarities as calculated from unencoded suffix
trees.
This shows our approach does accurately calculate the longest common
sub-string similarities on encoded suffix trees in a
privacy-preserving manner where the DOs do not need to reveal their
sensitive plain-text strings to any other party. As can also be
seen, BF based Dice coefficient similarities can be much higher than
their corresponding q-gram based similarities especially for string
pairs that have only few q-grams in common. This is because BF
encoding introduces collisions where different q-grams are hashed to
the same bit positions. Similarly, tabulation based hashing leads
to inaccurate similarities being calculated, where this approach
leads to encoded similarities that are both above and below the
actual Jaccard similarities calculated on unencoded q-gram sets.

\begin{figure}[!t]
  \centering
  \includegraphics[width=0.42\textwidth]
  {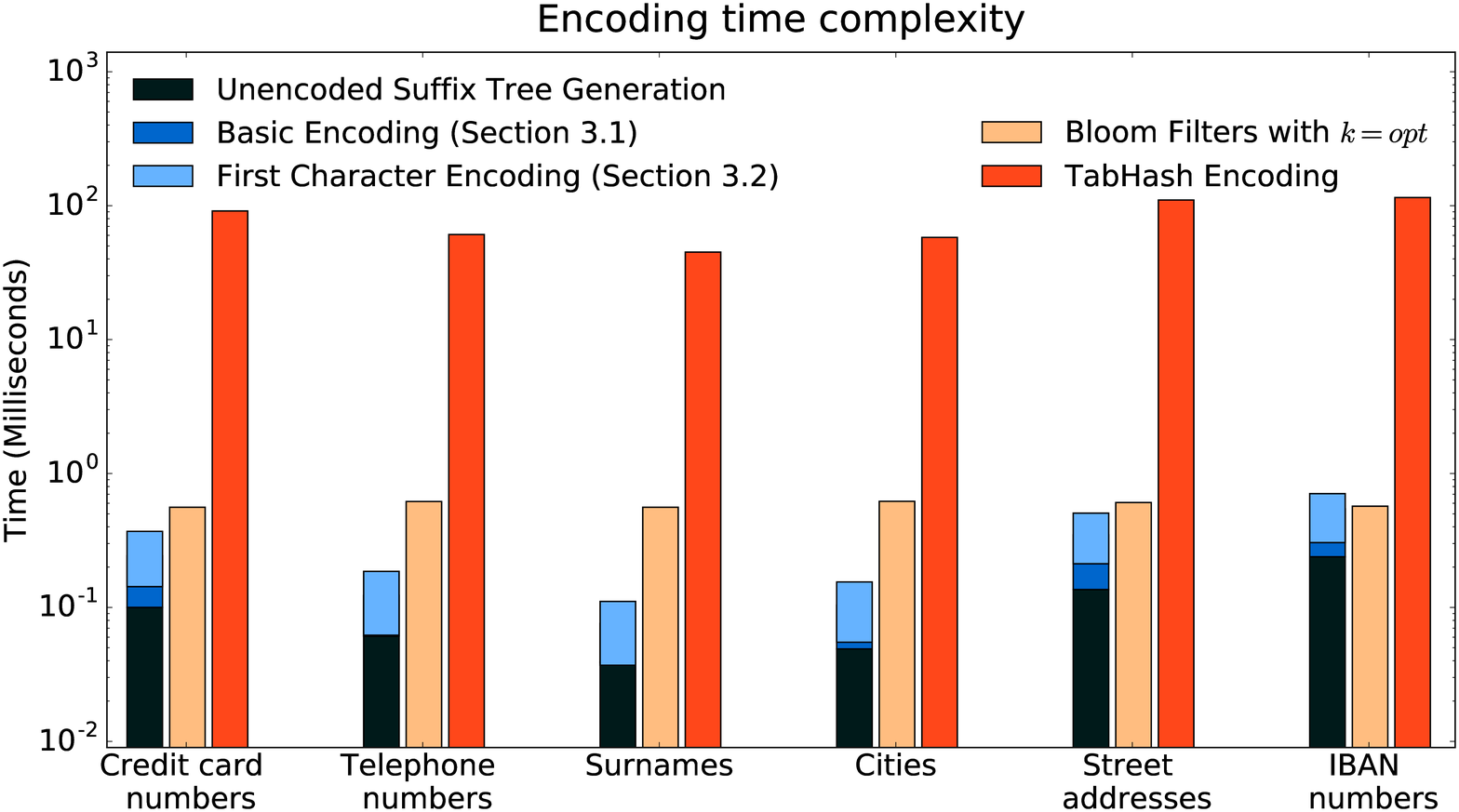}
  \hspace{10mm}
  \includegraphics[width=0.42\textwidth]
  {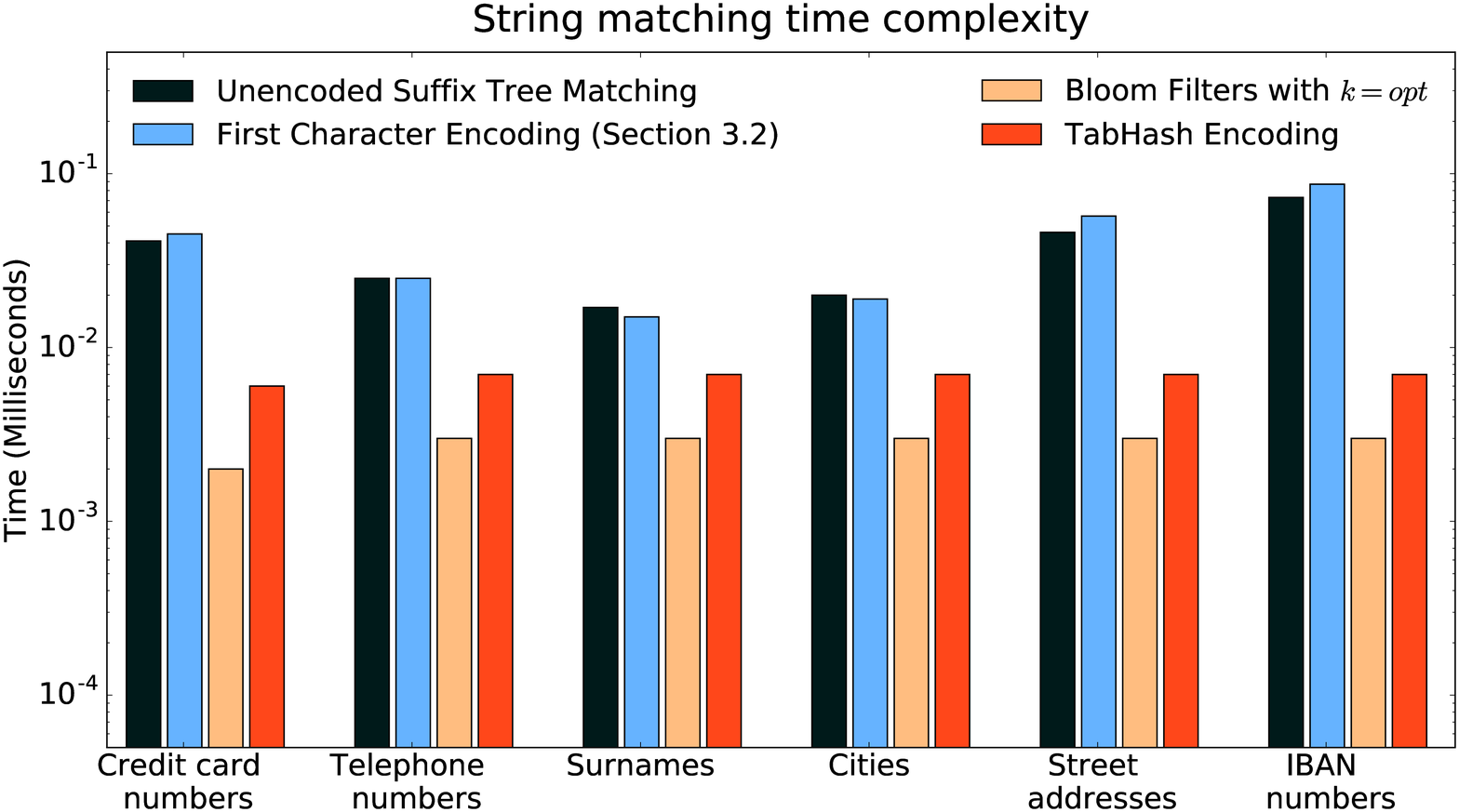}
  \caption{Comparison of run-times for encoding (left) and string
  matching (right) between our approach, Bloom filters, and
  tabulation based hashing (TabHash). Shown are average times for
  encoding one string and matching one string pair.
  } \label{fig:time-used}
\end{figure}

These issues will affect the similarities calculated between strings
for both BF encoding and tabulation based hashing, and therefore
affect the quality of matched strings and the resulting quality of
any follow-up analysis or investigation that is based on these
matched strings. Of serious concern would be if wrong high BF or
tabulation hashing similarities lead to falsely matched individuals
in the context of fraud detection or national security.

In Fig.~\ref{fig:freq-dist}, we show the frequency distributions of 
the first characters of strings between the original first character
distributions and the encoded first characters after our first
character encoding from Sect.~\ref{sec:first_encoding} has
been applied. As can be seen, our first character encoding method
results in more uniform or significantly changed frequency
distributions of the first characters in strings, where these
distributions depend on the value of $k$, the number of first
characters to use in the encoding. As we discussed in
Sect.~\ref{sec:accuracy_analysis}, the larger $k$ the more uniform
the frequency distributions of these first character encodings
become.


Finally, in Fig.~\ref{fig:time-used} we show run-times for encoding
and string matching. As can be seen, our approach to convert strings into
suffix trees and encoding them using chained hash encoding, as well
as re-hashing the first characters, is faster or equally fast as BF
encoding. Both our encoding approach and BF encoding are much faster
than tabulation hashing which requires significantly more hash
encodings. On the other hand, our encoding approach is around one to
almost two magnitudes slower in the string matching phase than the
very efficient bit array based baseline methods. This is expected
because our approach requires the individual comparison of hash
codes for each position in a suffix compared to the highly efficient
single bit-wise operations on bit arrays. We believe this is a price
worth paying given the accurate and privacy-preserving longest
common sub-string similarities our method can calculate.

\section{Conclusions and Future Work}
\label{sec:conclusions}

We have presented a novel privacy-preserving string matching
technique based on suffix trees that allows the accurate and
efficient calculation of longest common sub-string based string
similarities. Our approach encodes strings into suffix trees such
that no re-identification of the full input string is possible,
and neither can a frequency attack be mounted on individual
character encodings. The experimental evaluation has shown that
our approach results in the same string similarities as on
unencoded suffix trees, while commonly used Bloom filter
encoding and tabulation based hashing will lead to potentially
much higher or lower similarities between encoded strings. 

As future work we aim to conduct a more formal analysis of the
privacy of our approach and investigate different counter-measures
that can be applied upon our approach to reduce the amount of 
information that can be learned by an attacker by conducting a 
graph similarity analysis. 



\balance

\section*{Acknowledgements}

This work was partially funded by the Australian Research Council
under Discovery Project DP160101934. The authors like to thank Alex
Antic for discussions and contributions to the experimental design.



\bibliography{paper}

%
% ...

\end{document}